\documentclass[%
 aip,
 amsmath,amssymb,
 reprint,%
]{revtex4-1}

\usepackage{graphicx}
\usepackage{dcolumn}
\usepackage{bm}

\usepackage[utf8]{inputenc}
\usepackage[T1]{fontenc}
\usepackage{mathptmx}
\usepackage{etoolbox}
\usepackage{amsmath}
\usepackage{amssymb}
\usepackage{multirow}
\usepackage{diagbox} 
\usepackage{siunitx}
\usepackage{subcaption}

\makeatletter
\def\@email#1#2{%
 \endgroup
 \patchcmd{\titleblock@produce}
  {\frontmatter@RRAPformat}
  {\frontmatter@RRAPformat{\produce@RRAP{*#1\href{mailto:#2}{#2}}}\frontmatter@RRAPformat}
  {}{}
}%
\makeatother
\begin{document}

\preprint{AIP/123-QED}


\title{Remote Reactor Ranging via Antineutrino Oscillations}
\author{S. T. Wilson}
\email[Corresponding author: ]{s.t.wilson@sheffield.ac.uk}
\author{J. Armitage}
\author{C. Cotsford}
\author{N. Holland}
\affiliation{Department of Physics and Astronomy, University of Sheffield, S3 7RH, Sheffield, United Kingdom}
\author{J. G. Learned}
\affiliation{Department of Physics and Astronomy, University of Hawaii, Honolulu, HI 96822, USA}
\author{M. Malek}
\affiliation{Department of Physics and Astronomy, University of Sheffield, S3 7RH, Sheffield, United Kingdom}

\date{\today}

\begin{abstract}
Antineutrinos from nuclear reactors have the potential to be used for reactor monitoring in the mid- to far-field under certain conditions.
Antineutrinos are an unshieldable signal and carry information about the reactor core and the distance they travel.
Using gadolinium-doped water Cherenkov detectors for this purpose has been previously proposed alongside rate-only analyses.
As antineutrinos carry information about their distance of travel in their energy spectrum, the analyses can be extended to a spectral analysis to gain more knowledge about the detected core.
A Fourier transform analysis has been used to evaluate the distance between a proposed gadolinium-doped water-based liquid scintillator detector and a detected nuclear reactor.
Example cases are shown for a detector in Boulby Mine, near the Boulby Underground Laboratory in the UK, and six reactor sites in the UK and France.
The analysis shows potential to range reactors, but is strongly limited by the detector design.
It is concluded that the proposed water-based detector is not sufficient for ranging remote reactors in a reasonable time, but other detector designs show potential.
\end{abstract}

\maketitle

\section{Introduction}
\label{sec:intro}

Nuclear fission reactors are a key source of power generation in several parts of the world, with an increase in global capacity of 80\% predicted by 2050 by the International Atomic Energy Agency (IAEA) \cite{iaea2020predictions}.
However, alongside use in power generation, nuclear fission has been highly weaponized, and nuclear reactors are a key part of the process in producing the required material for proliferation.
This, combined with the predicted increase in global capacity and historical incidents of misuse (e.g. see \cite{Christensen2014}) and disaster, creates concern surrounding reactor operation.
In particular, diversion of material for plutonium production is of concern.
Several treaties, such as the Treaty on the Non-proliferation of Nuclear Weapons, and safeguarding methods are in place to verify the use of reactors, and identify any misuse \cite{IAEA-Safeguard}.

Safeguarding is typically performed by the IAEA, and consists predominantly of item accountancy and inspections, with current methods deemed generally suitable for the current fleet of power reactors.
However, there remains interest in the community in new techniques which provide additional information or less intrusive methods of reactor monitoring, and may be applicable to future scenarios and reactor types \cite{NuTools}.
One such method is the use of antineutrinos, produced in the decay chains of fission products, to monitor reactor power output and core composition.
Neutrino detectors have previously shown the ability to observe reactor power cycles and fuel evolution when situated in proximity to a reactor \cite{Bowden2009}, and studies into less-intrusive far-field detectors have shown potential for long distance observations \cite{Sensitivity2022, Akindele2023,Slava2022}.
Where inspections and item accountancy rely on access to reactors, neutrino observation could be performed without entering a reactor complex.
It is also less susceptible to false information from the host, as well as shielding due to the unshieldable nature of neutrinos.
This makes it attractive, and a potential tool to help prevent issues such as the 1994 DPRK nuclear crisis \cite{Christensen2014}.

Previous detection and studies have principally used the observed rate of antineutrino interactions in a detector.
However, further information can be obtained by applying spectral analysis techniques.
Neutrinos have a flavor associated with the lepton involved in the interaction they are produced in; electron, muon and tau flavors are the three in the Standard Model.
Once produced, neutrinos do not always remain in a single flavor.
They can oscillate in a distance- and energy-dependent manner, with both the emission and dominant detection mechanism for reactor neutrinos involving electron-flavor antineutrinos.
By obtaining the energy spectrum of electron antineutrinos from a source, information about their distance of travel can be determined alongside the reactor power and core composition.
This can be used to identify the source of a reactor signal, or verify a known signal by confirming the spectrum is as expected.
The work presented here builds on a previous study \cite{Sensitivity2022} that used reactor antineutrino rates by adding spectral analysis to an existing rate-only analysis to harness the flavor oscillation of the emitted reactor antineutrinos.
This has been applied to a hypothetical detector in Boulby Mine in the UK, near the Science \& Technology Facilities Council's Boulby Underground Laboratory, with a real reactor landscape used.
The aim of this study is to determine the utility of including spectral analysis, and test it on a previously developed detector design using real reactor signals.

The paper structure is as follows.
Section \ref{sec:reactor-neutrinos} details the emission, propagation and interaction of reactor antineutrinos.
Section \ref{sec:neo} presents a reactor antineutrino monitoring prototype detector and facility.
The simulation of the detector, signal and backgrounds are discussed in Sec. \ref{sec:sims}.
An existing data reduction is summarized in Sec. \ref{sec:data-reduction}, before a spectral analysis is presented in Sec. \ref{sec:analyses}.
The results of this analysis are presented in Sec. \ref{sec:results}, with consideration of alternative detector designs in Sec. \ref{sec:other_detectors}.
Discussion of the results and concluding remarks follow in Sec. \ref{sec:discussion} and \ref{sec:conclusion} respectively.

\section{Reactor antineutrinos}
\label{sec:reactor-neutrinos}

\subsection{Reactor antineutrino emission}

Nuclear power reactors emit an isotropic flux of antineutrinos of $\mathcal{O}$(10$^{20}$) s$^{-1}$ GW$_{\mathrm{th}}^{-1}$ due to the fission of $^{235}$U, $^{238}$U, $^{239}$Pu and $^{241}$Pu into neutron-rich nuclei \cite{Porta2010a}.
These nuclei undergo a series of $\beta$ decays until stability is reached, releasing an average of 6 antineutrinos per fission with energies up to approximately 10 MeV.
Despite the small cross-section of interaction, the large number of emitted antineutrinos produces an observable signal.
The emitted flux from the reactor depends on the reactor thermal power and core composition, and the fission processes occurring in the core.
The composition of the core, and its time evolution ($\textit{burnup}$) are determined by the reactor type.

The antineutrino spectrum of a fissioning isotope is given by
\begin{equation}
    \Phi_{\Bar{\nu}_e,i}(E_{\Bar{\nu}_e}) = P_\mathrm{th} \frac{p_i \lambda_{i}(E_{\Bar{\nu}_e})}{Q_{i}},
    \label{eq:antineutrino_emitted_flux}
\end{equation}
where $P_\mathrm{th}$ is the thermal power of the core, $p_i$ is the fraction of the thermal power resulting from the fission of isotope $i$, $Q_i$ is the average thermal energy emitted per fission and $\lambda_{i}(E_{\Bar{\nu}_e})$ is the antineutrino emission energy spectrum normalized to one fission for fissioning isotope $i$ as a function of antineutrino energy $E_{\Bar{\nu}_e}$.
$\lambda_{i}(E_{\Bar{\nu}_e})$ is given by
\begin{equation}
    \lambda_i(E_{\Bar{\nu}_e}) = \exp\Bigg( \sum\limits_{j=1}^6 a_j E^{j-1}_{\Bar{\nu}_e}\Bigg),
    \label{eq:spectrum}
\end{equation}
where the coefficients $a_j$ are fit parameters from the Huber-Mueller predictions \cite{Huber2011,Mueller2011}, which are derived from measurements of the $\beta$ spectra from nuclear fission.
The total antineutrino flux for a reactor is the summation of the contributions from individual isotopes.

The antineutrino flux from the fission of $^{239}$Pu is $\sim$65\% of the flux from $^{235}$U for the same thermal power output \cite{Christensen2013,Christensen2014}.
As $^{235}$U fissions and $^{239}$Pu accumulates, the reactor power remains consistent but the antineutrino flux drops.
This allows the antineutrino flux to be sensitive to the composition of the core, and measure a potential $^{239}$Pu accumulation.
The SONGS1 detector demonstrated this effect, observing both core burnup and reactor power outages in the measured flux \cite{Bowden2009} from the San Onofre Nuclear Generating Station.

\begin{figure}
    \centering
    \includegraphics[width=8.5cm]{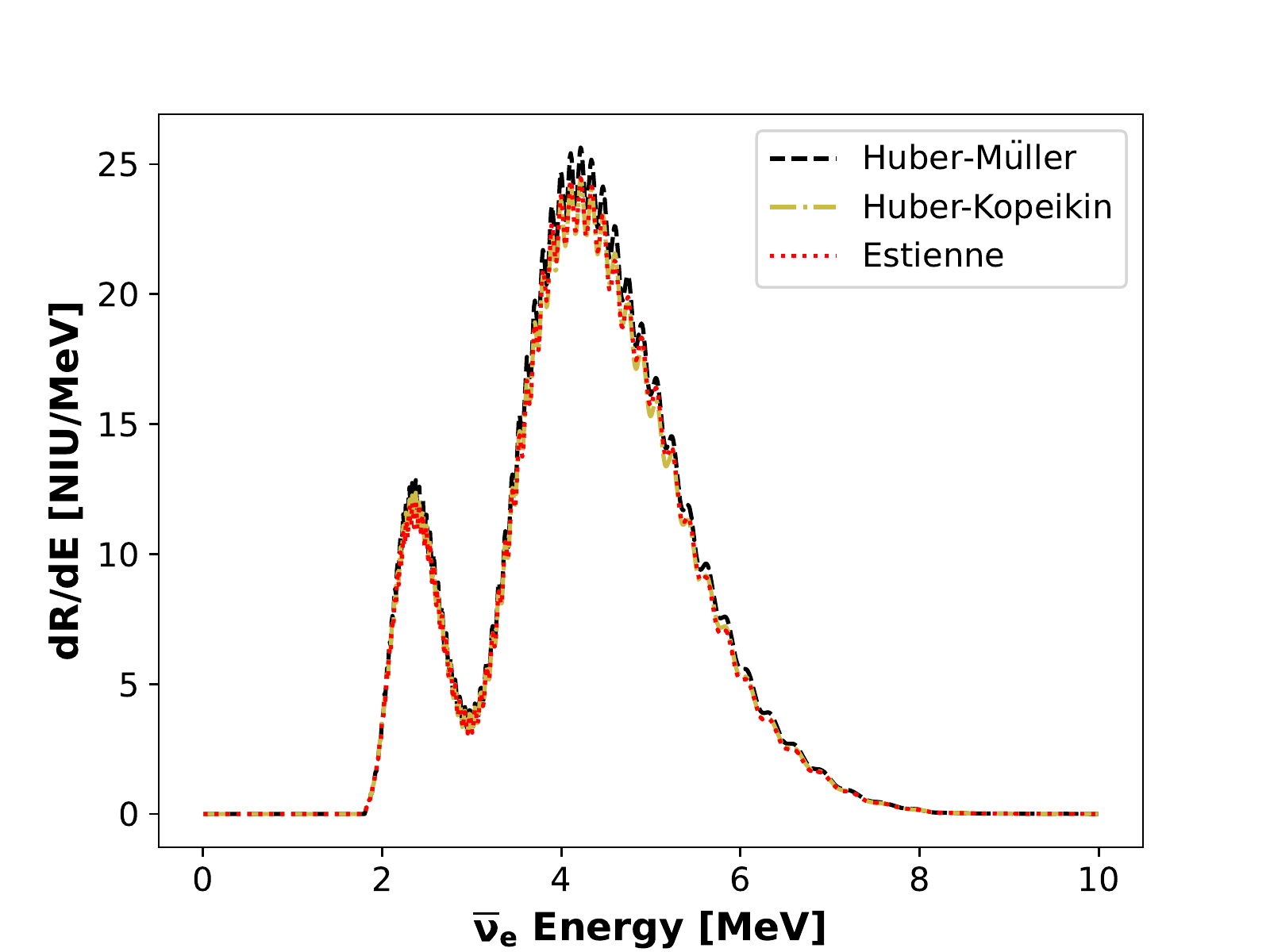}
    \caption{Comparison of reactor antineutrino emission models used to produce the expected antineutrino spectrum from the Heysham reactor complex, 149 km from a detector at Boulby. The Huber-Mueller model \cite{Huber2011,Mueller2011}, used in this work, has an excess in flux compared to data. The Estienne \cite{Estienne2019} and Huber-Kopeikin \cite{Kopeikin2021b} models are newer and correct this deficit.}
    \label{fig:Hey-emission-models}
\end{figure}

The reactor spectra used in a previous rate-only study\cite{Sensitivity2022}, which the work presented here builds on, use the Huber-Mueller predictions \cite{Huber2011,Mueller2011}.
Measurements across several experiments show a slight flux deficit when compared to these predictions, and an excess at 5 MeV \cite{Mention2011,An2017,Bak2019,DoubleChooz2020,Ko2017}.
More recent calculations that solve the reactor antineutrino flux deficit using new measurements of the beta decay spectra of the fissioning isotopes have since been developed \cite{Estienne2019,Kopeikin2021b}.
To maintain consistency with previous work \cite{Sensitivity2022}, the Huber-Mueller predictions have been used, with the anomaly considered as part of the uncertainty.
This is expected to have a small effect on the spectral shape, shown in Fig. \ref{fig:Hey-emission-models}, which is of relevance to the spectral analyses presented.


\subsection{Reactor antineutrino propagation}

\begin{figure}
    \centering
    \includegraphics[width=8.5cm]{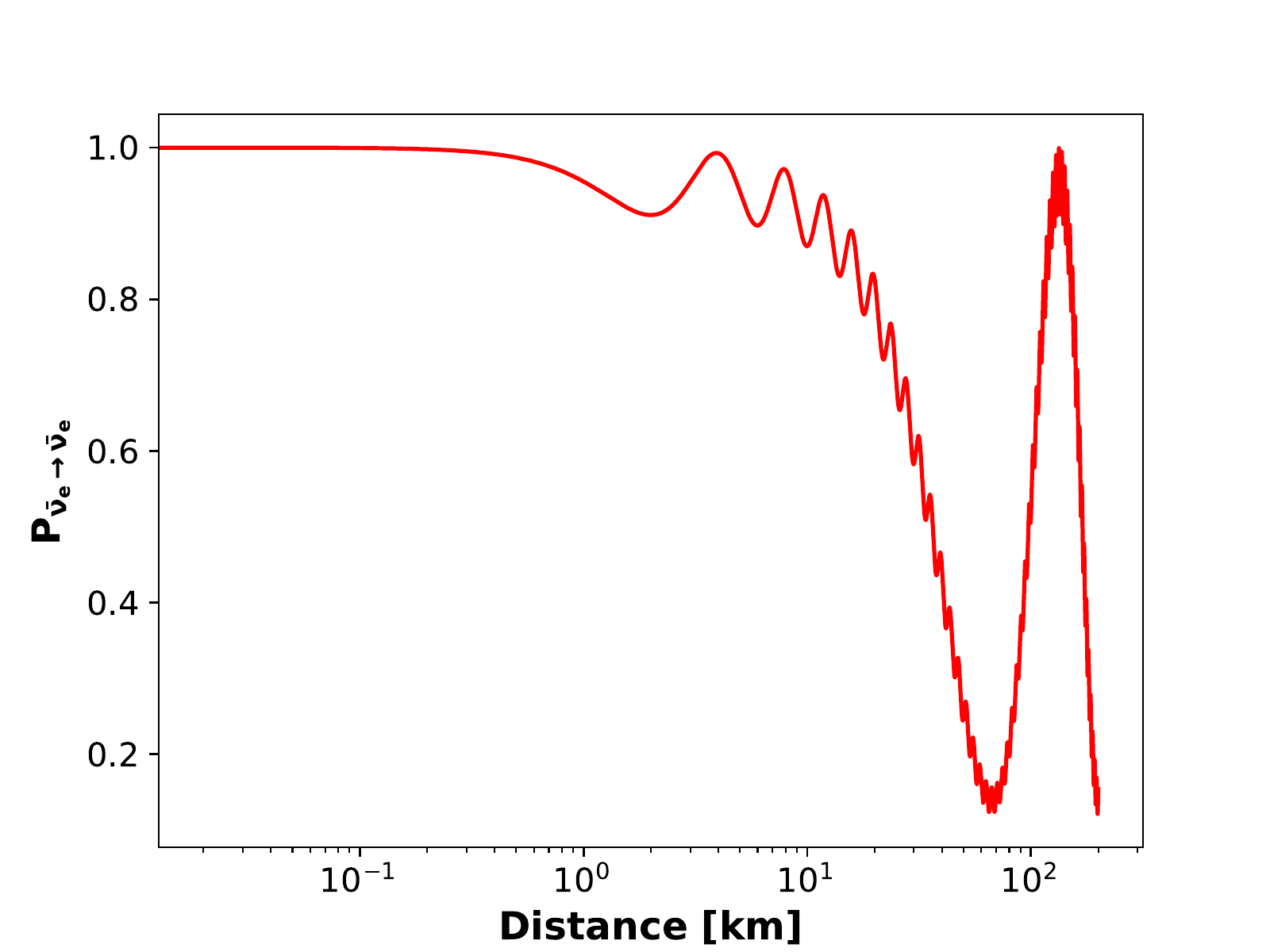}
    \caption{Survival probability of 4 MeV electron antineutrinos with distance of travel.}
    \label{fig:survival-prob}
\end{figure}

The observed flux of antineutrinos depends heavily on neutrino oscillation.
Reactor antineutrinos are emitted as electron flavor and inverse beta decay (IBD), the main detection mechanism, is sensitive to electron flavor, so the survival probability of these antineutrinos is of relevance.
In the situation where the PMNS mixing matrix with three neutrino mass eigenstates is used, the survival probability of electron antineutrinos is given by
\begin{equation}
    \begin{aligned}
    P_{\bar{\nu}_e \rightarrow \bar{\nu}_e} = 1 - \cos^4(\theta_{13})\sin^2(2\theta_{12})\sin^2\Big(\frac{1.27\Delta m^{^{2}}_{21}L}{E_{\bar{\nu}} }\Big) \\
    - \cos^2(\theta_{12})\sin^2(2\theta_{13})\sin^2\Big(\frac{1.27\Delta m^{^{2}}_{31}L}{E_{\bar{\nu}} }\Big) \\
    - \sin^2(\theta_{12})\sin^2(2\theta_{13})\sin^2\Big(\frac{1.27\Delta m^{^{2}}_{32}L}{E_{\bar{\nu}} }\Big),
    \end{aligned}
    \label{eq:antineu_survival}
\end{equation}
where $\Delta m^2_{ij} = m^2_j - m^2_i$ is the mass-squared difference between mass eigenstates $i$ and $j$, $\theta_{ij}$ is the mixing angle between states $i$ and $j$, $L$ is the distance of travel in km, and $E_{\bar{\nu}}$ is the antineutrino energy in GeV.
The survival probability of 4 MeV electron antineutrinos is shown in Fig. \ref{fig:survival-prob}.
Two oscillations are visible, those due to the $\theta_{12}$ and $m^2_{21}$ terms and those due to the $\theta_{13}$ and $m^2_{31}$ terms.
The small-amplitude, high-frequency oscillations starting around 1 km are due to $\theta_{13}$ and $m^2_{31}$ terms.
The low-frequency, high-amplitude oscillations that cause the large trough followed by a peak at around 100 km are caused by the $\theta_{12}$ and $m^2_{21}$ terms.
Neutrino oscillations will not only change the shape of the spectrum, they also reduce the total observable flux through the conversion of electron flavor antineutrinos to muon and tau flavor, which do not undergo IBD.

\subsection{Reactor antineutrino detection}

The dominant mechanism of antineutrino interaction in a hydrogenous material is IBD \cite{IBDcross-sec-new}, in which an electron antineutrino interacts with a proton to produce a positron and a neutron:

\begin{equation*}
    \bar{\nu}_e + p \rightarrow e^+ + n.
\end{equation*}.

The cross-section of IBD is $\mathcal{O}$(10$^{-44}$)$E_ep_e$ cm$^2$, where $E_e$ and $p_e$ are the positron energy and momentum respectively.
At the time of the previous study \cite{Sensitivity2022}, the most accurate cross-section in the MeV to GeV range was given by Strumia and Vissani \cite{Strumia2003}.
An updated cross-section with reduced uncertainty has since been recalculated \cite{IBDcross-sec-new}.
For consistency, and due to the negligible change at the energies of interest, the cross-section used in this work is from Strumia and Vissani \cite{Strumia2003}.

During IBD, the positron and neutron are produced in a pair.
The positron is detected directly whereas the neutron first thermalizes and then captures, leading to prompt and delayed signal components in coincidence.
This coincident-pair of signals can be used to discriminate against backgrounds by correlating the position and time of the neutron capture to the positron detection.

Due to the kinematics of IBD, the positron carries information about the energy of the incoming antineutrino.
The threshold energy, $E_{\mathrm{thr}}$, for IBD in the laboratory frame can be determined from
\begin{equation}
    E_{\mathrm{thr}} = \frac{(m_n + m_e)^2 - m^2_p}{2m_p} \approx 1.8 ~\mathrm{MeV},
    \label{eq:ibd_threshold}
\end{equation}
where $m_n$, $m_p$ and $m_e$ are the neutron, proton and positron rest masses.
The antineutrino energy, $E_{\bar{\nu}_e}$, can then be determined from the positron energy through
\begin{equation}
    E_{\bar{\nu}_e} = E_{e^+} + E_{\mathrm{thr}} - m_e.
    \label{eq:ibd_nue_energy}
\end{equation}

The positron direction is almost isotropic, with a slight bias in the backwards direction.
The neutron, however, takes most of the antineutrino's momentum and its initial direction is largely parallel to the incoming antineutrino's direction.
The neutron will take a random walk from near the point of emission and thermalize through successive scatterings in the detector medium.
Once thermalized, the neutron will capture on hydrogen or another nucleus added as a neutron capture agent, such as gadolinium.
The capture and subsequent radiative de-excitation produces a second signal, occurring a short time and distance from the initial positron signal, creating a coincident-pair signal in the detector.
This time and distance are dependent on the detector medium.

\begin{figure}
    \centering
    \includegraphics[width=8.5cm]{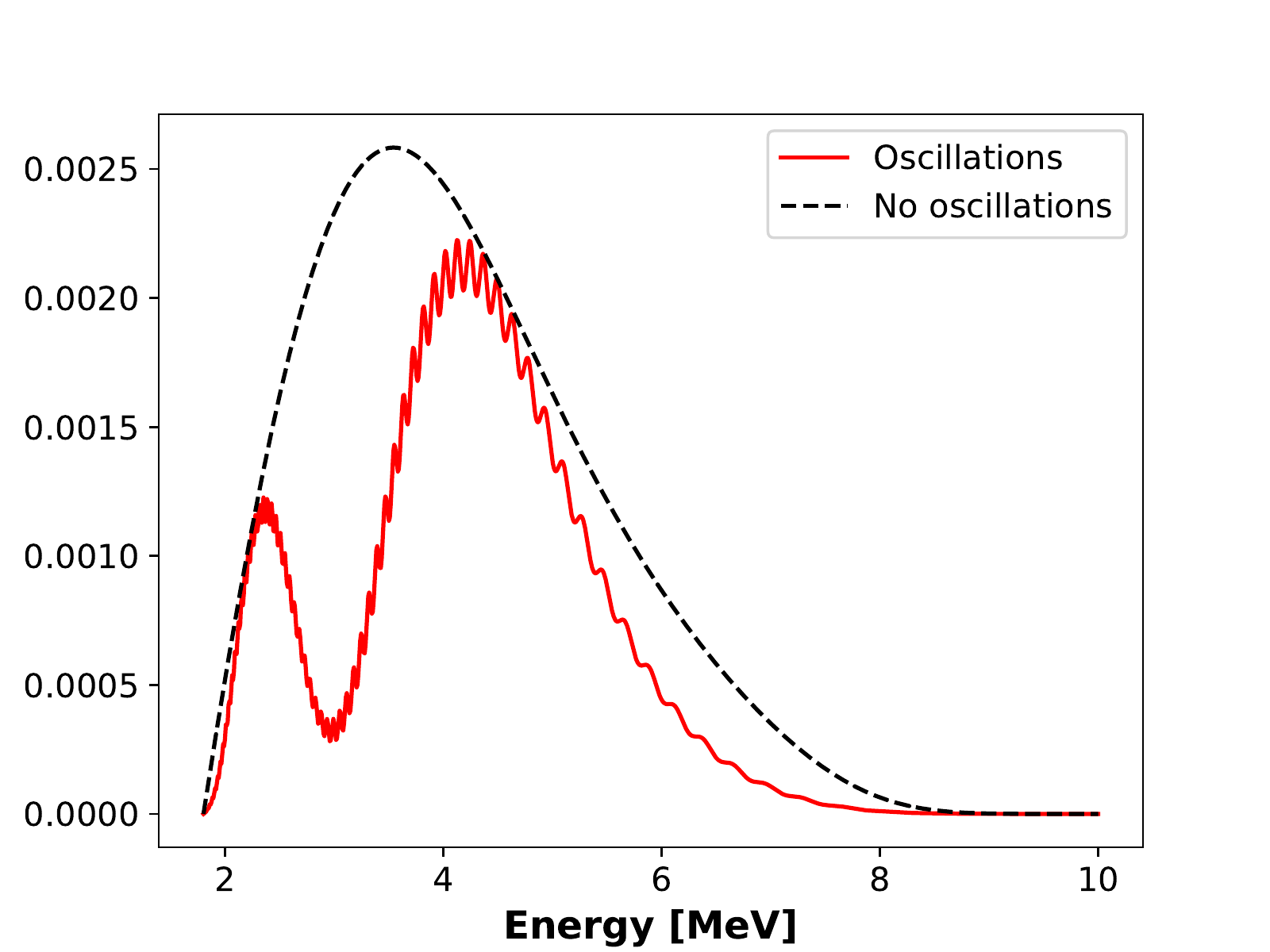}
    \caption{Observable reactor antineutrino spectrum at a 150 km standoff with oscillations (red solid) and no oscillations (black dashed). Both spectra have been normalized to the total flux of the no oscillation case, showing the reduction in the flux of electron antineutrinos due to oscillation.}
    \label{fig:osc-non-osc}
\end{figure}

The final observable antineutrino flux at energy $E_{\bar{\nu}}$ is given by
\begin{equation}
	f(E_{\bar{\nu}} \big\vert L ) = \phi(E_{\bar{\nu}} )\sigma(E_{\bar{\nu}} )P(L,E_{\bar{\nu}}),
	\label{eq:pdf}
\end{equation}
where  $\phi(E_{\bar{\nu}})$ is the reactor emission flux (Eq. \ref{eq:antineutrino_emitted_flux}), $\sigma(E_{\bar{\nu}} )$ is the IBD cross-section and $P(L,E_{\bar{\nu}})$ is the survival probability at distance $L$ (Eq. \ref{eq:antineu_survival}).
The spectrum produced from this model, and the effect of neutrino oscillation, is shown in Fig. \ref{fig:osc-non-osc}.

\section{Reactor antineutrino monitoring prototype}
\label{sec:neo}

The Advanced Instrumentation Testbed-Neutrino Experiment One (AIT-NEO) facility and detector was a proposed project to demonstrate the application of antineutrino monitoring for non-proliferation purposes.
AIT is the facility designed to develop new technologies, where NEO was the planned first experiment to be housed in the facility.
Several detector designs were considered, with the options considered previously all being upright cylinders with a water-based fill and gadolinium doping \cite{Slava2022,Akindele2023,Sensitivity2022}.

\begin{figure}[htb]
    \includegraphics[width=8.5cm]{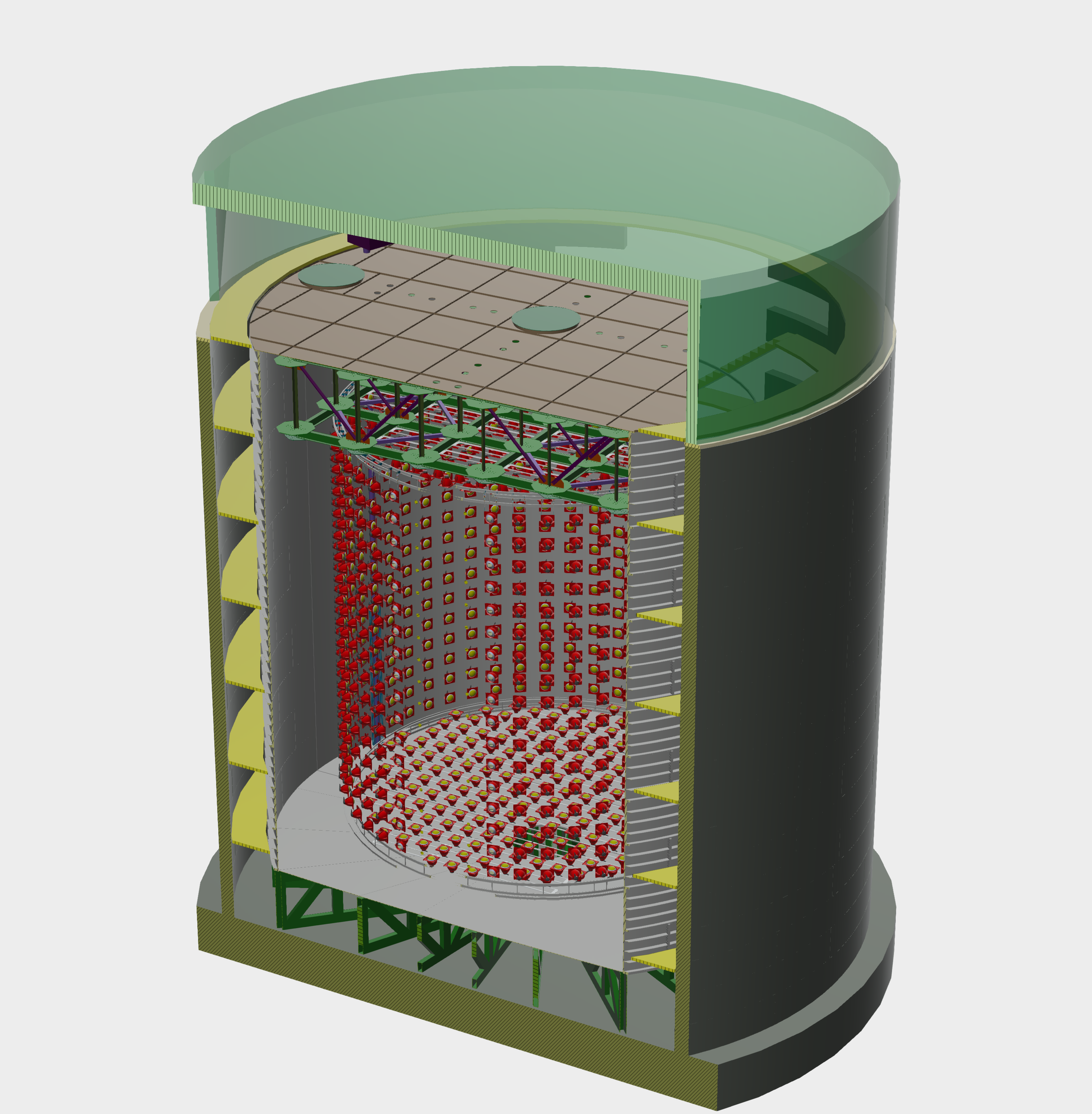}
  \caption{Schematic of the detector design by Jan Boissevain (University of Pennsylvania), showing the tank supported on a steel truss structure and inner PMT support structure.}
  \label{fig:detector-schematic}
\end{figure}

The investigated detector design here is a hypothetical 22 m height and diameter right cylinder, water-based Cherenkov detector, located 1100 m underground (2800 m.w.e, $\sim$ 10$^6$ muon attenuation versus surface \cite{Robinson2003}) near to Boulby Underground Laboratory in a 25 m height and diameter cavern.
The detector considered has a 9 m inner detector region, surrounded by 4600 photomultiplier tubes (PMTs) for a 15\% coverage facing inwards, and a 2 m passive buffer region surrounding this to attenuate external backgrounds.
This gives a fiducial volume of approximately 4.5 kT.
A schematic of the detector used is shown in Fig.~\ref{fig:detector-schematic}.

The detector is filled with water-based liquid scintillator (WbLS).
WbLS is a water and scintillator mix, providing higher light yields and lower energy thresholds than water, but longer light attenuation lengths than scintillators \cite{Yeh2011,Zsoldos2022}.
WbLS in principle allows the scalability and lower cost of water to be combined with some of the performance properties of scintillators.
As part of AIT, the first kilotonne-scale deployment of WbLS was proposed, with a significant benefit from a low energy threshold being found for reactor antineutrino monitoring \cite{Sensitivity2022}.
The cocktail used contains liquid scintillator at a concentration of 1\% w/w, giving a light yield of 100 photons/MeV and an approximately 18\% energy resolution at its 1 MeV energy threshold.
This is a significant improvement over Super-Kamiokande's pure water energy performance, in particular its 3.5 MeV threshold \cite{Suzuki2019}.
By lowering the energy threshold through the addition of scintillator, the rate of observable reactor IBD events can be doubled.
Proposals have been made for WbLS detectors up to 100 kT, significantly larger than the largest liquid scintillator detector \cite{JUNO2022}, with energy resolutions reaching $\frac{6\%}{\sqrt{MeV}}$ and sub-MeV energy thresholds \cite{Askins2020}.

The detector fill is also gadolinium doped at a concentration of 0.1\% w/w.
This provides a boost in neutron capture and detection efficiency \cite{Beacom2003,ABE2021}, allowing the neutron components of IBD to be more easily observed.

The expected reactor landscape around Boulby is used for this study.
There are two reactor types considered in this landscape; they are Advanced Gas-cooled Reactors (AGRs) and Pressurized Water Reactors (PWRs).
Both use enriched uranium as fuel, with the AGRs using graphite as a moderator and the PWRs using water.
The AGRs are older and dominate the UK reactor fleet, with the nearest three complexes to Boulby being AGRs.
These are split into two generations, AGR-1 and AGR-2, with similar designs.
The PWRs are more common in France, with newer reactors in the UK also using PWR designs.

\begin{figure}[htb]
    \includegraphics[width=8.5cm]{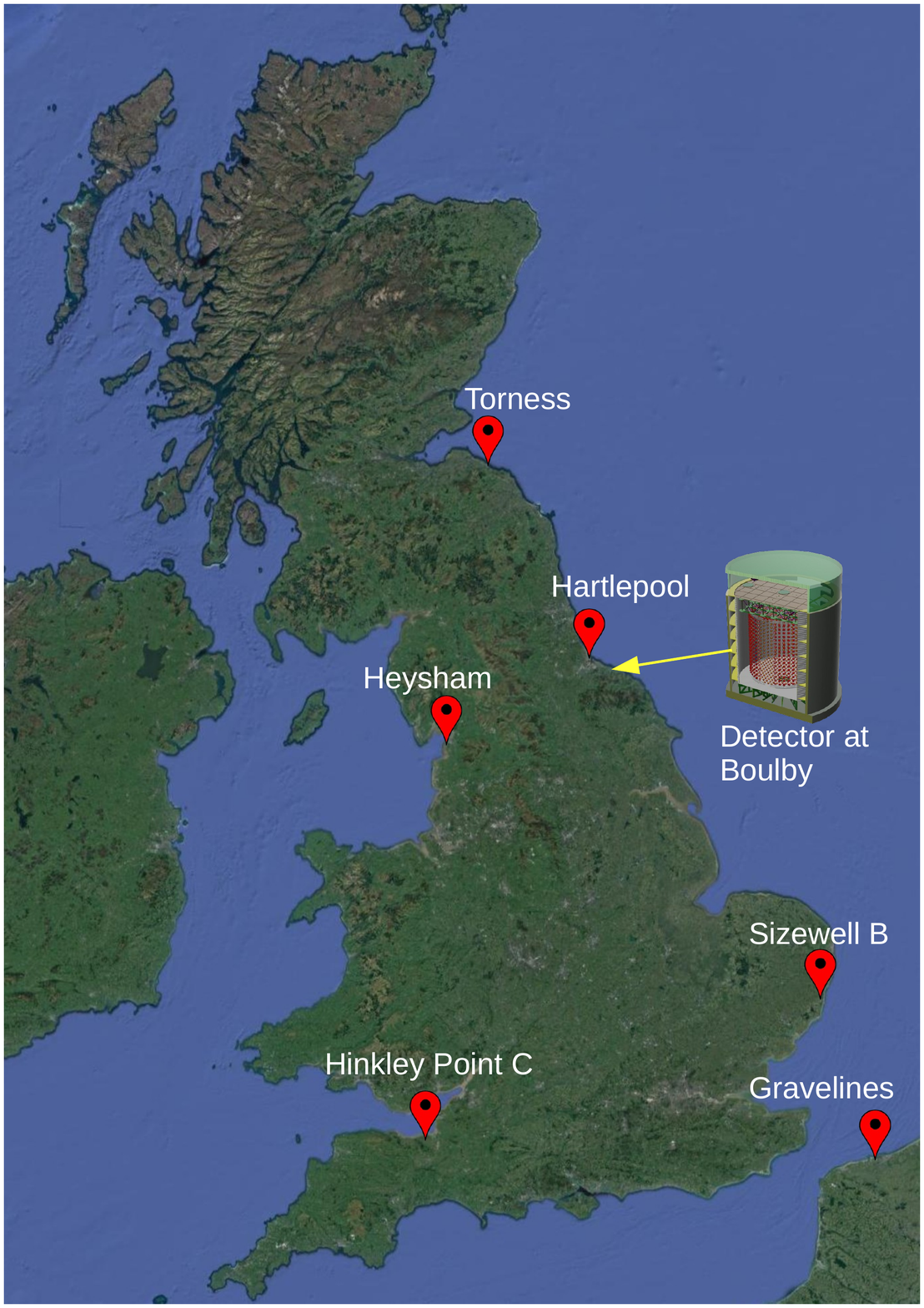}
  \caption[Map]{Map showing the location of the detector at Boulby and the reactor sites studied.~\cite{Google2022}}
  \label{fig:map}
\end{figure}

Table~\ref{tab:reactors} shows the reactor sites considered for this study, along with their type, standoff distance, approximate signal rate after data reduction and expected decommissioning date. At the time of this study, the UK's AGR fleet was due for decommissioning, with the first generation AGR-1 cores by 2026 followed by the second generation AGR-2 fleet by 2028. Hinkley Point C (a PWR) was also expected to come online by around 2030 \cite{EDFEnergy,EDFhinkley}. Their locations on a map are shown in Fig.~\ref{fig:map}. Sizewell B, a PWR, was undergoing review for an extension beyond its initially planned end date of 2035 \cite{EDFsizewell}.

\begin{table*}[htb]
    \centering
    \begin{tabular}{lcccccc}
        \hline\hline
        {Signal} & {Number} & {Type} & {Standoff }  & {Rate} & {Decommissioning}  \\
                    &   {of cores} &  & {distance [km]} & {[per day]} & {date}      \\
        \hline
        Hartlepool          & 2   & AGR-1 &  26             & 7.65 & 2026 \cite{EDFEnergy}  \\
        Heysham 1           & 2   & AGR-1 & 149             & 0.20 & 2026  \cite{EDFEnergy} \\
        Heysham 2           & 2   & AGR-2 & 149             & 0.23 & 2028  \cite{EDFEnergy} \\
        Torness             & 2   & AGR-2 & 187             & 0.13 & 2028  \cite{EDFEnergy} \\
        Sizewell-B          & 1   & PWR   & 306             & 0.045 & after 2035 \cite{EDFsizewell}  \\
        Hinkley Point C     & 2   & PWR   & 404             & 0.089 & 2090  \cite{EDFhinkley} \\
        Gravelines (France) & 6   & PWR   & 441             & 0.089 & 2031  \cite{ASN} \\
        \hline\hline
    \end{tabular}
    \caption{The reactor type, standoff distance, approximate signal rate after data reduction and expected decommissioning date for the reactors considered in this study. Sizewell B was under review for a long term extension beyond 2035 at the time of this study.}
    \label{tab:reactors}
\end{table*}

\section{Simulations}
\label{sec:sims}

Full Monte Carlo (MC) detector simulations were performed using RAT-PAC \cite{Seibert2014}.
RAT-PAC is based on the physics simulation framework Geant4 \cite{AGOSTINELLI2003,ALLISON2016}, the CLHEP physics library \cite{LONNBLAD1994}, the Generic Liquid-scintillator Anti-Neutrino Detector (GLG4sim or $\textit{GenericLAND}$) Geant4 simulation for neutrino physics \cite{glg4sim} and the ROOT data-analysis framework \cite{BRUN1997}.

RAT-PAC models the detector response event-by-event, including triggering and data acquisition (DAQ), with custom GLG4sim MC event generators used to produce the signal and background particles.
Light emission from the detector medium and PMT response are handled by GLG4sim, with the propagation of particles and light performed by Geant4.

The models used for MC simulations of the WbLS \cite{MEV_WbLS_perf} are based on time profile measurements of the scintillation light \cite{CherenkovScintillation,WbLSTimeResponse}, and light yield and scattering measurements of gadolinium-doped WbLS \cite{Gabriel2022}.

The detector is simulated in a cavern surrounded by a 2 m rock layer, which has been shown to be sufficient to sample all backgrounds external to the detector when combined with the detector's passive buffer region and fiducial volume in previous studies \cite{Sensitivity2022}.
The large-scale, complex structures such as the support structures in the detector are simplified to approximate positions and volumes to allow their background contributions to be considered.

\subsection{Signal}

The antineutrino signal options considered are the power reactors in the UK's fleet, as well as Gravelines in northern France, detailed in Table \ref{tab:reactors}.
The signals are simulated by sampling spectral and angular distributions based on the reactor emission flux, antineutrino oscillations during travel and the IBD cross-section to produce an expected antineutrino signal distribution for each reactor.
A custom event generator is used to produce the positron and neutron from an antineutrino interaction in a correlated pair, with their kinematics defined by the Stumia and Vissani cross-section model \cite{Strumia2003}.
The mean time and distance between positron and neutron signals in simulated IBD is 28 $\mu$s and 6 cm respectively, which matches expectation for a 0.1\% gadolinium concentration \cite{Marti2020}.

The Hartlepool complex is simulated as two individual cores, Heysham is split into its two dual-core sites representing the Heysham 1 AGR-1 and Heysham 2 AGR-2 sites, and all other complexes are simulated as a single signal.
Spectra are taken from geoneutrinos.org \cite{Dye2015}, which uses IAEA data for the year 2020 \cite{pris}.
The spectra use the mid-cycle core composition.
For AGRs, due to their short refueling cycle geared toward optimal power generation, there is only a small amount of fuel evolution, so using the mid-cycle core composition is a good approximation.

\subsection{Background}

A large number of backgrounds are simulated, which are detailed further by Kneale $\textit{et al.}$ \cite{Sensitivity2022}, using custom event generators.

Given the coincident-pair nature of the IBD signal, the overall background is suppressed by requiring the detection of both the prompt and delayed signals in a time and distance characteristic of the positron and neutron-capture signals.
However, several background sources can mimic the correlated signal expected from IBD, some accidentally and some through their own inherent correlation.

The key correlated background sources are produced by comic ray muons; they are fast neutrons produced by spallation in the rock surrounding the detector and neutron evaporation along the muon track, and radionuclides produced primarily in hadronic showers from the spallation of oxygen in the detector fill \cite{Li2014}.

Fast neutrons are commonly produced in multiplicity, and can generate neutron pairs in the detector volume.
When they thermalize and capture, this pair can mimic the IBD signal.
The spectrum and multiplicity used is from comparison between data and simulations \cite{Wang2001,Mei2006}, as is the angular distribution \cite{Mei2006}.
The first neutron is produced largely along the path of the muon, and so is generally downward-going.
The secondary neutrons from neutron evaporation are emitted isotropically.
Neutrons with energies less than 10 MeV are insufficiently penetrating to reach the inner detector volume, and so are neglected.
The mean time between neutrons in a pair is 20 $\mu$s in simulation, and the mean distance is 80 cm, which is similar to IBD.

The radionuclides of concern for reactor antineutrino detection are the $\beta$-neutron emitters, in particular $^9$Li and $^{17}$N due to their high yields and branching ratios, long half lives and $\beta$ endpoint energies.
These decay via the emission of both a $\beta$ particle and a neutron, producing the same correlation in time and space as an IBD event.
Due to the low yield, consistent with zero in Super-Kamiokande \cite{Zhang2016}, $^8$He is neglected.

The expected rate of each radionuclide isotope is calculated using the muon flux in Boulby mine \cite{Robinson2003}, $\Phi_\mu = (4.09 \pm 0.15) \times 10^{-8} ~ \mathrm{cm}^{-2}\mathrm{s}^{-1}$, and assumes the muon travels the vertical height of the detector.
The $\beta$-n decay rates are calculated using
\begin{multline}
    R_\mathrm{iso}(\mathrm{s}^{-1}) = R_\mu(\mathrm{s}^{-1}) \times L_\mu(\mathrm{cm}) \times Y_\mathrm{iso}(\mu^{-1}\mathrm{g}^{-1}\mathrm{cm}^2) \\ \times br \times \rho(\mathrm{g}~\mathrm{cm}^{-3}) \times \bigg(\frac{E_{\mu,\mathrm{Boulby}}}{E_{\mu,\mathrm{Super-Kamiokande}}}\bigg)^\alpha,
\end{multline}
where $R_\mu = \Phi_\mu \times$ tank surface area, $L_\mu$ is the muon path length, $Y_\mathrm{iso}$ is the isotope yield, $br$ is the branching ratio for the $\beta$-n decay, $\rho = 1 ~ \mathrm{g} ~ \mathrm{cm}^{-3}$ for water and $E_\mu$ is the average muon energy.
The path length in Super-Kamiokande is taken as the vertical height of the detector \cite{Li2014}, making the assumption all muons are downward going, which sets a conservative upper limit.
The same assumption is made for NEO, and rates are calculated for the full detector volume.
A depth-related correction to the average muon energy $E_\mu^\alpha$ is applied \cite{Mei2006}, where $\alpha = 0.73 \pm 0.10$, as higher-energy muons will survive to greater depths on average.

Other antineutrino sources create a background to the reactor signal as they also interact via IBD.
Other reactors are the dominant source of background IBD at Boulby; which reactors are part of the background depends on which reactor is the target of observation.
All reactors more distant to Boulby than Gravelines are amalgamated into a single world reactor background, as these are present in the background for all reactor targets considered.
The reactors from Gravelines and closer are added to the background individually where appropriate based on the expected decommissioning dates and target reactor signal.
Antineutrinos from uranium and thorium in the Earth's mantle, termed $\textit{geoneutrinos}$, are the other source of IBD background.
Geoneutrinos are produced via the $\beta$ decay of $^{238}$U and $^{232}$Th in the Earth's mantle, with the antineutrino spectra obtained from geoneutrinos.org \cite{Dye2015}.

Uncorrelated events from radioactive $\beta$ decays in the detector material and environment can mimic a correlated pair through accidental coincidences.
These events are typically low in energy and, with the dominant sources originating outside or coming from the edge of the detector, are mostly observed toward the outer edges of the detector.
Their total rate is very high in a large detector, $\mathcal{O}$(MHz), but most events can be rejected through energy and position cuts, and by specifying both a prompt and delayed event must occur within a small time and distance of each other.
The rate of coincidences is highly dependent on the analysis cuts used to remove backgrounds.
The largest source of concern for these decays is the PMTs, in particular their glass.
The rates for the radioactive decays in the detector materials and surrounding cavern are taken from a combination of experimental data, material assays and theory \cite{Marti2020,HASELSCHWARDT2019,Hamamatsu2020,Zhang2016b,Zhang2016,ARAUJO2012,Li2014,TOI,TUNL,JOLLET2020,Robinson2003,Wang2001,Mei2006,Tang2006,Sutanto2020,ARAUJO2012,Battat2014}.

The total rate of background components is shown in Table \ref{tab:22m-background-rates}.
The uncorrelated single $\beta$s dominate the overall rate, but are suppressed significantly by the requirement of particles being in a pair.
The fast neutron rate is for individual neutrons, but they often interact close to each other and mimic pairs.

\begin{table}
    \centering
    \begin{tabular}{lc}
        \hline\hline
        Component & Pair Rate [Hz]\\
        \hline
        $^{17}$N $\beta$-n & 1.99 $\times$ 10$^{-5}$\\
        $^{9}$Li $\beta$-n & 3.25 $\times$ 10$^{-5}$\\
        Reactor IBD & 1.47 $\times$ 10$^{-5}$\\
        Geoneutrino IBD & 2.60 $\times$ 10$^{-6}$\\
        \hline
         & Single Rate [Hz]\\
        \hline
        Fast neutrons & 3.22 $\times$ 10$^{-2}$\\
        Uncorrelated single $\beta$ & 3.57 $\times$ 10$^{7}$\\
        \hline\hline
    \end{tabular}
    \caption{The raw background rates in the 22 m NEO design. The reactor IBD is for all reactors further away than Gravelines.}
    \label{tab:22m-background-rates}
\end{table}

The uncertainties on signals and backgrounds are shown in Table \ref{tab:uncert}.
Uncertainties limit detector sensitivity by making it difficult to distinguish between components based on expected rates of interaction for each component.

\begin{table}
    \centering
    \begin{tabular}{cc}
    \hline\hline
    Component & Uncertainty\\
    \hline
    Hartlepool & 2.5\% \cite{Dye2015}\\
    Heysham & 2.0\% \cite{Dye2015}\\
    Torness & 2.6\% \cite{Dye2015}\\
    Sizewell B & 2.75\% \cite{Dye2015}\\
    Hinkley Point C & 3.0\% \cite{Dye2015}\\
    Gravelines & 3.4\% \cite{Dye2015}\\
    World Reactor & 6.0\% \cite{Dye2015}\\
    Geoneutrinos & 25\% \cite{Dye2015}\\
    $^9$Li & 0.2\% \cite{Sensitivity2022}\\
    $^{17}$N & 0.2\% \cite{Sensitivity2022}\\
    Fast Neutrons & 27\% \cite{Mei2006}\\
    \hline\hline
    \end{tabular}
    \caption[Uncertainties]{Uncertainties on signals and backgrounds.}
    \label{tab:uncert}
\end{table}

\section{Data Reduction}
\label{sec:data-reduction}

The simulated data are passed through an event reconstruction and data reduction pathway typical of neutrino detectors.
The position of interactions is reconstructed from the PMT response using an adapted version of BONSAI \cite{Smy2007}, which is used in Super-Kamiokande for low energy events, and the backgrounds suppressed using Likelihood Event Analysis for Reactor Neutrinos (LEARN).
Further detail is given on the full data reduction in Kneale $\textit{et al.}$ \cite{Sensitivity2022}.

To remove large chains of neutrons induced by fast neutron interactions, a multiplicity cut of two is applied to the data.
This accepts pairs and single events, but rejects hadronic showers and neutron clouds.
The accepted events are passed to the LEARN data reduction.

\begin{figure}
    \centering
    \includegraphics[width=8.5cm]{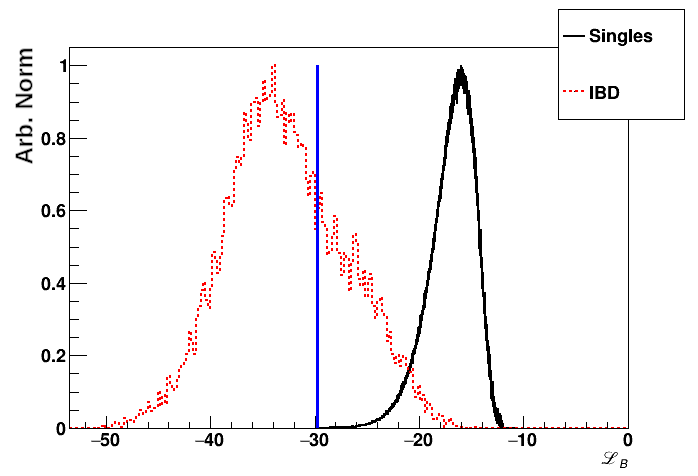}
    \caption{Log likelihood for single uncorrelated triggers due to radioactivity (solid black) and from neutrons in IBD (dashed red). The vertical blue line corresponds to the minimum log likelihood for the radioactive singles.}
    \label{fig:likelihood}
\end{figure}

LEARN uses a likelihood analysis to remove uncorrelated events due to radioactivity, suppressing them to a level consistent with zero, and accept pairs of detected particles which have the characteristics of IBD.
This can include fast neutron pairs and radionuclides that are $\beta$-neutron emitters, as well as IBD interactions.
The likelihood for an event is determined from probability density functions (PDFs) for the event's detected position, energy, and time in relation to the previously detected event.
A neutron from IBD will have a very different distribution to a single event from a radioactive decay.
By comparing the likelihood of an event to the distribution of likelihoods for the uncorrelated event background, it can be determined if an event is the second in a pair and the pair can be accepted.
The log likelihood distributions for neutrons from IBD and single uncorrelated events from radioactive decays can be seen in Fig. \ref{fig:likelihood}.

An AdaBoost \cite{scikit-learn} machine learning algorithm is used to suppress fast neutrons, which have a high uncertainty and therefore significantly limit detector sensitivity, by comparing the differences in location, light yield and reconstruction quality between pairs with two neutrons and pairs with an electron or positron.
This technique is approximately 94\% efficient at removing fast neutrons, and has a negligible effect on the signal rate.

Energy cuts are then used to suppress the low-energy geoneutrinos and high-energy radionuclides, in particular $^9$Li which has a high $\beta$ endpoint energy of $\sim$ 10 MeV compared to positrons from reactor neutrinos which peak below 4 MeV.

Finally, a post-muon analytical veto is used to reject long-lived radionuclides from muon-induced spallation by applying a dead time to a limited transverse distance around an observed muon track.

The detection efficiency for each event type after each step in the data reduction is shown in Table \ref{tab:remaining_events_v2}.

\begin{table*}[!htb]
    \centering
    \begin{tabular}{lccccc}
    \hline\hline
    \multirow{2}{2.5em}{Signal} & \multicolumn{5}{c}{Detection Efficiency After Step [\%]} \\
    \cline{2-6}
    & Multiplicity & Likelihood & Machine Learning & Energy Cuts & Muon Veto\\
    \hline
    Reactor IBD & 88 & 88 & 52 & 33 & 33\\
    Geoneutrinos & 47 & 47 & 37 & 13 & 13\\
    $^9$Li & 99 & 99 & 54 & 21 & 14\\
    $^{17}$N & 72 & 72 & 47 & 22 & 19\\
    Fast Neutrons & 0.12 & 0.12 & 2.1 $\times$ 10$^{-4}$ & 2.1 $\times$ 10$^{-4}$ & 2.1 $\times$ 10$^{-4}$\\
    \hline\hline
    \end{tabular}
    \caption{Detection efficiency of key event types after each step in the data reduction.}
    \label{tab:remaining_events_v2}
\end{table*}

\subsection{Production of spectra}

Three scenarios are considered depending on the target reactor signal.
They are:
\begin{itemize}
    \item unknown backgrounds,
    \item known backgrounds with uncertainties,
    \item known backgrounds with no uncertainties.
\end{itemize}
These are chosen to test the limits of the detector's sensitivity and determine the most distant power reactor that can be accurately ranged from Boulby using NEO.

The output from the data reduction applied to simulated data is in the form of spectra for all events that pass background suppression, split into individual sources.
These can be combined into a single ``observed" spectrum, or remain as separate components.
This allows the option of background contributions to be unknown or known, and therefore be subtracted from the spectrum if known.

If a background is known, it is assumed it is known at the rate that passes data reduction with a known spectral shape, allowing it to be removed from the total spectrum.
This is a hypothetical scenario, as it would involve background-specific measurements and further analysis to confirm the background rate and spectrum.
By allowing for backgrounds to be known, and subtracted from the spectrum, their impact on the ability for the detector to range a reactor can be determined.
In some scenarios, in particular more distant reactors, it is expected that backgrounds dominate and will be the limiting factor for a spectral analysis.

When uncertainties are applied, they are at the level in Table \ref{tab:uncert}.
Uncertainties are assumed to be Gaussian distributed about the observed rates after data reduction.
To include uncertainties, the distribution for each source is drawn from for each energy bin to give the fluctuations in the spectrum due to uncertainties.
For the situation of unknown backgrounds, these fluctuations are added to the spectrum for each source after data reduction, before the spectra are summed to give a total observed spectrum including uncertainties.
When it is assumed backgrounds are known, and can therefore be subtracted from the observed spectrum, only the fluctuations are added to the signal spectrum.
This creates a signal with fluctuations due to uncertainties.

To understand the impact of these uncertainties the uncertainties are drawn one hundred times to give one hundred spectra.
Each spectrum is equivalent to making a single observation of a signal, with the number of data points in the sample being related to the length of the observation.
Due to the Gaussian nature of the uncertainties, each observation will differ.
The analysis is performed on each observation individually, and the determined ranges used to find a mean range and uncertainty.

\begin{figure}
    \centering
    \includegraphics[width=8.5cm]{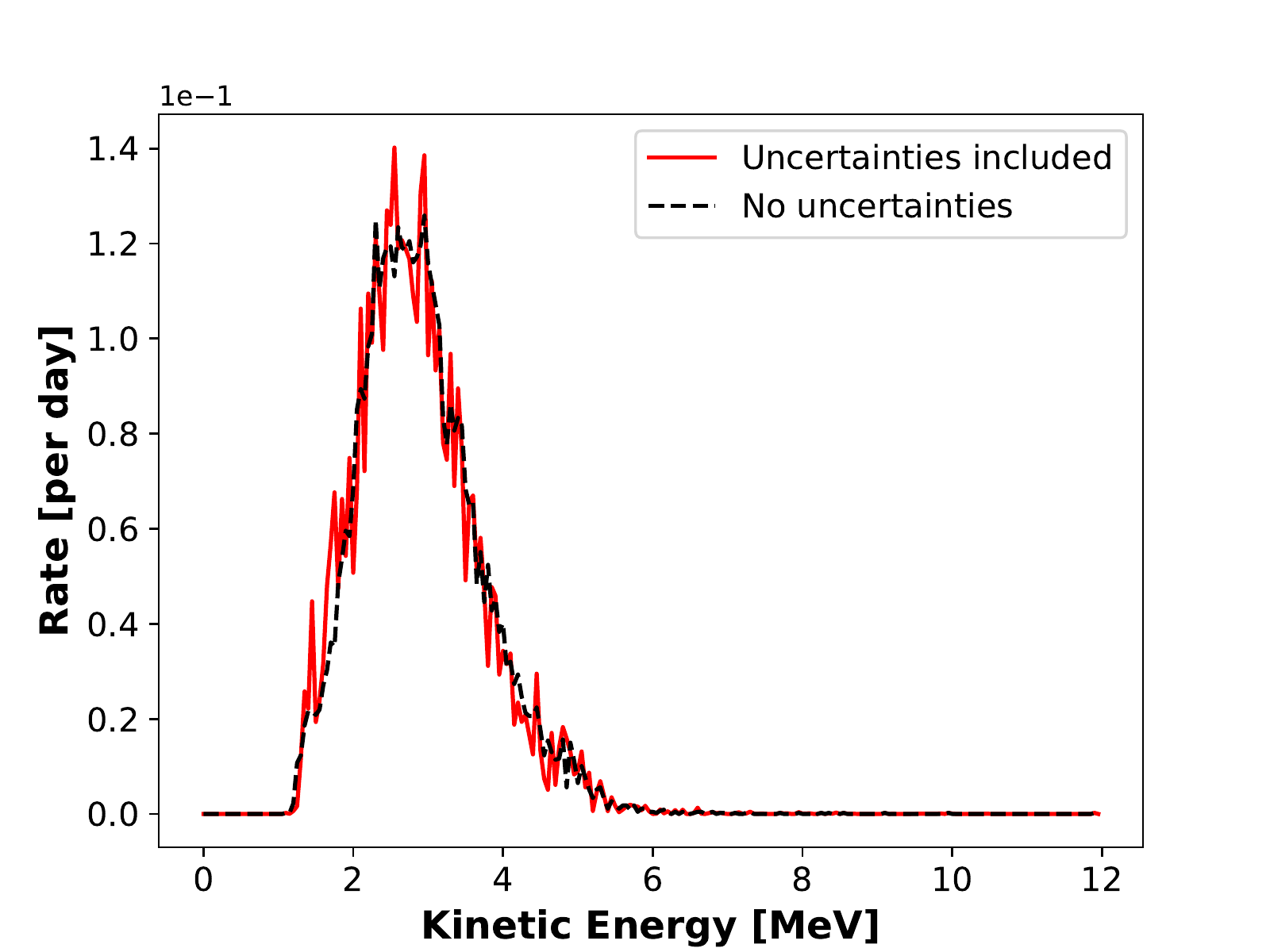}
    \caption{Simulated positron energy spectrum after data reduction for the Heysham 2 cores with reconstructed energy (both) and background uncertainties (red, solid).}
    \label{fig:hey-spec_reco}
\end{figure}

Further to these background scenarios, the effects of detector energy resolution are considered by using the number of PMT hits for a detected prompt interaction to determine the energy.
A linear fit between collected charge and true positron or electron energy is used to convert the PMT response to approximate particle energy.
The effect of this, alongside background uncertainties, on the spectrum for Heysham 2 can be seen in Fig. \ref{fig:hey-spec_reco}.
The low energy events, when compared the spectra in Fig. \ref{fig:Hey-emission-models}, are not efficiently reconstructed due to the lower light yield, resulting in only the higher energy peak.
The simulated true energy of particles is used to remove detector effects.
By comparing the results from reactor ranging when true energy is used to when reconstructed energy is used to produce the detected positron spectrum, the extent to which the energy resolution of this detector limits the ranging sensitivity can be obtained.

The combination of background scenarios and energy resolution options allows confirmation on the limiting factor for each signal i.e. whether the background, detector or analysis limits sensitivity.

\section{Spectral analysis}
\label{sec:analyses}






As shown in Eq. \ref{eq:antineu_survival}, the oscillation probability of one neutrino flavor state to another is proportional to $\sin ^2(\frac{1.27\Delta m_{ij}^2L}{E_{\bar{\nu}}})$.
As such, the oscillation of neutrino flavor is sinusoidally dependent on the distance of travel and energy of the neutrino.
The kinetic energy of the positrons from IBD can be measured and the antineutrino energy determined from this, meaning an Fourier Transform (FT) can be used to switch from antineutrino energy to the distance of travel.
The application of Fourier transforms has been proposed for measuring oscillations from reactor antineutrinos previously \cite{FCTFST}, with applications in detectors such as JUNO being considered \cite{Ciuffoli2014}.
In these cases, the energy spectrum is detected and the distance is known, with the oscillation parameters being determined from analysis.
In the situations considered in this work, the distance is the unknown quantity being determined from analysis.

As the oscillation probability depends on $\sin ^2(\frac{1.27\Delta m_{ij}^2L}{E_{\bar{\nu}}})$, the identity $\sin ^2(\theta) = \frac{1-\cos(2\theta)}{2}$ can be used to express the FT as
\begin{equation}
    \mathrm{FCT}(L) \propto \int_{\frac{1}{E_{max}}}^{\frac{1}{E_{min}}} f(L,E_{\bar{\nu}}) \cos \bigg(2 \times \frac{ 1.27 \Delta m_{ij}^2 L}{E_{\bar{\nu}}}\bigg) d\frac{1}{E_{\bar{\nu}}}.
    \label{eq:FCT}
\end{equation}
Here, Eq. \ref{eq:FCT} is defined as a Fourier Cosine Transform (FCT), and $f(L,E_{\bar{\nu}})$ is the model in Eq. \ref{eq:pdf}. A $\frac{\pi}{2}$ phase shift can be applied for a Fourier Sine Transform (FST)
\begin{equation}
    \mathrm{FST}(L) \propto \int_{\frac{1}{E_{max}}}^{\frac{1}{E_{min}}} f(L,E_{\bar{\nu}}) \sin \bigg(2 \times \frac{ 1.27 \Delta m_{ij}^2 L}{E_{\bar{\nu}}}\bigg) d\frac{1}{E_{\bar{\nu}}}.
    \label{eq:FST}
\end{equation}
The splitting of the FT into an FCT and FST consistent with the previous proposal \cite{FCTFST}, as each transform highlights spectral features differently.

\begin{figure}[htb]
    \centering
    \includegraphics[width=8.5cm]{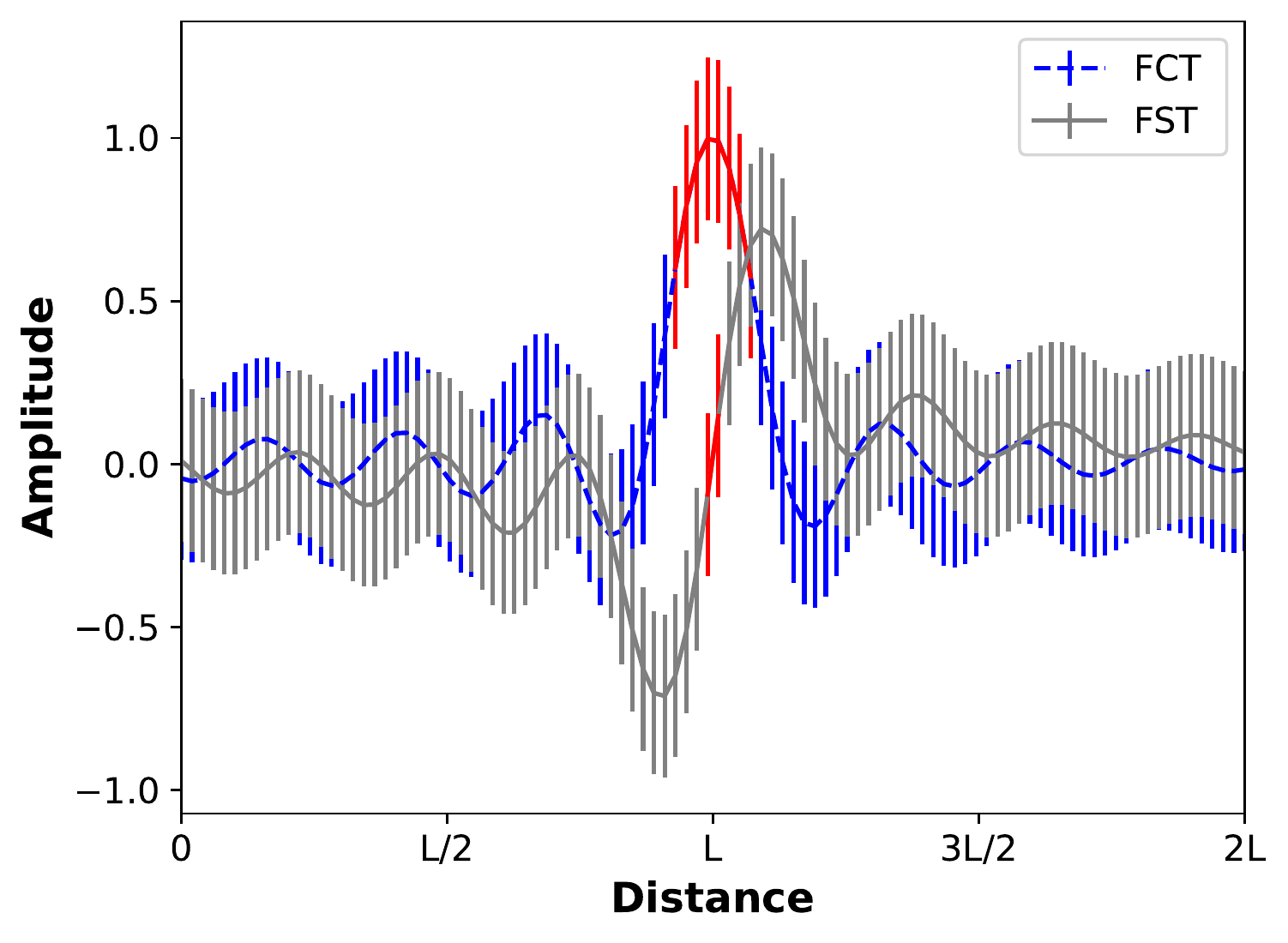}
    \caption{The combination of an FCT (blue dashed) and FST (grey solid) allows the area of interest (red solid) to be narrowed down to reduce uncertainties by comparing where the maxima of the FCT and zeroes of the FST occur at matching distances.}
    \label{fig:FCTFST}
\end{figure}

The difference between the FCT and FST due to the phase shift can be used to improve the precision of the analysis by combining the two.
As the distance is varied, where the peak amplitude of the FCT is the determined range, the zero amplitude of the FST are the points of interest due to the phase shift.
The FCT can be used to determine the region of interest, with all possible distances within uncertainty of the peak amplitude being considered.
This provides constraints to the FST, which can be used to determine the final distance and uncertainties by searching for a value of zero within the constraints set by the FCT.
The FST has its steepest gradient at the region of interest, which reduces the potential distances that fall within the uncertainty and hence reduces the uncertainty.
Fig. \ref{fig:FCTFST} shows how the FCT and FST can be used in combination to reduce the possible ranges responsible for the detected spectrum by only considering the regions in which the peak of the FCT matches a zero-amplitude point of the FST.

\begin{figure}[htb]
 \begin{subfigure}[b]{0.495\textwidth}
  \includegraphics[width=\textwidth]{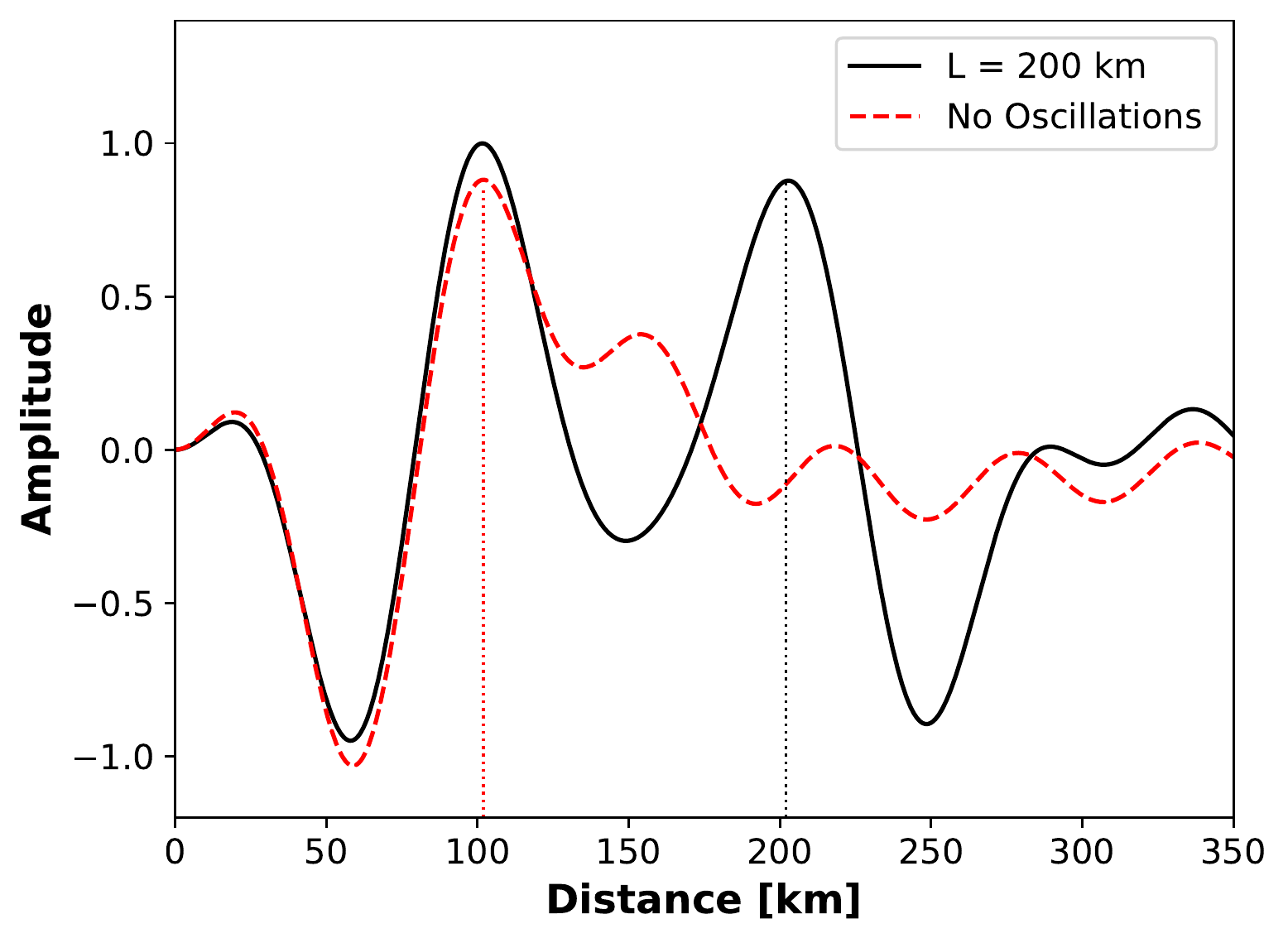}
  \caption{}
  \label{fig:FCT-osc}
 \end{subfigure}
 \hfill
 \begin{subfigure}[b]{0.495\textwidth}
 \includegraphics[width=\textwidth]{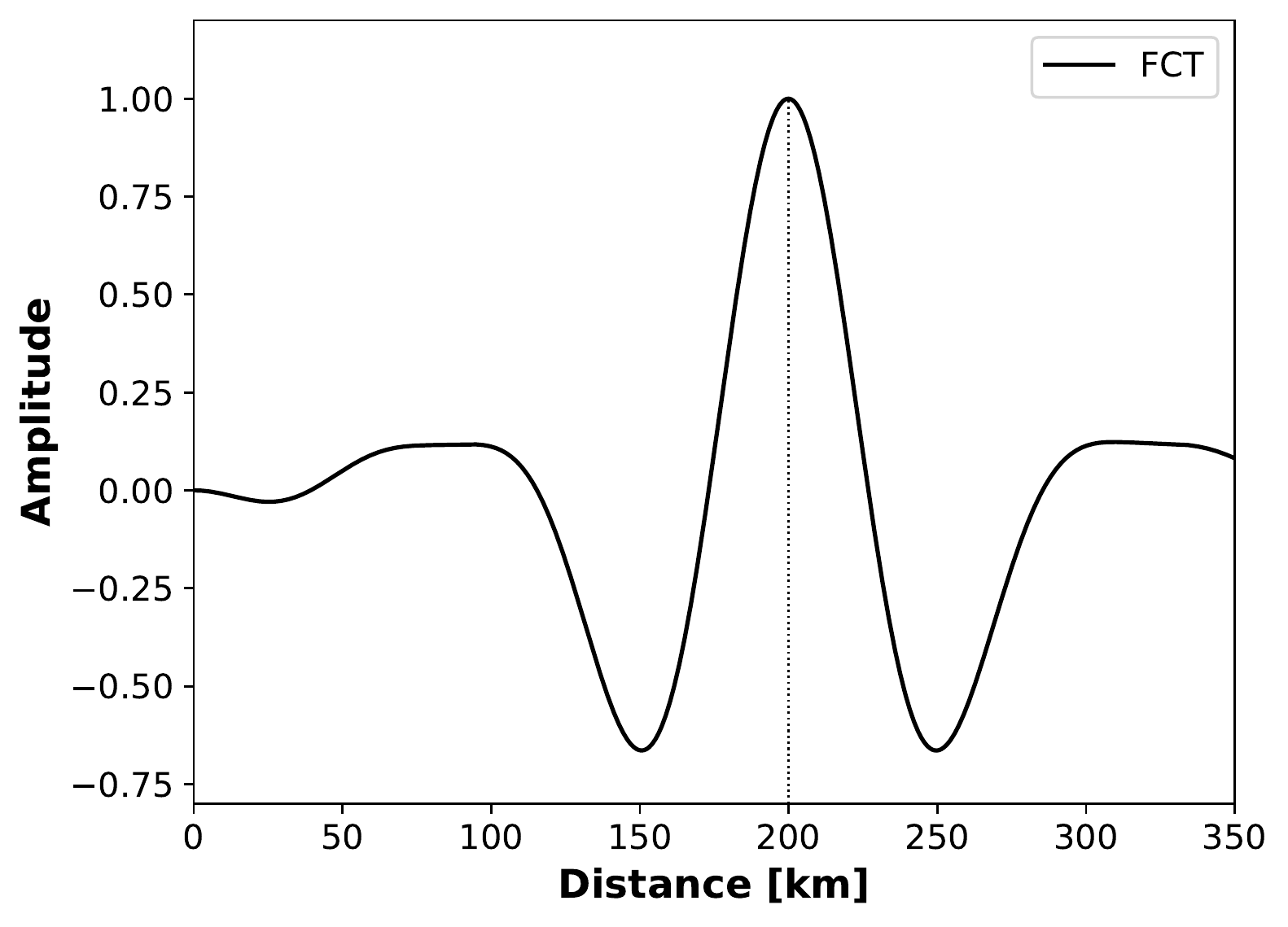}
 \caption{}
 \label{fig:FCT-osc-sub}
 \end{subfigure}
 \caption{(a) Comparison of the FCT for oscillations (black solid) and no oscillations (red dashed) in the reactor spectrum model for a 200 km standoff distance, and (b) the subtraction of the no oscillation situation from the original reactor model for the same reactor standoff. The reactor model has peaks for 100 km and 200 km before the subtraction, and only the expected peak at 200 km after subtraction.}
 \label{fig:FCT-osc-subtract}
\end{figure}

Both Eq. \ref{eq:FCT} and Eq. \ref{eq:FST} include terms not associated with neutrino oscillations within the term $f(L,E_{\bar{\nu}})$.
To isolate the oscillation terms, a spectrum where no oscillations are assumed is simulated i.e., the model in Eq. \ref{eq:pdf} with the survival probability set to one.
An FT is performed on this spectrum, and it is then subtracted from the one performed on the simulated data.
The effect of this can be seen clearly in Fig. \ref{fig:FCT-osc-subtract}, where the peak associated with factors not related to neutrino oscillations is removed.

\begin{figure}[htb]
    \centering
    \includegraphics[width=8.5cm]{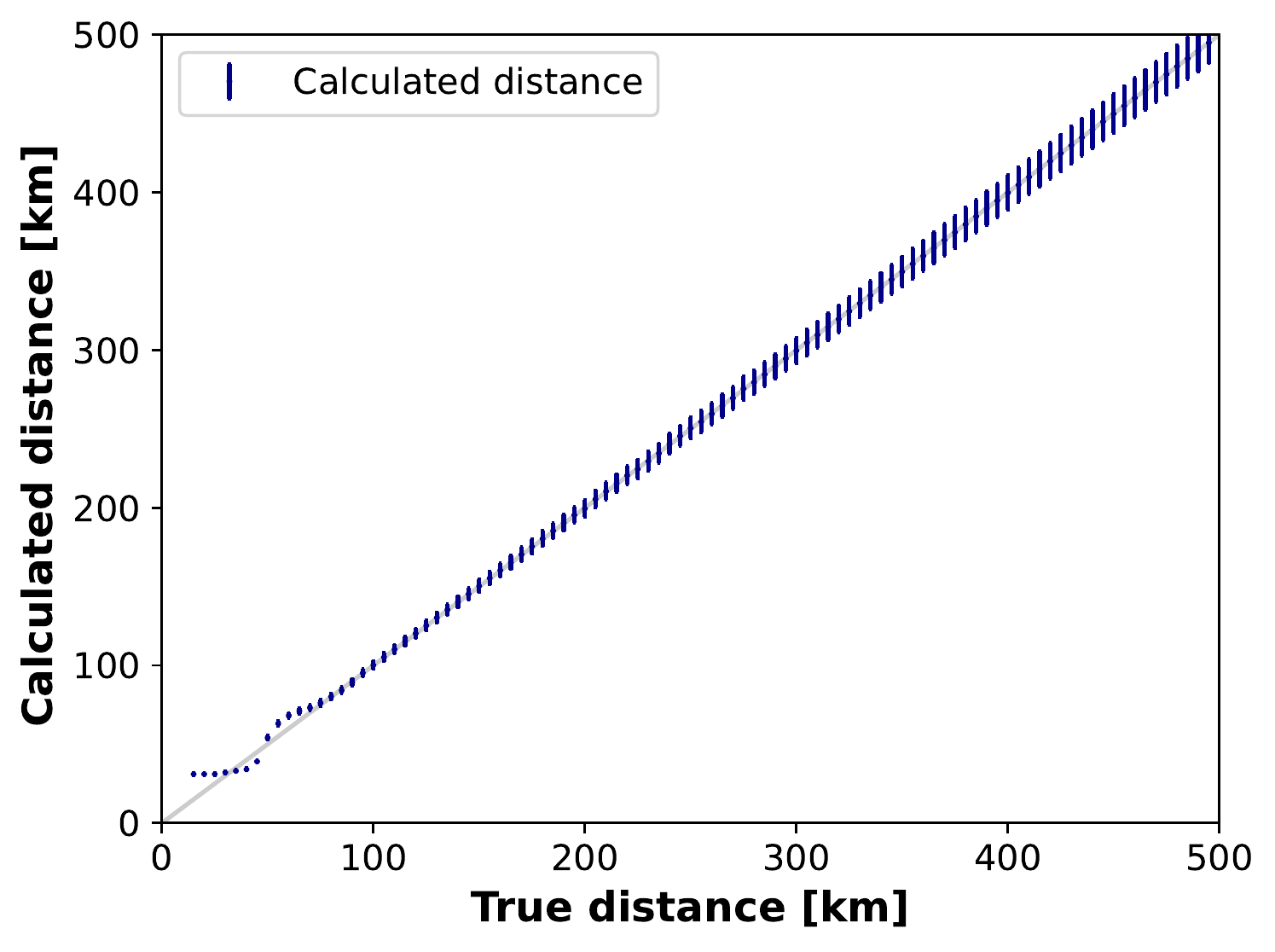}
    \caption{The calculated range of reactor signals with true distance using the Fourier transform analysis applied to Eq. \ref{eq:pdf}. The FT relies on resolving the $\theta_{12}$ oscillations, which are not obviously present at ranges below 100 km, as the $\theta_{13}$ oscillations are smaller than the detector's energy resolution.}
    \label{fig:FT-range-limit}
\end{figure}

Due to the detector resolution, only the $\theta_{12}$ oscillation pattern can be resolved.
As such, the FTs are normalized to the $\theta_{12}$ term, and $\theta_{13}$ and $\theta_{23}$ are neglected.
This creates a lower limit to the range that can be observed with this method, as at least a significant part of one full wavelength of the oscillation pattern must be visible in the spectrum for an FT to work.
Fig. \ref{fig:survival-prob} shows how the larger changes in electron antineutrino survival probability due to $\theta_{12}$ do not occur until approaching a 100 km distance of travel.
The oscillations due to $\theta_{13}$, that occur at short distances, produce much smaller changes in the survival probability.
A lower limit of approximately 80 km, due to the requirement of a full wavelength of the oscillation pattern, can be seen in Fig. \ref{fig:FT-range-limit} when the analysis is applied to the model in Eq. \ref{eq:pdf}.

\section{Results}
\label{sec:results}

Due to the detector's energy resolution giving a minimum distance that the Fourier transform is effective at (Fig. \ref{fig:FT-range-limit}), the Hartlepool complex could not be ranged using this detector at Boulby.
While the 26 km standoff results in a signal dominated spectrum, there are no clear features from the $\theta_{12}$ oscillations that can be used for ranging.

The FT method is applied in four possible scenarios on five reactor complexes.
The scenarios are combinations of including uncertainties and detector energy resolution, all of which assume background rates are known.
The target reactors used are those more distant than Hartlepool, starting at Heysham, under the assumption that the Hartlepool cores are decommissioned and do not contribute to the background.
As these reactors are at large distances, the signals are small.
Any assumption that all backgrounds after data reduction are unknown and indistinguishable from the signal would mean that ranging is impossible.
As such, the two known background scenarios are used, in which it is assumed that background contributions to the total spectrum could be removed, either completely or leaving some uncertainties in the remaining signal.

The results of the FT method shown in Table \ref{tab:results-ft} show that for reactors at large distances, the range can be determined when the detector's energy resolution is accounted for.
However, reactors beyond 300 km do not have a large enough signal to be ranged effectively when background uncertainties are included.
The two nearer reactor sites, AGRs at Heysham and Torness, can be ranged close to the true value when uncertainties are included.
Heysham 1 \& 2 can be ranged to the same level as Heysham 2 in isolation as the spectral shapes are the same, but the required time is lower due to the increased signal rate.

\begin{table*}[!htb]
    \centering
    \begin{tabular}{l|ccccc}
    \hline\hline
        \multirow{2}{2.5em}{Situation} & \multicolumn{5}{c}{Range (km)}\\
        \cline{2-6}
        & {Heysham}  & {Torness} & {Sizewell B} & {Hinkley Point C} & {Gravelines}\\
        \hline
        True Range & 149 & 187 & 304 & 404 & 441 \\
        \hline
        No Uncertainties, True Energy & 148 $\pm$ 4 & 188 $\pm$ 5 & 306 $\pm$ 8 &  403 $\pm$ 11 & 440 $\pm$ 11 \\
        No Uncertainties, Reconstructed Energy & 157 $\pm$ 4 & 195 $\pm$ 5 & 307 $\pm$ 8 &  397 $\pm$ 11 & 432 $\pm$ 11 \\
        Uncertainties, True Energy & 156 $\pm$ 6 & 177 $\pm$ 10 & \diagbox[innerwidth=5em, height=0.5\line]{}{} & \diagbox[innerwidth=5em, height=0.5\line]{}{} & \diagbox[innerwidth=5em, height=0.5\line]{}{}\\  
        Uncertainties, Reconstructed Energy & 155 $\pm$ 5 & 171 $\pm$ 9 & \diagbox[innerwidth=5em, height=0.5\line]{}{} & \diagbox[innerwidth=5em, height=0.5\line]{}{} & \diagbox[innerwidth=5em, height=0.5\line]{}{}\\
        \hline\hline
        Time (Years) & 21 & 74 & 200 & 100 & 100\\
    \hline\hline
    \end{tabular}
    \caption[Results summary FT]{Determined distance in km. Both the true energy and reconstructed energy are considered, as is the inclusion of uncertainties. Situations with a slash are deemed impossible to range due to background uncertainties dominating. The time to determine the quotes ranges are based on requiring enough events to produce a complete spectrum.}
    \label{tab:results-ft}
\end{table*}

\begin{figure}[htb]
    \centering
    \includegraphics[width=8.5cm]{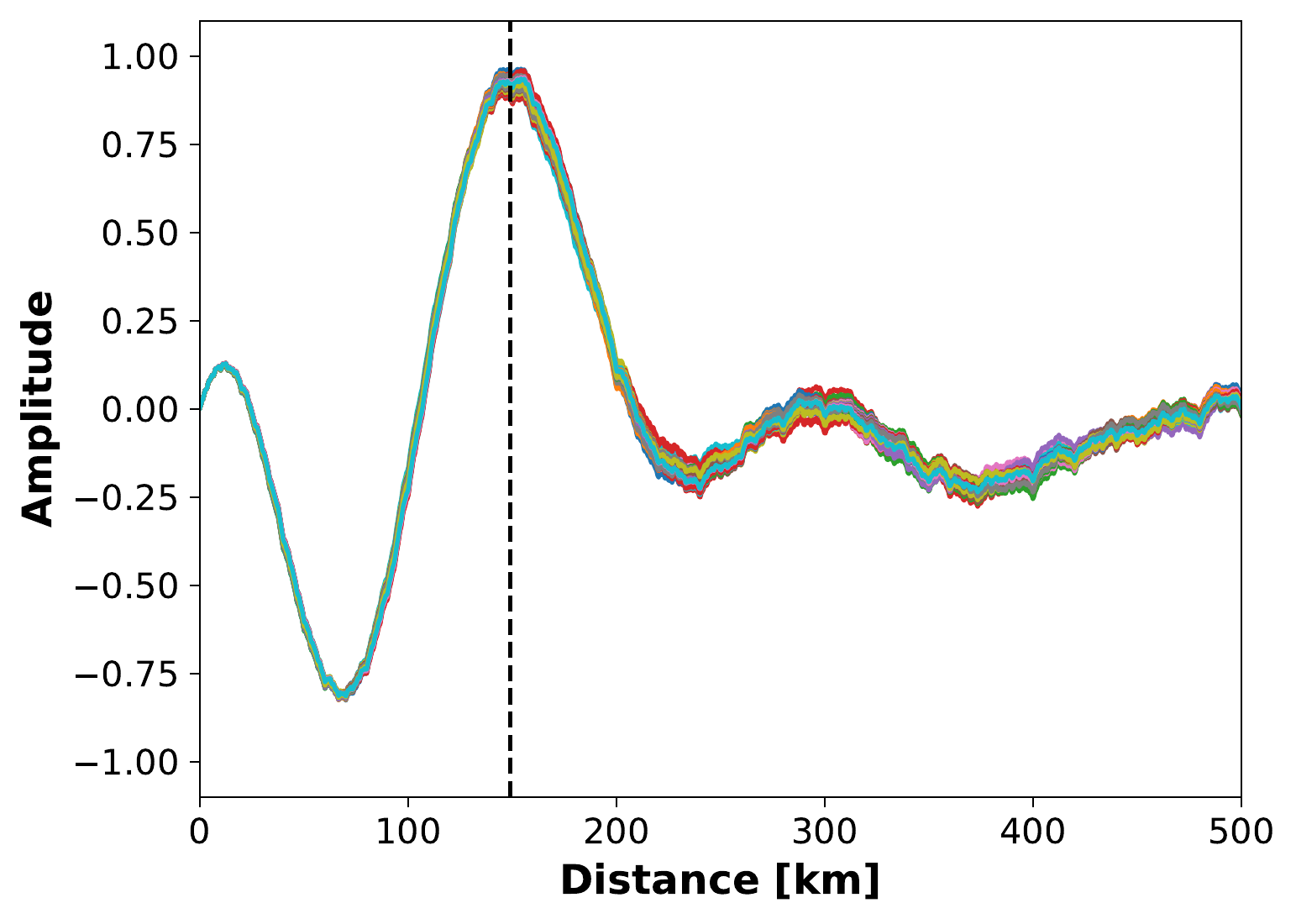}
    \caption{FCT analysis of Heysham 2 repeated 100 times to show the effect of fluctuations due to uncertainties. The vertical dashed line at 149 km is the true distance to Heysham 2 from Boulby.}
    \label{fig:hey2_fct_uncert}
\end{figure}

Uncertainties are determined by repeating the FT, sampling from the uncertainty distributions separately each time.
This produces a range of possible FTs, which fluctuate due to uncertainties.
These fluctuations can be seen in the FCT for Heysham 2 in Fig. \ref{fig:hey2_fct_uncert}, where the spectrum has been produced one hundred times and the analysis repeated.
This results in a range of possible amplitudes for every distance for both the FCT and FST.
The range of amplitudes is then used to determine the uncertainty on the amplitudes from the FTs.
The uncertainty on the distance is then determined by the range of distances that fall within the uncertainty of the amplitudes accepted by both the FCT and FST.

\begin{figure}[!htb]
\includegraphics[width=8.5cm]{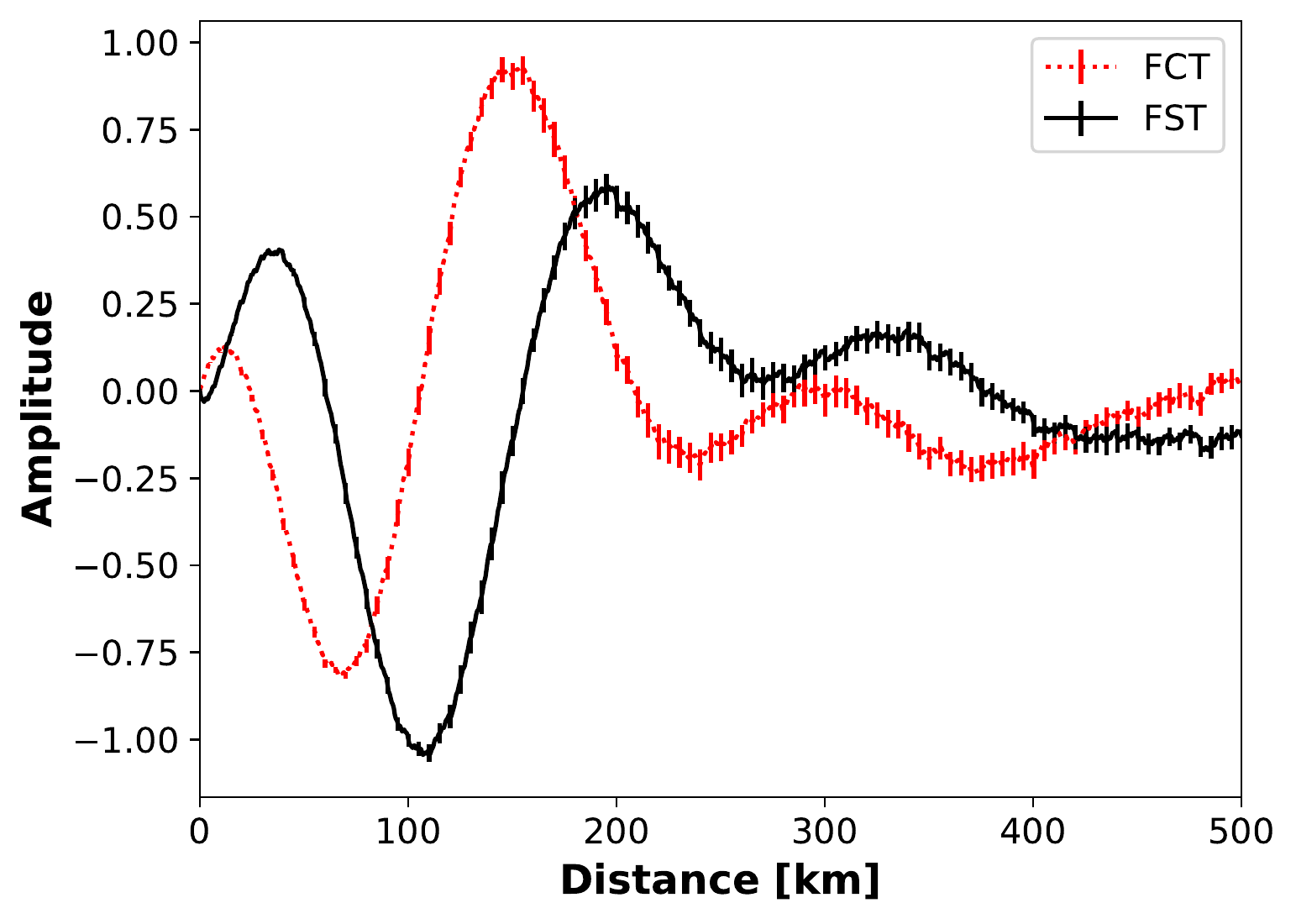}
\centering
\caption{The Fourier transform, in sine (black solid) and cosine (red dotted), for Heysham 2 with background uncertainties and energy resolution effects.}
\label{fig:hey-FT}
\end{figure}

The FT for Heysham 2 with uncertainties and with energy reconstruction applied is shown in Fig. \ref{fig:hey-FT}. The maximum for the FCT yields an accurate range, but with an uncertainty of $\pm$ 15 km. The FST is able to reduce this uncertainty significantly to $\pm$ 6 km, shown in Table \ref{tab:results-ft}.

Due to the low event rates for the distant reactors, it takes over 20 years of observation time to be able to range the Heysham complex, and significantly longer for the more distant reactors.

\section{Alternative detectors}
\label{sec:other_detectors}

The JUNO liquid scintillator detector aims to use multiple reactors at known distances to measure neutrino oscillation parameters \cite{JUNO2022} by applying Fourier cosine and sine transforms.
This is similar to this work, but with a fixed distance and floating oscillation parameters, so the same techniques can be applied by switching the free parameters.
The JUNO detector is expected to have a 73\% IBD detection efficiency \cite{An2016}, and is significantly larger than the AIT-NEO proposal at 20 kT.
Before the consideration of improved energy performance due to the use of liquid scintillator allowing better analysis cuts to be made, the time required to range Heysham reduces to just over 2 years.
Improvements to analysis due to better energy resolution could further lower this.

By using a scintillator-filled detector with a good energy resolution, closer reactors could potentially be considered as the $\theta_{13}$ oscillations could be resolved.
JUNO will be observing reactors at 53 km \cite{JUNO2022}.
Closer reactors would mean a significantly larger antineutrino flux for the same reactor power.
A reactor at 50 km could be ranged in around 1 year using a NEO-sized detector filled with liquid scintillator if $\theta_{13}$ oscillations could be resolved, and a reactor at 100 km could be ranged in 4 years.
A 20 kT volume of liquid scintillator could range a 50 km reactor in 4 months, and a 100 km reactor in 1 year.

The proposal made for the Theia WbLS detector includes a higher scintillation concentration than the 1\% used here, and a very high photocoverage \cite{Askins2020}.
As such, the detection efficiency could be as high as 95\% with neutron tagging.
Two detector sizes have been proposed, 25 kT with a 17.5 kT fidicual volume, and a 100 kT detector with a 70 kT fiducial volume.
From this, the smaller Theia detector could range the Heysham complex in 2 years, and the larger detector in under 6 months.
Theia is actively considering reactors at 1000s km standoffs \cite{Zsoldos2022}.
This does not take into account the improved energy resolution due to the higher scintillation concentration and light collection efficiency.
Again, further improvement could potentially be obtained here.

\section{Discussion}
\label{sec:discussion}

The results of the Fourier transform indicate that this detector is insufficient to range the UK's reactor fleet.
The nearest reactor, 26 km away at Hartlepool, is too close for $\theta_{12}$ oscillation features to be useful, and the detector does not have the necessary energy resolution to resolve the $\theta_{13}$ features.
The more distant reactors, at over 150 km, do not produce a large enough signal in this detector design to be ranged in a practical time, with the Heysham complex requiring over 20 years even when backgrounds are known and subtracted from the total spectrum.

However, given enough data to build a spectrum, the analysis is capable of ranging the source of the spectrum even when background uncertainties are included on small signals.
This indicates the analysis itself has potential; this type of analysis is being considered for reactor neutrino oscillation experiments such as JUNO \cite{FCTFST,Ciuffoli2014}.
By considering the improved performance of other detector designs, such as JUNO or the proposed Theia WbLS-filled detector, this analysis could have some utility.
Based on increased volumes and detection efficiencies, both JUNO and the smaller Theia design could reduce the required observation time by an order of magnitude.
This is before improved energy resolution is considered.
JUNO is expected to have a resolution of 3\% at 1 MeV \cite{An2016}, and Theia could have 7\% at 1 MeV \cite{Askins2020}; the NEO detector considered has a resolution of 18\% at 1 MeV.
The substantially better energy resolutions provided by other detector designs could allow better data reduction, and the observation of the $\theta_{13}$ oscillation features.
This allows closer reactors that give much larger signals to be considered, and would boost the rate of the more distant reactor measurements.

Beyond improving the detector design, a relaxing of the level of measurement precision may allow a ranging analysis to occur more quickly.
The results quoted are the best results where the most precise and accurate values are obtained based on the data reduction and rate-only analysis previously performed.
However, if instead of a precise measurement, a determination of whether the observed spectrum is consistent with expectation for the expected reactor operation is acceptable, a quicker result may be obtained.
In this case, rather than matching an exact model, showing the spectrum does not match the expected shape or FT would be a positive result, particularly if it is in combination with an observation of excess signal from the rate-only analysis.
The FT analysis could also be applied to remove signals from known reactors from the data.
If their distance and rate are known, using the FT analysis could allow spectra due to known reactor operations to be subtracted from the observed data.
This would then highlight unexpected reactor signals, a key goal of using antineutrinos for non-proliferation.

The limiting factor in the situations presented is the detector.
The detector is not large enough to detect a sufficient rate of events from more distant reactors, meaning a range cannot be determined in a practical observation time.
To do this, the detector would need to double its detection efficiency after data reduction and be significantly larger, at around 20 kT.
NEO also has an insufficient energy resolution to resolve the $\theta_{13}$ oscillation features, which creates a minimum distance of around 80 km that can be ranged.
As such, the more local reactors such as Hartlepool at 26 km cannot be ranged.
Hartlepool could be ranged with this detector in under 2 years based on its event rate if the oscillation features could be resolved.
To be able to range nearer reactors, it is likely a pure liquid scintillator or higher concentration of scintillator in WbLS would be needed.

Separating the detector's limitations from the analysis, the results show that the FT analysis shows some promise.
By considering other detector designs, a more timely ranging could potentially occur.
This would allow spectral information to be used in addition to the existing rate-determination techniques.
Applying spectral analysis to a detector with better performance, and using the analysis in tandem with other information, means this technique may be applicable for non-proliferation.
Using Fourier transforms to analyze reactor antineutrino oscillations in this way, but with the distance fixed and the oscillation parameters free, is considered for the JUNO liquid scintillator detector.

It should be noted that the scenarios tested all use reactors with enriched uranium as fuel, and uses limited fuel evolution in the generation of spectra.
While for the AGRs, which are all reactors that could be ranged when uncertainties on all backgrounds are considered, this is not expected to change the overall results, other reactor types may need a more comprehensive inclusion of core burnup.
The analysis could be applied to any neutrino detector and any reactor landscape, and assuming the fuel type and evolution is appropriately handled in modeling expected spectra, the analysis should still perform in a similar fashion.

\section{Conclusion}
\label{sec:conclusion}

A Fourier transform spectral analysis has been developed and applied to an existing data reduction for reactor antineutrino monitoring in the mid- to far-field.
The analysis is applied to a previously developed detector design placed underground in Boulby Mine, and the UK reactor landscape is simulated to provide a realistic test.

The detector is unable to practically range any of the reactors in the UK's fleet.
However, the analysis shows potential when other detector designs are considered.
By applying this technique to a different detector design, such as JUNO or Theia, ranging could be performed in a more reasonable time.
This could allow it to be an extra tool for non-proliferation as it provides extra information on top of existing analysis techniques.
For ranging to be possible in the near- to mid-field, a better energy resolution is needed; for ranging far-field reactors, a much larger detector with a better detection efficiency is needed.


\begin{acknowledgments}

This work was supported in the U.K. by the Science and Technology Facilities Council (STFC).

This work was performed under the auspices of the U.S. Department of Energy (DOE) by
Lawrence Livermore National Laboratory under contract DE-AC52-07NA27344, and supported
by the DOE National Nuclear Security Administration under Award Number DE-NA0000979.
LLNL-JRNL-850296.

This work was supported by the University of Sheffield Institutional Open Access Fund.
For the purpose of open access, the author has applied a Creative Commons Attribution (CC BY) licence to any Author Accepted Manuscript version arising.

The authors would like to thank L. Kneale for her input on the simulation and analysis of data (see \cite{Sensitivity2022}) as well as her review of the work. Thanks also go to T. Appleyard for her early work on the LEARN analysis, and A. Scarff for his regular review of the work.

\end{acknowledgments}

\section*{Author contributions}

The initial proposal of both reactor ranging in general and the application of a Fourier transform came from J. G. Learned.

Monte Carlo simulations produced by S. Wilson with input from L. Kneale.

The LEARN analysis for data reduction developed by S. Wilson with input from J. Armitage, N. Holland and T. Appleyard. Sections of the data reduction, such as the analytic post-muon veto and radionuclide calculations were developed by L. Kneale.

The initial ranging concept was worked on by J. Armitage, with the Fourier transform being developed by C. Cotsford.

\bibliography{aps}

\begin{thebibliography}{66}%
\makeatletter
\providecommand \@ifxundefined [1]{%
 \@ifx{#1\undefined}
}%
\providecommand \@ifnum [1]{%
 \ifnum #1\expandafter \@firstoftwo
 \else \expandafter \@secondoftwo
 \fi
}%
\providecommand \@ifx [1]{%
 \ifx #1\expandafter \@firstoftwo
 \else \expandafter \@secondoftwo
 \fi
}%
\providecommand \natexlab [1]{#1}%
\providecommand \enquote  [1]{``#1''}%
\providecommand \bibnamefont  [1]{#1}%
\providecommand \bibfnamefont [1]{#1}%
\providecommand \citenamefont [1]{#1}%
\providecommand \href@noop [0]{\@secondoftwo}%
\providecommand \href [0]{\begingroup \@sanitize@url \@href}%
\providecommand \@href[1]{\@@startlink{#1}\@@href}%
\providecommand \@@href[1]{\endgroup#1\@@endlink}%
\providecommand \@sanitize@url [0]{\catcode `\\12\catcode `\$12\catcode `\&12\catcode `\#12\catcode `\^12\catcode `\_12\catcode `\%12\relax}%
\providecommand \@@startlink[1]{}%
\providecommand \@@endlink[0]{}%
\providecommand \url  [0]{\begingroup\@sanitize@url \@url }%
\providecommand \@url [1]{\endgroup\@href {#1}{\urlprefix }}%
\providecommand \urlprefix  [0]{URL }%
\providecommand \Eprint [0]{\href }%
\providecommand \doibase [0]{http://dx.doi.org/}%
\providecommand \selectlanguage [0]{\@gobble}%
\providecommand \bibinfo  [0]{\@secondoftwo}%
\providecommand \bibfield  [0]{\@secondoftwo}%
\providecommand \translation [1]{[#1]}%
\providecommand \BibitemOpen [0]{}%
\providecommand \bibitemStop [0]{}%
\providecommand \bibitemNoStop [0]{.\EOS\space}%
\providecommand \EOS [0]{\spacefactor3000\relax}%
\providecommand \BibitemShut  [1]{\csname bibitem#1\endcsname}%
\let\auto@bib@innerbib\@empty
\bibitem [{\citenamefont {{International Atomic Energy Agency}}(2020)}]{iaea2020predictions}%
  \BibitemOpen
  \bibfield  {author} {\bibinfo {author} {\bibnamefont {{International Atomic Energy Agency}}},\ }\href {https://www.iaea.org/publications/14786/energy-electricity-and-nuclear-power-estimates-for-the-period-up-to-2050} {\emph {\bibinfo {title} {Energy, Electricity and Nuclear Power Estimates for the Period up to 2050}}},\ \bibinfo {series} {Reference Data Series}\ No.~\bibinfo {number} {1}\ (\bibinfo  {publisher} {IAEA},\ \bibinfo {address} {Vienna},\ \bibinfo {year} {2020})\BibitemShut {NoStop}%
\bibitem [{\citenamefont {Christensen}, \citenamefont {Huber},\ and\ \citenamefont {Jaffke}(2015)}]{Christensen2014}%
  \BibitemOpen
  \bibfield  {author} {\bibinfo {author} {\bibfnamefont {E.}~\bibnamefont {Christensen}}, \bibinfo {author} {\bibfnamefont {P.}~\bibnamefont {Huber}}, \ and\ \bibinfo {author} {\bibfnamefont {P.}~\bibnamefont {Jaffke}},\ }\bibfield  {title} {\enquote {\bibinfo {title} {Antineutrino reactor safeguards: A case study of the {DPRK} 1994 nuclear crisis},}\ }\href {\doibase 10.1080/08929882.2015.996076} {\bibfield  {journal} {\bibinfo  {journal} {Sci. Glob. Secur.}\ }\textbf {\bibinfo {volume} {23}},\ \bibinfo {pages} {20--47} (\bibinfo {year} {2015})}\BibitemShut {NoStop}%
\bibitem [{\citenamefont {{International Atomic Energy Agency}}(1997)}]{IAEA-Safeguard}%
  \BibitemOpen
  \bibfield  {author} {\bibinfo {author} {\bibnamefont {{International Atomic Energy Agency}}},\ }\href {https://www.iaea.org/sites/default/files/infcirc540c.pdf} {\enquote {\bibinfo {title} {Model protocol additional to the agreements(s) between the state(s) and the international atomic energy agency for the application of safeguards},}\ } (\bibinfo {year} {1997})\BibitemShut {NoStop}%
\bibitem [{\citenamefont {Akindele}\ \emph {et~al.}(2021)\citenamefont {Akindele}, \citenamefont {Bowden}, \citenamefont {Carr}, \citenamefont {Conant}, \citenamefont {Diwan}, \citenamefont {Erickson}, \citenamefont {Foxe}, \citenamefont {Goldblum}, \citenamefont {Huber}, \citenamefont {Jovanovic}, \citenamefont {Link}, \citenamefont {Littlejohn}, \citenamefont {Mumm},\ and\ \citenamefont {Newby}}]{NuTools}%
  \BibitemOpen
  \bibfield  {author} {\bibinfo {author} {\bibfnamefont {T.}~\bibnamefont {Akindele}}, \bibinfo {author} {\bibfnamefont {N.}~\bibnamefont {Bowden}}, \bibinfo {author} {\bibfnamefont {R.}~\bibnamefont {Carr}}, \bibinfo {author} {\bibfnamefont {A.~J.}\ \bibnamefont {Conant}}, \bibinfo {author} {\bibfnamefont {M.}~\bibnamefont {Diwan}}, \bibinfo {author} {\bibfnamefont {A.}~\bibnamefont {Erickson}}, \bibinfo {author} {\bibfnamefont {M.~P.}\ \bibnamefont {Foxe}}, \bibinfo {author} {\bibfnamefont {B.~L.}\ \bibnamefont {Goldblum}}, \bibinfo {author} {\bibfnamefont {P.}~\bibnamefont {Huber}}, \bibinfo {author} {\bibfnamefont {I.}~\bibnamefont {Jovanovic}}, \bibinfo {author} {\bibfnamefont {J.~M.}\ \bibnamefont {Link}}, \bibinfo {author} {\bibfnamefont {B.~R.}\ \bibnamefont {Littlejohn}}, \bibinfo {author} {\bibfnamefont {H.~P.}\ \bibnamefont {Mumm}}, \ and\ \bibinfo {author} {\bibfnamefont {J.}~\bibnamefont {Newby}},\ }\bibfield  {title} {\enquote {\bibinfo {title} {Nu tools: Exploring practical roles for neutrinos
  in nuclear energy and security},}\ }\href {\doibase 10.2172/1826602} {\  (\bibinfo {year} {2021}),\ 10.2172/1826602}\BibitemShut {NoStop}%
\bibitem [{\citenamefont {Bowden}\ \emph {et~al.}(2009)\citenamefont {Bowden}, \citenamefont {Bernstein}, \citenamefont {Dazeley}, \citenamefont {Svoboda}, \citenamefont {Misner},\ and\ \citenamefont {Palmer}}]{Bowden2009}%
  \BibitemOpen
  \bibfield  {author} {\bibinfo {author} {\bibfnamefont {N.~S.}\ \bibnamefont {Bowden}}, \bibinfo {author} {\bibfnamefont {A.}~\bibnamefont {Bernstein}}, \bibinfo {author} {\bibfnamefont {S.}~\bibnamefont {Dazeley}}, \bibinfo {author} {\bibfnamefont {R.}~\bibnamefont {Svoboda}}, \bibinfo {author} {\bibfnamefont {A.}~\bibnamefont {Misner}}, \ and\ \bibinfo {author} {\bibfnamefont {T.}~\bibnamefont {Palmer}},\ }\bibfield  {title} {\enquote {\bibinfo {title} {Observation of the isotopic evolution of pressurized water reactor fuel using an antineutrino detector},}\ }\href {\doibase 10.1063/1.3080251} {\bibfield  {journal} {\bibinfo  {journal} {J. App. Phys.}\ }\textbf {\bibinfo {volume} {105}},\ \bibinfo {pages} {064902} (\bibinfo {year} {2009})}\BibitemShut {NoStop}%
\bibitem [{\citenamefont {Kneale}\ \emph {et~al.}(2023)\citenamefont {Kneale}, \citenamefont {Wilson}, \citenamefont {Appleyard}, \citenamefont {Armitage}, \citenamefont {Holland},\ and\ \citenamefont {Malek}}]{Sensitivity2022}%
  \BibitemOpen
  \bibfield  {author} {\bibinfo {author} {\bibfnamefont {L.}~\bibnamefont {Kneale}}, \bibinfo {author} {\bibfnamefont {S.}~\bibnamefont {Wilson}}, \bibinfo {author} {\bibfnamefont {T.}~\bibnamefont {Appleyard}}, \bibinfo {author} {\bibfnamefont {J.}~\bibnamefont {Armitage}}, \bibinfo {author} {\bibfnamefont {N.}~\bibnamefont {Holland}}, \ and\ \bibinfo {author} {\bibfnamefont {M.}~\bibnamefont {Malek}},\ }\bibfield  {title} {\enquote {\bibinfo {title} {Sensitivity of an antineutrino monitor for remote nuclear reactor discovery},}\ }\href {\doibase 10.1103/PhysRevApplied.20.034073} {\bibfield  {journal} {\bibinfo  {journal} {Phys. Rev. Appl.}\ }\textbf {\bibinfo {volume} {20}},\ \bibinfo {pages} {034073} (\bibinfo {year} {2023})}\BibitemShut {NoStop}%
\bibitem [{\citenamefont {Akindele}\ \emph {et~al.}(2023)\citenamefont {Akindele}, \citenamefont {Bernstein}, \citenamefont {Bergevin}, \citenamefont {Dazeley}, \citenamefont {Sutanto}, \citenamefont {Mullen},\ and\ \citenamefont {Hecla}}]{Akindele2023}%
  \BibitemOpen
  \bibfield  {author} {\bibinfo {author} {\bibfnamefont {O.}~\bibnamefont {Akindele}}, \bibinfo {author} {\bibfnamefont {A.}~\bibnamefont {Bernstein}}, \bibinfo {author} {\bibfnamefont {M.}~\bibnamefont {Bergevin}}, \bibinfo {author} {\bibfnamefont {S.}~\bibnamefont {Dazeley}}, \bibinfo {author} {\bibfnamefont {F.}~\bibnamefont {Sutanto}}, \bibinfo {author} {\bibfnamefont {A.}~\bibnamefont {Mullen}}, \ and\ \bibinfo {author} {\bibfnamefont {J.}~\bibnamefont {Hecla}},\ }\bibfield  {title} {\enquote {\bibinfo {title} {Exclusion and verification of remote nuclear reactors with a 1-kiloton $\mathrm{Gd}$-doped water detector},}\ }\href {\doibase 10.1103/PhysRevApplied.19.034060} {\bibfield  {journal} {\bibinfo  {journal} {Phys. Rev. Appl.}\ }\textbf {\bibinfo {volume} {19}},\ \bibinfo {pages} {034060} (\bibinfo {year} {2023})}\BibitemShut {NoStop}%
\bibitem [{\citenamefont {Li}\ \emph {et~al.}(2022)\citenamefont {Li}, \citenamefont {Dazeley}, \citenamefont {Bergevin},\ and\ \citenamefont {Bernstein}}]{Slava2022}%
  \BibitemOpen
  \bibfield  {author} {\bibinfo {author} {\bibfnamefont {V.~A.}\ \bibnamefont {Li}}, \bibinfo {author} {\bibfnamefont {S.~A.}\ \bibnamefont {Dazeley}}, \bibinfo {author} {\bibfnamefont {M.}~\bibnamefont {Bergevin}}, \ and\ \bibinfo {author} {\bibfnamefont {A.}~\bibnamefont {Bernstein}},\ }\bibfield  {title} {\enquote {\bibinfo {title} {Scalability of gadolinium-doped-water cherenkov detectors for nuclear nonproliferation},}\ }\href {\doibase 10.1103/PhysRevApplied.18.034059} {\bibfield  {journal} {\bibinfo  {journal} {Phys. Rev. Appl.}\ }\textbf {\bibinfo {volume} {18}},\ \bibinfo {pages} {034059} (\bibinfo {year} {2022})}\BibitemShut {NoStop}%
\bibitem [{\citenamefont {Porta}\ \emph {et~al.}(2010)\citenamefont {Porta}, \citenamefont {Bui}, \citenamefont {Cribier}, \citenamefont {Fallot}, \citenamefont {Fechner}, \citenamefont {Giot}, \citenamefont {Lasserre}, \citenamefont {Letourneau}, \citenamefont {Lhuillier}, \citenamefont {Martino} \emph {et~al.}}]{Porta2010a}%
  \BibitemOpen
  \bibfield  {author} {\bibinfo {author} {\bibfnamefont {A.}~\bibnamefont {Porta}}, \bibinfo {author} {\bibfnamefont {V.~M.}\ \bibnamefont {Bui}}, \bibinfo {author} {\bibfnamefont {M.}~\bibnamefont {Cribier}}, \bibinfo {author} {\bibfnamefont {M.}~\bibnamefont {Fallot}}, \bibinfo {author} {\bibfnamefont {M.}~\bibnamefont {Fechner}}, \bibinfo {author} {\bibfnamefont {L.}~\bibnamefont {Giot}}, \bibinfo {author} {\bibfnamefont {T.}~\bibnamefont {Lasserre}}, \bibinfo {author} {\bibfnamefont {A.}~\bibnamefont {Letourneau}}, \bibinfo {author} {\bibfnamefont {D.}~\bibnamefont {Lhuillier}}, \bibinfo {author} {\bibfnamefont {J.}~\bibnamefont {Martino}},  \emph {et~al.},\ }\bibfield  {title} {\enquote {\bibinfo {title} {Reactor neutrino detection for non-proliferation with the nucifer experiment},}\ }\href {\doibase 10.1109/TNS.2009.2035119} {\bibfield  {journal} {\bibinfo  {journal} {IEEE Trans. Nucl. Sci.}\ }\textbf {\bibinfo {volume} {57}},\ \bibinfo {pages} {2732--2739} (\bibinfo {year} {2010})}\BibitemShut
  {NoStop}%
\bibitem [{\citenamefont {Huber}(2011)}]{Huber2011}%
  \BibitemOpen
  \bibfield  {author} {\bibinfo {author} {\bibfnamefont {P.}~\bibnamefont {Huber}},\ }\bibfield  {title} {\enquote {\bibinfo {title} {Determination of antineutrino spectra from nuclear reactors},}\ }\href {\doibase 10.1103/PhysRevC.84.024617} {\bibfield  {journal} {\bibinfo  {journal} {Phys. Rev. C}\ }\textbf {\bibinfo {volume} {84}},\ \bibinfo {pages} {024617} (\bibinfo {year} {2011})}\BibitemShut {NoStop}%
\bibitem [{\citenamefont {Mueller}\ \emph {et~al.}(2011)\citenamefont {Mueller}, \citenamefont {Lhuillier}, \citenamefont {Fallot}, \citenamefont {Letourneau}, \citenamefont {Cormon}, \citenamefont {Fechner}, \citenamefont {Giot}, \citenamefont {Lasserre}, \citenamefont {Martino}, \citenamefont {Mention}, \citenamefont {Porta},\ and\ \citenamefont {Yermia}}]{Mueller2011}%
  \BibitemOpen
  \bibfield  {author} {\bibinfo {author} {\bibfnamefont {T.~A.}\ \bibnamefont {Mueller}}, \bibinfo {author} {\bibfnamefont {D.}~\bibnamefont {Lhuillier}}, \bibinfo {author} {\bibfnamefont {M.}~\bibnamefont {Fallot}}, \bibinfo {author} {\bibfnamefont {A.}~\bibnamefont {Letourneau}}, \bibinfo {author} {\bibfnamefont {S.}~\bibnamefont {Cormon}}, \bibinfo {author} {\bibfnamefont {M.}~\bibnamefont {Fechner}}, \bibinfo {author} {\bibfnamefont {L.}~\bibnamefont {Giot}}, \bibinfo {author} {\bibfnamefont {T.}~\bibnamefont {Lasserre}}, \bibinfo {author} {\bibfnamefont {J.}~\bibnamefont {Martino}}, \bibinfo {author} {\bibfnamefont {G.}~\bibnamefont {Mention}}, \bibinfo {author} {\bibfnamefont {A.}~\bibnamefont {Porta}}, \ and\ \bibinfo {author} {\bibfnamefont {F.}~\bibnamefont {Yermia}},\ }\bibfield  {title} {\enquote {\bibinfo {title} {Improved predictions of reactor antineutrino spectra},}\ }\href {\doibase 10.1103/PhysRevC.83.054615} {\bibfield  {journal} {\bibinfo  {journal} {Phys. Rev. C}\ }\textbf {\bibinfo
  {volume} {83}},\ \bibinfo {pages} {054615} (\bibinfo {year} {2011})}\BibitemShut {NoStop}%
\bibitem [{\citenamefont {Christensen}, \citenamefont {Huber},\ and\ \citenamefont {Jaffke}(2013)}]{Christensen2013}%
  \BibitemOpen
  \bibfield  {author} {\bibinfo {author} {\bibfnamefont {E.}~\bibnamefont {Christensen}}, \bibinfo {author} {\bibfnamefont {P.}~\bibnamefont {Huber}}, \ and\ \bibinfo {author} {\bibfnamefont {P.}~\bibnamefont {Jaffke}},\ }\bibfield  {title} {\enquote {\bibinfo {title} {Antineutrino reactor safeguards - a case study},}\ }\href {\doibase 10.48550/ARXIV.1312.1959} {\  (\bibinfo {year} {2013}),\ 10.48550/ARXIV.1312.1959},\ \Eprint {http://arxiv.org/abs/1312.1959} {arXiv:1312.1959 [physics.ins-det]} \BibitemShut {NoStop}%
\bibitem [{\citenamefont {Mention}\ \emph {et~al.}(2011)\citenamefont {Mention}, \citenamefont {Fechner}, \citenamefont {Lasserre}, \citenamefont {Mueller}, \citenamefont {Lhuillier}, \citenamefont {Cribier},\ and\ \citenamefont {Letourneau}}]{Mention2011}%
  \BibitemOpen
  \bibfield  {author} {\bibinfo {author} {\bibfnamefont {G.}~\bibnamefont {Mention}}, \bibinfo {author} {\bibfnamefont {M.}~\bibnamefont {Fechner}}, \bibinfo {author} {\bibfnamefont {T.}~\bibnamefont {Lasserre}}, \bibinfo {author} {\bibfnamefont {T.~A.}\ \bibnamefont {Mueller}}, \bibinfo {author} {\bibfnamefont {D.}~\bibnamefont {Lhuillier}}, \bibinfo {author} {\bibfnamefont {M.}~\bibnamefont {Cribier}}, \ and\ \bibinfo {author} {\bibfnamefont {A.}~\bibnamefont {Letourneau}},\ }\bibfield  {title} {\enquote {\bibinfo {title} {Reactor antineutrino anomaly},}\ }\href {\doibase 10.1103/PhysRevD.83.073006} {\bibfield  {journal} {\bibinfo  {journal} {Phys. Rev. D}\ }\textbf {\bibinfo {volume} {83}},\ \bibinfo {pages} {073006} (\bibinfo {year} {2011})}\BibitemShut {NoStop}%
\bibitem [{\citenamefont {An}\ \emph {et~al.}(2017)\citenamefont {An}, \citenamefont {Balantekin}, \citenamefont {Band}, \citenamefont {Bishai}, \citenamefont {Blyth}, \citenamefont {Cao}, \citenamefont {Cao}, \citenamefont {Cao}, \citenamefont {Chan}, \citenamefont {Chang}, \citenamefont {Chang}, \citenamefont {Chen}, \citenamefont {Chen}, \citenamefont {Chen}, \citenamefont {Chen}, \citenamefont {Chen}, \citenamefont {Cheng}, \citenamefont {Cheng}, \citenamefont {Cherwinka}, \citenamefont {Chu}, \citenamefont {Chukanov}, \citenamefont {Cummings}, \citenamefont {Ding}, \citenamefont {Diwan}, \citenamefont {Dolgareva}, \citenamefont {Dove}, \citenamefont {Dwyer}, \citenamefont {Edwards}, \citenamefont {Gill}, \citenamefont {Gonchar}, \citenamefont {Gong}, \citenamefont {Gong}, \citenamefont {Grassi}, \citenamefont {Gu}, \citenamefont {Guo}, \citenamefont {Guo}, \citenamefont {Guo}, \citenamefont {Guo}, \citenamefont {Hackenburg}, \citenamefont {Hans}, \citenamefont {He}, \citenamefont {Heeger}, \citenamefont
  {Heng}, \citenamefont {Higuera}, \citenamefont {Hsiung}, \citenamefont {Hu}, \citenamefont {Hu}, \citenamefont {Huang}, \citenamefont {Huang}, \citenamefont {Huang}, \citenamefont {Huang}, \citenamefont {Huber}, \citenamefont {Huo}, \citenamefont {Hussain}, \citenamefont {Jaffe}, \citenamefont {Jen}, \citenamefont {Ji}, \citenamefont {Ji}, \citenamefont {Jiao}, \citenamefont {Johnson}, \citenamefont {Jones}, \citenamefont {Kang}, \citenamefont {Kettell}, \citenamefont {Khan}, \citenamefont {Kohn}, \citenamefont {Kramer}, \citenamefont {Kwan}, \citenamefont {Kwok}, \citenamefont {Langford}, \citenamefont {Lau}, \citenamefont {Lebanowski}, \citenamefont {Lee}, \citenamefont {Lee}, \citenamefont {Lei}, \citenamefont {Leitner}, \citenamefont {Leung}, \citenamefont {Li}, \citenamefont {Li}, \citenamefont {Li}, \citenamefont {Li}, \citenamefont {Li}, \citenamefont {Li}, \citenamefont {Li}, \citenamefont {Li}, \citenamefont {Li}, \citenamefont {Li}, \citenamefont {Li}, \citenamefont {Li}, \citenamefont {Liang},
  \citenamefont {Lin}, \citenamefont {Lin}, \citenamefont {Lin}, \citenamefont {Lin}, \citenamefont {Lin}, \citenamefont {Ling}, \citenamefont {Link}, \citenamefont {Littenberg}, \citenamefont {Littlejohn}, \citenamefont {Liu}, \citenamefont {Liu}, \citenamefont {Loh}, \citenamefont {Lu}, \citenamefont {Lu}, \citenamefont {Lu}, \citenamefont {Luk}, \citenamefont {Ma}, \citenamefont {Ma}, \citenamefont {Ma}, \citenamefont {Malyshkin}, \citenamefont {Martinez~Caicedo}, \citenamefont {McDonald}, \citenamefont {McKeown}, \citenamefont {Mitchell}, \citenamefont {Nakajima}, \citenamefont {Napolitano}, \citenamefont {Naumov}, \citenamefont {Naumova}, \citenamefont {Ngai}, \citenamefont {Ochoa-Ricoux}, \citenamefont {Olshevskiy}, \citenamefont {Pan}, \citenamefont {Park}, \citenamefont {Patton}, \citenamefont {Pec}, \citenamefont {Peng}, \citenamefont {Pinsky}, \citenamefont {Pun}, \citenamefont {Qi}, \citenamefont {Qi}, \citenamefont {Qian}, \citenamefont {Qiu}, \citenamefont {Raper}, \citenamefont {Ren},
  \citenamefont {Rosero}, \citenamefont {Roskovec}, \citenamefont {Ruan}, \citenamefont {Steiner}, \citenamefont {Stoler}, \citenamefont {Sun}, \citenamefont {Tang}, \citenamefont {Taychenachev}, \citenamefont {Treskov}, \citenamefont {Tsang}, \citenamefont {Tull}, \citenamefont {Viaux}, \citenamefont {Viren}, \citenamefont {Vorobel}, \citenamefont {Wang}, \citenamefont {Wang}, \citenamefont {Wang}, \citenamefont {Wang}, \citenamefont {Wang}, \citenamefont {Wang}, \citenamefont {Wang}, \citenamefont {Wang}, \citenamefont {Wang}, \citenamefont {Wang}, \citenamefont {Wei}, \citenamefont {Wen}, \citenamefont {Whisnant}, \citenamefont {White}, \citenamefont {Whitehead}, \citenamefont {Wise}, \citenamefont {Wong}, \citenamefont {Wong}, \citenamefont {Worcester}, \citenamefont {Wu}, \citenamefont {Wu}, \citenamefont {Wu}, \citenamefont {Xia}, \citenamefont {Xia}, \citenamefont {Xing}, \citenamefont {Xu}, \citenamefont {Xu}, \citenamefont {Xue}, \citenamefont {Yang}, \citenamefont {Yang}, \citenamefont {Yang},
  \citenamefont {Yang}, \citenamefont {Yang}, \citenamefont {Yang}, \citenamefont {Ye}, \citenamefont {Ye}, \citenamefont {Yeh}, \citenamefont {Young}, \citenamefont {Yu}, \citenamefont {Zeng}, \citenamefont {Zhan}, \citenamefont {Zhang}, \citenamefont {Zhang}, \citenamefont {Zhang}, \citenamefont {Zhang}, \citenamefont {Zhang}, \citenamefont {Zhang}, \citenamefont {Zhang}, \citenamefont {Zhang}, \citenamefont {Zhang}, \citenamefont {Zhang}, \citenamefont {Zhang}, \citenamefont {Zhang}, \citenamefont {Zhang}, \citenamefont {Zhao}, \citenamefont {Zhou}, \citenamefont {Zhuang},\ and\ \citenamefont {Zou}}]{An2017}%
  \BibitemOpen
  \bibfield  {author} {\bibinfo {author} {\bibfnamefont {F.~P.}\ \bibnamefont {An}}, \bibinfo {author} {\bibfnamefont {A.~B.}\ \bibnamefont {Balantekin}}, \bibinfo {author} {\bibfnamefont {H.~R.}\ \bibnamefont {Band}}, \bibinfo {author} {\bibfnamefont {M.}~\bibnamefont {Bishai}}, \bibinfo {author} {\bibfnamefont {S.}~\bibnamefont {Blyth}}, \bibinfo {author} {\bibfnamefont {D.}~\bibnamefont {Cao}}, \bibinfo {author} {\bibfnamefont {G.~F.}\ \bibnamefont {Cao}}, \bibinfo {author} {\bibfnamefont {J.}~\bibnamefont {Cao}}, \bibinfo {author} {\bibfnamefont {Y.~L.}\ \bibnamefont {Chan}}, \bibinfo {author} {\bibfnamefont {J.~F.}\ \bibnamefont {Chang}}, \bibinfo {author} {\bibfnamefont {Y.}~\bibnamefont {Chang}}, \bibinfo {author} {\bibfnamefont {H.~S.}\ \bibnamefont {Chen}}, \bibinfo {author} {\bibfnamefont {Q.~Y.}\ \bibnamefont {Chen}}, \bibinfo {author} {\bibfnamefont {S.~M.}\ \bibnamefont {Chen}}, \bibinfo {author} {\bibfnamefont {Y.~X.}\ \bibnamefont {Chen}}, \bibinfo {author} {\bibfnamefont {Y.}~\bibnamefont
  {Chen}}, \bibinfo {author} {\bibfnamefont {J.}~\bibnamefont {Cheng}}, \bibinfo {author} {\bibfnamefont {Z.~K.}\ \bibnamefont {Cheng}}, \bibinfo {author} {\bibfnamefont {J.~J.}\ \bibnamefont {Cherwinka}}, \bibinfo {author} {\bibfnamefont {M.~C.}\ \bibnamefont {Chu}}, \bibinfo {author} {\bibfnamefont {A.}~\bibnamefont {Chukanov}}, \bibinfo {author} {\bibfnamefont {J.~P.}\ \bibnamefont {Cummings}}, \bibinfo {author} {\bibfnamefont {Y.~Y.}\ \bibnamefont {Ding}}, \bibinfo {author} {\bibfnamefont {M.~V.}\ \bibnamefont {Diwan}}, \bibinfo {author} {\bibfnamefont {M.}~\bibnamefont {Dolgareva}}, \bibinfo {author} {\bibfnamefont {J.}~\bibnamefont {Dove}}, \bibinfo {author} {\bibfnamefont {D.~A.}\ \bibnamefont {Dwyer}}, \bibinfo {author} {\bibfnamefont {W.~R.}\ \bibnamefont {Edwards}}, \bibinfo {author} {\bibfnamefont {R.}~\bibnamefont {Gill}}, \bibinfo {author} {\bibfnamefont {M.}~\bibnamefont {Gonchar}}, \bibinfo {author} {\bibfnamefont {G.~H.}\ \bibnamefont {Gong}}, \bibinfo {author} {\bibfnamefont {H.}~\bibnamefont
  {Gong}}, \bibinfo {author} {\bibfnamefont {M.}~\bibnamefont {Grassi}}, \bibinfo {author} {\bibfnamefont {W.~Q.}\ \bibnamefont {Gu}}, \bibinfo {author} {\bibfnamefont {L.}~\bibnamefont {Guo}}, \bibinfo {author} {\bibfnamefont {X.~H.}\ \bibnamefont {Guo}}, \bibinfo {author} {\bibfnamefont {Y.~H.}\ \bibnamefont {Guo}}, \bibinfo {author} {\bibfnamefont {Z.}~\bibnamefont {Guo}}, \bibinfo {author} {\bibfnamefont {R.~W.}\ \bibnamefont {Hackenburg}}, \bibinfo {author} {\bibfnamefont {S.}~\bibnamefont {Hans}}, \bibinfo {author} {\bibfnamefont {M.}~\bibnamefont {He}}, \bibinfo {author} {\bibfnamefont {K.~M.}\ \bibnamefont {Heeger}}, \bibinfo {author} {\bibfnamefont {Y.~K.}\ \bibnamefont {Heng}}, \bibinfo {author} {\bibfnamefont {A.}~\bibnamefont {Higuera}}, \bibinfo {author} {\bibfnamefont {Y.~B.}\ \bibnamefont {Hsiung}}, \bibinfo {author} {\bibfnamefont {B.~Z.}\ \bibnamefont {Hu}}, \bibinfo {author} {\bibfnamefont {T.}~\bibnamefont {Hu}}, \bibinfo {author} {\bibfnamefont {E.~C.}\ \bibnamefont {Huang}}, \bibinfo
  {author} {\bibfnamefont {H.~X.}\ \bibnamefont {Huang}}, \bibinfo {author} {\bibfnamefont {X.~T.}\ \bibnamefont {Huang}}, \bibinfo {author} {\bibfnamefont {Y.~B.}\ \bibnamefont {Huang}}, \bibinfo {author} {\bibfnamefont {P.}~\bibnamefont {Huber}}, \bibinfo {author} {\bibfnamefont {W.}~\bibnamefont {Huo}}, \bibinfo {author} {\bibfnamefont {G.}~\bibnamefont {Hussain}}, \bibinfo {author} {\bibfnamefont {D.~E.}\ \bibnamefont {Jaffe}}, \bibinfo {author} {\bibfnamefont {K.~L.}\ \bibnamefont {Jen}}, \bibinfo {author} {\bibfnamefont {X.~P.}\ \bibnamefont {Ji}}, \bibinfo {author} {\bibfnamefont {X.~L.}\ \bibnamefont {Ji}}, \bibinfo {author} {\bibfnamefont {J.~B.}\ \bibnamefont {Jiao}}, \bibinfo {author} {\bibfnamefont {R.~A.}\ \bibnamefont {Johnson}}, \bibinfo {author} {\bibfnamefont {D.}~\bibnamefont {Jones}}, \bibinfo {author} {\bibfnamefont {L.}~\bibnamefont {Kang}}, \bibinfo {author} {\bibfnamefont {S.~H.}\ \bibnamefont {Kettell}}, \bibinfo {author} {\bibfnamefont {A.}~\bibnamefont {Khan}}, \bibinfo {author}
  {\bibfnamefont {S.}~\bibnamefont {Kohn}}, \bibinfo {author} {\bibfnamefont {M.}~\bibnamefont {Kramer}}, \bibinfo {author} {\bibfnamefont {K.~K.}\ \bibnamefont {Kwan}}, \bibinfo {author} {\bibfnamefont {M.~W.}\ \bibnamefont {Kwok}}, \bibinfo {author} {\bibfnamefont {T.~J.}\ \bibnamefont {Langford}}, \bibinfo {author} {\bibfnamefont {K.}~\bibnamefont {Lau}}, \bibinfo {author} {\bibfnamefont {L.}~\bibnamefont {Lebanowski}}, \bibinfo {author} {\bibfnamefont {J.}~\bibnamefont {Lee}}, \bibinfo {author} {\bibfnamefont {J.~H.~C.}\ \bibnamefont {Lee}}, \bibinfo {author} {\bibfnamefont {R.~T.}\ \bibnamefont {Lei}}, \bibinfo {author} {\bibfnamefont {R.}~\bibnamefont {Leitner}}, \bibinfo {author} {\bibfnamefont {J.~K.~C.}\ \bibnamefont {Leung}}, \bibinfo {author} {\bibfnamefont {C.}~\bibnamefont {Li}}, \bibinfo {author} {\bibfnamefont {D.~J.}\ \bibnamefont {Li}}, \bibinfo {author} {\bibfnamefont {F.}~\bibnamefont {Li}}, \bibinfo {author} {\bibfnamefont {G.~S.}\ \bibnamefont {Li}}, \bibinfo {author} {\bibfnamefont
  {Q.~J.}\ \bibnamefont {Li}}, \bibinfo {author} {\bibfnamefont {S.}~\bibnamefont {Li}}, \bibinfo {author} {\bibfnamefont {S.~C.}\ \bibnamefont {Li}}, \bibinfo {author} {\bibfnamefont {W.~D.}\ \bibnamefont {Li}}, \bibinfo {author} {\bibfnamefont {X.~N.}\ \bibnamefont {Li}}, \bibinfo {author} {\bibfnamefont {X.~Q.}\ \bibnamefont {Li}}, \bibinfo {author} {\bibfnamefont {Y.~F.}\ \bibnamefont {Li}}, \bibinfo {author} {\bibfnamefont {Z.~B.}\ \bibnamefont {Li}}, \bibinfo {author} {\bibfnamefont {H.}~\bibnamefont {Liang}}, \bibinfo {author} {\bibfnamefont {C.~J.}\ \bibnamefont {Lin}}, \bibinfo {author} {\bibfnamefont {G.~L.}\ \bibnamefont {Lin}}, \bibinfo {author} {\bibfnamefont {S.}~\bibnamefont {Lin}}, \bibinfo {author} {\bibfnamefont {S.~K.}\ \bibnamefont {Lin}}, \bibinfo {author} {\bibfnamefont {Y.~C.}\ \bibnamefont {Lin}}, \bibinfo {author} {\bibfnamefont {J.~J.}\ \bibnamefont {Ling}}, \bibinfo {author} {\bibfnamefont {J.~M.}\ \bibnamefont {Link}}, \bibinfo {author} {\bibfnamefont {L.}~\bibnamefont
  {Littenberg}}, \bibinfo {author} {\bibfnamefont {B.~R.}\ \bibnamefont {Littlejohn}}, \bibinfo {author} {\bibfnamefont {J.~L.}\ \bibnamefont {Liu}}, \bibinfo {author} {\bibfnamefont {J.~C.}\ \bibnamefont {Liu}}, \bibinfo {author} {\bibfnamefont {C.~W.}\ \bibnamefont {Loh}}, \bibinfo {author} {\bibfnamefont {C.}~\bibnamefont {Lu}}, \bibinfo {author} {\bibfnamefont {H.~Q.}\ \bibnamefont {Lu}}, \bibinfo {author} {\bibfnamefont {J.~S.}\ \bibnamefont {Lu}}, \bibinfo {author} {\bibfnamefont {K.~B.}\ \bibnamefont {Luk}}, \bibinfo {author} {\bibfnamefont {X.~Y.}\ \bibnamefont {Ma}}, \bibinfo {author} {\bibfnamefont {X.~B.}\ \bibnamefont {Ma}}, \bibinfo {author} {\bibfnamefont {Y.~Q.}\ \bibnamefont {Ma}}, \bibinfo {author} {\bibfnamefont {Y.}~\bibnamefont {Malyshkin}}, \bibinfo {author} {\bibfnamefont {D.~A.}\ \bibnamefont {Martinez~Caicedo}}, \bibinfo {author} {\bibfnamefont {K.~T.}\ \bibnamefont {McDonald}}, \bibinfo {author} {\bibfnamefont {R.~D.}\ \bibnamefont {McKeown}}, \bibinfo {author} {\bibfnamefont
  {I.}~\bibnamefont {Mitchell}}, \bibinfo {author} {\bibfnamefont {Y.}~\bibnamefont {Nakajima}}, \bibinfo {author} {\bibfnamefont {J.}~\bibnamefont {Napolitano}}, \bibinfo {author} {\bibfnamefont {D.}~\bibnamefont {Naumov}}, \bibinfo {author} {\bibfnamefont {E.}~\bibnamefont {Naumova}}, \bibinfo {author} {\bibfnamefont {H.~Y.}\ \bibnamefont {Ngai}}, \bibinfo {author} {\bibfnamefont {J.~P.}\ \bibnamefont {Ochoa-Ricoux}}, \bibinfo {author} {\bibfnamefont {A.}~\bibnamefont {Olshevskiy}}, \bibinfo {author} {\bibfnamefont {H.~R.}\ \bibnamefont {Pan}}, \bibinfo {author} {\bibfnamefont {J.}~\bibnamefont {Park}}, \bibinfo {author} {\bibfnamefont {S.}~\bibnamefont {Patton}}, \bibinfo {author} {\bibfnamefont {V.}~\bibnamefont {Pec}}, \bibinfo {author} {\bibfnamefont {J.~C.}\ \bibnamefont {Peng}}, \bibinfo {author} {\bibfnamefont {L.}~\bibnamefont {Pinsky}}, \bibinfo {author} {\bibfnamefont {C.~S.~J.}\ \bibnamefont {Pun}}, \bibinfo {author} {\bibfnamefont {F.~Z.}\ \bibnamefont {Qi}}, \bibinfo {author} {\bibfnamefont
  {M.}~\bibnamefont {Qi}}, \bibinfo {author} {\bibfnamefont {X.}~\bibnamefont {Qian}}, \bibinfo {author} {\bibfnamefont {R.~M.}\ \bibnamefont {Qiu}}, \bibinfo {author} {\bibfnamefont {N.}~\bibnamefont {Raper}}, \bibinfo {author} {\bibfnamefont {J.}~\bibnamefont {Ren}}, \bibinfo {author} {\bibfnamefont {R.}~\bibnamefont {Rosero}}, \bibinfo {author} {\bibfnamefont {B.}~\bibnamefont {Roskovec}}, \bibinfo {author} {\bibfnamefont {X.~C.}\ \bibnamefont {Ruan}}, \bibinfo {author} {\bibfnamefont {H.}~\bibnamefont {Steiner}}, \bibinfo {author} {\bibfnamefont {P.}~\bibnamefont {Stoler}}, \bibinfo {author} {\bibfnamefont {J.~L.}\ \bibnamefont {Sun}}, \bibinfo {author} {\bibfnamefont {W.}~\bibnamefont {Tang}}, \bibinfo {author} {\bibfnamefont {D.}~\bibnamefont {Taychenachev}}, \bibinfo {author} {\bibfnamefont {K.}~\bibnamefont {Treskov}}, \bibinfo {author} {\bibfnamefont {K.~V.}\ \bibnamefont {Tsang}}, \bibinfo {author} {\bibfnamefont {C.~E.}\ \bibnamefont {Tull}}, \bibinfo {author} {\bibfnamefont {N.}~\bibnamefont
  {Viaux}}, \bibinfo {author} {\bibfnamefont {B.}~\bibnamefont {Viren}}, \bibinfo {author} {\bibfnamefont {V.}~\bibnamefont {Vorobel}}, \bibinfo {author} {\bibfnamefont {C.~H.}\ \bibnamefont {Wang}}, \bibinfo {author} {\bibfnamefont {M.}~\bibnamefont {Wang}}, \bibinfo {author} {\bibfnamefont {N.~Y.}\ \bibnamefont {Wang}}, \bibinfo {author} {\bibfnamefont {R.~G.}\ \bibnamefont {Wang}}, \bibinfo {author} {\bibfnamefont {W.}~\bibnamefont {Wang}}, \bibinfo {author} {\bibfnamefont {X.}~\bibnamefont {Wang}}, \bibinfo {author} {\bibfnamefont {Y.~F.}\ \bibnamefont {Wang}}, \bibinfo {author} {\bibfnamefont {Z.}~\bibnamefont {Wang}}, \bibinfo {author} {\bibfnamefont {Z.}~\bibnamefont {Wang}}, \bibinfo {author} {\bibfnamefont {Z.~M.}\ \bibnamefont {Wang}}, \bibinfo {author} {\bibfnamefont {H.~Y.}\ \bibnamefont {Wei}}, \bibinfo {author} {\bibfnamefont {L.~J.}\ \bibnamefont {Wen}}, \bibinfo {author} {\bibfnamefont {K.}~\bibnamefont {Whisnant}}, \bibinfo {author} {\bibfnamefont {C.~G.}\ \bibnamefont {White}}, \bibinfo
  {author} {\bibfnamefont {L.}~\bibnamefont {Whitehead}}, \bibinfo {author} {\bibfnamefont {T.}~\bibnamefont {Wise}}, \bibinfo {author} {\bibfnamefont {H.~L.~H.}\ \bibnamefont {Wong}}, \bibinfo {author} {\bibfnamefont {S.~C.~F.}\ \bibnamefont {Wong}}, \bibinfo {author} {\bibfnamefont {E.}~\bibnamefont {Worcester}}, \bibinfo {author} {\bibfnamefont {C.-H.}\ \bibnamefont {Wu}}, \bibinfo {author} {\bibfnamefont {Q.}~\bibnamefont {Wu}}, \bibinfo {author} {\bibfnamefont {W.~J.}\ \bibnamefont {Wu}}, \bibinfo {author} {\bibfnamefont {D.~M.}\ \bibnamefont {Xia}}, \bibinfo {author} {\bibfnamefont {J.~K.}\ \bibnamefont {Xia}}, \bibinfo {author} {\bibfnamefont {Z.~Z.}\ \bibnamefont {Xing}}, \bibinfo {author} {\bibfnamefont {J.~L.}\ \bibnamefont {Xu}}, \bibinfo {author} {\bibfnamefont {Y.}~\bibnamefont {Xu}}, \bibinfo {author} {\bibfnamefont {T.}~\bibnamefont {Xue}}, \bibinfo {author} {\bibfnamefont {C.~G.}\ \bibnamefont {Yang}}, \bibinfo {author} {\bibfnamefont {H.}~\bibnamefont {Yang}}, \bibinfo {author} {\bibfnamefont
  {L.}~\bibnamefont {Yang}}, \bibinfo {author} {\bibfnamefont {M.~S.}\ \bibnamefont {Yang}}, \bibinfo {author} {\bibfnamefont {M.~T.}\ \bibnamefont {Yang}}, \bibinfo {author} {\bibfnamefont {Y.~Z.}\ \bibnamefont {Yang}}, \bibinfo {author} {\bibfnamefont {M.}~\bibnamefont {Ye}}, \bibinfo {author} {\bibfnamefont {Z.}~\bibnamefont {Ye}}, \bibinfo {author} {\bibfnamefont {M.}~\bibnamefont {Yeh}}, \bibinfo {author} {\bibfnamefont {B.~L.}\ \bibnamefont {Young}}, \bibinfo {author} {\bibfnamefont {Z.~Y.}\ \bibnamefont {Yu}}, \bibinfo {author} {\bibfnamefont {S.}~\bibnamefont {Zeng}}, \bibinfo {author} {\bibfnamefont {L.}~\bibnamefont {Zhan}}, \bibinfo {author} {\bibfnamefont {C.}~\bibnamefont {Zhang}}, \bibinfo {author} {\bibfnamefont {C.~C.}\ \bibnamefont {Zhang}}, \bibinfo {author} {\bibfnamefont {H.~H.}\ \bibnamefont {Zhang}}, \bibinfo {author} {\bibfnamefont {J.~W.}\ \bibnamefont {Zhang}}, \bibinfo {author} {\bibfnamefont {Q.~M.}\ \bibnamefont {Zhang}}, \bibinfo {author} {\bibfnamefont {R.}~\bibnamefont {Zhang}},
  \bibinfo {author} {\bibfnamefont {X.~T.}\ \bibnamefont {Zhang}}, \bibinfo {author} {\bibfnamefont {Y.~M.}\ \bibnamefont {Zhang}}, \bibinfo {author} {\bibfnamefont {Y.~X.}\ \bibnamefont {Zhang}}, \bibinfo {author} {\bibfnamefont {Y.~M.}\ \bibnamefont {Zhang}}, \bibinfo {author} {\bibfnamefont {Z.~J.}\ \bibnamefont {Zhang}}, \bibinfo {author} {\bibfnamefont {Z.~Y.}\ \bibnamefont {Zhang}}, \bibinfo {author} {\bibfnamefont {Z.~P.}\ \bibnamefont {Zhang}}, \bibinfo {author} {\bibfnamefont {J.}~\bibnamefont {Zhao}}, \bibinfo {author} {\bibfnamefont {L.}~\bibnamefont {Zhou}}, \bibinfo {author} {\bibfnamefont {H.~L.}\ \bibnamefont {Zhuang}}, \ and\ \bibinfo {author} {\bibfnamefont {J.~H.}\ \bibnamefont {Zou}} (\bibinfo {collaboration} {Daya Bay Collaboration}),\ }\bibfield  {title} {\enquote {\bibinfo {title} {Evolution of the reactor antineutrino flux and spectrum at {Daya Bay}},}\ }\href {\doibase 10.1103/PhysRevLett.118.251801} {\bibfield  {journal} {\bibinfo  {journal} {Phys. Rev. Lett.}\ }\textbf {\bibinfo
  {volume} {118}},\ \bibinfo {pages} {251801} (\bibinfo {year} {2017})}\BibitemShut {NoStop}%
\bibitem [{\citenamefont {Bak}\ \emph {et~al.}(2019)\citenamefont {Bak}, \citenamefont {Choi}, \citenamefont {Jang}, \citenamefont {Jang}, \citenamefont {Jeon}, \citenamefont {Joo}, \citenamefont {Ju}, \citenamefont {Jung}, \citenamefont {Kim}, \citenamefont {Kim}, \citenamefont {Kim}, \citenamefont {Kim}, \citenamefont {Kim}, \citenamefont {Kim}, \citenamefont {Kwon}, \citenamefont {Lee}, \citenamefont {Lee}, \citenamefont {Lee}, \citenamefont {Lim}, \citenamefont {Moon}, \citenamefont {Pac}, \citenamefont {Park}, \citenamefont {Rott}, \citenamefont {Seo}, \citenamefont {Seo}, \citenamefont {Seo}, \citenamefont {Shin}, \citenamefont {Yang}, \citenamefont {Yoo},\ and\ \citenamefont {Yu}}]{Bak2019}%
  \BibitemOpen
  \bibfield  {author} {\bibinfo {author} {\bibfnamefont {G.}~\bibnamefont {Bak}}, \bibinfo {author} {\bibfnamefont {J.~H.}\ \bibnamefont {Choi}}, \bibinfo {author} {\bibfnamefont {H.~I.}\ \bibnamefont {Jang}}, \bibinfo {author} {\bibfnamefont {J.~S.}\ \bibnamefont {Jang}}, \bibinfo {author} {\bibfnamefont {S.~H.}\ \bibnamefont {Jeon}}, \bibinfo {author} {\bibfnamefont {K.~K.}\ \bibnamefont {Joo}}, \bibinfo {author} {\bibfnamefont {K.}~\bibnamefont {Ju}}, \bibinfo {author} {\bibfnamefont {D.~E.}\ \bibnamefont {Jung}}, \bibinfo {author} {\bibfnamefont {J.~G.}\ \bibnamefont {Kim}}, \bibinfo {author} {\bibfnamefont {J.~H.}\ \bibnamefont {Kim}}, \bibinfo {author} {\bibfnamefont {J.~Y.}\ \bibnamefont {Kim}}, \bibinfo {author} {\bibfnamefont {S.~B.}\ \bibnamefont {Kim}}, \bibinfo {author} {\bibfnamefont {S.~Y.}\ \bibnamefont {Kim}}, \bibinfo {author} {\bibfnamefont {W.}~\bibnamefont {Kim}}, \bibinfo {author} {\bibfnamefont {E.}~\bibnamefont {Kwon}}, \bibinfo {author} {\bibfnamefont {D.~H.}\ \bibnamefont {Lee}},
  \bibinfo {author} {\bibfnamefont {H.~G.}\ \bibnamefont {Lee}}, \bibinfo {author} {\bibfnamefont {Y.~C.}\ \bibnamefont {Lee}}, \bibinfo {author} {\bibfnamefont {I.~T.}\ \bibnamefont {Lim}}, \bibinfo {author} {\bibfnamefont {D.~H.}\ \bibnamefont {Moon}}, \bibinfo {author} {\bibfnamefont {M.~Y.}\ \bibnamefont {Pac}}, \bibinfo {author} {\bibfnamefont {Y.~S.}\ \bibnamefont {Park}}, \bibinfo {author} {\bibfnamefont {C.}~\bibnamefont {Rott}}, \bibinfo {author} {\bibfnamefont {H.}~\bibnamefont {Seo}}, \bibinfo {author} {\bibfnamefont {J.~W.}\ \bibnamefont {Seo}}, \bibinfo {author} {\bibfnamefont {S.~H.}\ \bibnamefont {Seo}}, \bibinfo {author} {\bibfnamefont {C.~D.}\ \bibnamefont {Shin}}, \bibinfo {author} {\bibfnamefont {J.~Y.}\ \bibnamefont {Yang}}, \bibinfo {author} {\bibfnamefont {J.}~\bibnamefont {Yoo}}, \ and\ \bibinfo {author} {\bibfnamefont {I.}~\bibnamefont {Yu}} (\bibinfo {collaboration} {RENO Collaboration}),\ }\bibfield  {title} {\enquote {\bibinfo {title} {Fuel-composition dependent reactor antineutrino
  yield at {RENO}},}\ }\href {\doibase 10.1103/PhysRevLett.122.232501} {\bibfield  {journal} {\bibinfo  {journal} {Phys. Rev. Lett.}\ }\textbf {\bibinfo {volume} {122}},\ \bibinfo {pages} {232501} (\bibinfo {year} {2019})}\BibitemShut {NoStop}%
\bibitem [{\citenamefont {{The Double Chooz Collaboration}}(2020)}]{DoubleChooz2020}%
  \BibitemOpen
  \bibfield  {author} {\bibinfo {author} {\bibnamefont {{The Double Chooz Collaboration}}},\ }\bibfield  {title} {\enquote {\bibinfo {title} {Double chooz {$\theta_{13}$} measurement via total neutron capture detection},}\ }\href {\doibase 10.1038/s41567-020-0831-y} {\bibfield  {journal} {\bibinfo  {journal} {Nat. Phys.}\ }\textbf {\bibinfo {volume} {16}},\ \bibinfo {pages} {558--564} (\bibinfo {year} {2020})}\BibitemShut {NoStop}%
\bibitem [{\citenamefont {Ko}\ \emph {et~al.}(2017)\citenamefont {Ko}, \citenamefont {Kim}, \citenamefont {Kim}, \citenamefont {Han}, \citenamefont {Jang}, \citenamefont {Jeon}, \citenamefont {Joo}, \citenamefont {Kim}, \citenamefont {Kim}, \citenamefont {Kim}, \citenamefont {Lee}, \citenamefont {Lee}, \citenamefont {Lee}, \citenamefont {Oh}, \citenamefont {Park}, \citenamefont {Park}, \citenamefont {Park}, \citenamefont {Seo}, \citenamefont {Siyeon},\ and\ \citenamefont {Sun}}]{Ko2017}%
  \BibitemOpen
  \bibfield  {author} {\bibinfo {author} {\bibfnamefont {Y.~J.}\ \bibnamefont {Ko}}, \bibinfo {author} {\bibfnamefont {B.~R.}\ \bibnamefont {Kim}}, \bibinfo {author} {\bibfnamefont {J.~Y.}\ \bibnamefont {Kim}}, \bibinfo {author} {\bibfnamefont {B.~Y.}\ \bibnamefont {Han}}, \bibinfo {author} {\bibfnamefont {C.~H.}\ \bibnamefont {Jang}}, \bibinfo {author} {\bibfnamefont {E.~J.}\ \bibnamefont {Jeon}}, \bibinfo {author} {\bibfnamefont {K.~K.}\ \bibnamefont {Joo}}, \bibinfo {author} {\bibfnamefont {H.~J.}\ \bibnamefont {Kim}}, \bibinfo {author} {\bibfnamefont {H.~S.}\ \bibnamefont {Kim}}, \bibinfo {author} {\bibfnamefont {Y.~D.}\ \bibnamefont {Kim}}, \bibinfo {author} {\bibfnamefont {J.}~\bibnamefont {Lee}}, \bibinfo {author} {\bibfnamefont {J.~Y.}\ \bibnamefont {Lee}}, \bibinfo {author} {\bibfnamefont {M.~H.}\ \bibnamefont {Lee}}, \bibinfo {author} {\bibfnamefont {Y.~M.}\ \bibnamefont {Oh}}, \bibinfo {author} {\bibfnamefont {H.~K.}\ \bibnamefont {Park}}, \bibinfo {author} {\bibfnamefont {H.~S.}\ \bibnamefont
  {Park}}, \bibinfo {author} {\bibfnamefont {K.~S.}\ \bibnamefont {Park}}, \bibinfo {author} {\bibfnamefont {K.~M.}\ \bibnamefont {Seo}}, \bibinfo {author} {\bibfnamefont {K.}~\bibnamefont {Siyeon}}, \ and\ \bibinfo {author} {\bibfnamefont {G.~M.}\ \bibnamefont {Sun}} (\bibinfo {collaboration} {NEOS Collaboration}),\ }\bibfield  {title} {\enquote {\bibinfo {title} {Sterile neutrino search at the {NEOS} experiment},}\ }\href {\doibase 10.1103/PhysRevLett.118.121802} {\bibfield  {journal} {\bibinfo  {journal} {Phys. Rev. Lett.}\ }\textbf {\bibinfo {volume} {118}},\ \bibinfo {pages} {121802} (\bibinfo {year} {2017})}\BibitemShut {NoStop}%
\bibitem [{\citenamefont {Estienne}\ \emph {et~al.}(2019)\citenamefont {Estienne}, \citenamefont {Fallot}, \citenamefont {Algora}, \citenamefont {Briz-Monago}, \citenamefont {Bui}, \citenamefont {Cormon}, \citenamefont {Gelletly}, \citenamefont {Giot}, \citenamefont {Guadilla}, \citenamefont {Jordan}, \citenamefont {Le~Meur}, \citenamefont {Porta}, \citenamefont {Rice}, \citenamefont {Rubio}, \citenamefont {Ta\'{\i}n}, \citenamefont {Valencia},\ and\ \citenamefont {Zakari-Issoufou}}]{Estienne2019}%
  \BibitemOpen
  \bibfield  {author} {\bibinfo {author} {\bibfnamefont {M.}~\bibnamefont {Estienne}}, \bibinfo {author} {\bibfnamefont {M.}~\bibnamefont {Fallot}}, \bibinfo {author} {\bibfnamefont {A.}~\bibnamefont {Algora}}, \bibinfo {author} {\bibfnamefont {J.}~\bibnamefont {Briz-Monago}}, \bibinfo {author} {\bibfnamefont {V.~M.}\ \bibnamefont {Bui}}, \bibinfo {author} {\bibfnamefont {S.}~\bibnamefont {Cormon}}, \bibinfo {author} {\bibfnamefont {W.}~\bibnamefont {Gelletly}}, \bibinfo {author} {\bibfnamefont {L.}~\bibnamefont {Giot}}, \bibinfo {author} {\bibfnamefont {V.}~\bibnamefont {Guadilla}}, \bibinfo {author} {\bibfnamefont {D.}~\bibnamefont {Jordan}}, \bibinfo {author} {\bibfnamefont {L.}~\bibnamefont {Le~Meur}}, \bibinfo {author} {\bibfnamefont {A.}~\bibnamefont {Porta}}, \bibinfo {author} {\bibfnamefont {S.}~\bibnamefont {Rice}}, \bibinfo {author} {\bibfnamefont {B.}~\bibnamefont {Rubio}}, \bibinfo {author} {\bibfnamefont {J.~L.}\ \bibnamefont {Ta\'{\i}n}}, \bibinfo {author} {\bibfnamefont {E.}~\bibnamefont
  {Valencia}}, \ and\ \bibinfo {author} {\bibfnamefont {A.-A.}\ \bibnamefont {Zakari-Issoufou}},\ }\bibfield  {title} {\enquote {\bibinfo {title} {Updated summation model: An improved agreement with the {Daya Bay} antineutrino fluxes},}\ }\href {\doibase 10.1103/PhysRevLett.123.022502} {\bibfield  {journal} {\bibinfo  {journal} {Phys. Rev. Lett.}\ }\textbf {\bibinfo {volume} {123}},\ \bibinfo {pages} {022502} (\bibinfo {year} {2019})}\BibitemShut {NoStop}%
\bibitem [{\citenamefont {Kopeikin}, \citenamefont {Skorokhvatov},\ and\ \citenamefont {Titov}(2021)}]{Kopeikin2021b}%
  \BibitemOpen
  \bibfield  {author} {\bibinfo {author} {\bibfnamefont {V.}~\bibnamefont {Kopeikin}}, \bibinfo {author} {\bibfnamefont {M.}~\bibnamefont {Skorokhvatov}}, \ and\ \bibinfo {author} {\bibfnamefont {O.}~\bibnamefont {Titov}},\ }\bibfield  {title} {\enquote {\bibinfo {title} {Reevaluating reactor antineutrino spectra with new measurements of the ratio between {$^{235}$U} and {$^{239}$Pu} {$\beta$} spectra},}\ }\href {\doibase 10.1103/PhysRevD.104.L071301} {\bibfield  {journal} {\bibinfo  {journal} {Phys. Rev. D}\ }\textbf {\bibinfo {volume} {104}},\ \bibinfo {pages} {L071301} (\bibinfo {year} {2021})}\BibitemShut {NoStop}%
\bibitem [{\citenamefont {Ricciardi}, \citenamefont {Vignaroli},\ and\ \citenamefont {Vassani}(2022)}]{IBDcross-sec-new}%
  \BibitemOpen
  \bibfield  {author} {\bibinfo {author} {\bibfnamefont {G.}~\bibnamefont {Ricciardi}}, \bibinfo {author} {\bibfnamefont {N.}~\bibnamefont {Vignaroli}}, \ and\ \bibinfo {author} {\bibfnamefont {F.}~\bibnamefont {Vassani}},\ }\bibfield  {title} {\enquote {\bibinfo {title} {An accurate evaluation of electron (anti-)neutrino scattering on nucleons},}\ }\href {\doibase 10.1007/JHEP08(2022)212} {\bibfield  {journal} {\bibinfo  {journal} {J. High. Energ. Phys}\ }\textbf {\bibinfo {volume} {2022}},\ \bibinfo {pages} {212} (\bibinfo {year} {2022})}\BibitemShut {NoStop}%
\bibitem [{\citenamefont {Strumia}\ and\ \citenamefont {Vissani}(2003)}]{Strumia2003}%
  \BibitemOpen
  \bibfield  {author} {\bibinfo {author} {\bibfnamefont {A.}~\bibnamefont {Strumia}}\ and\ \bibinfo {author} {\bibfnamefont {F.}~\bibnamefont {Vissani}},\ }\bibfield  {title} {\enquote {\bibinfo {title} {Precise quasielastic neutrino/nucleon cross-section},}\ }\href {\doibase https://doi.org/10.1016/S0370-2693(03)00616-6} {\bibfield  {journal} {\bibinfo  {journal} {Phys. Lett. B}\ }\textbf {\bibinfo {volume} {564}},\ \bibinfo {pages} {42--54} (\bibinfo {year} {2003})}\BibitemShut {NoStop}%
\bibitem [{\citenamefont {Robinson}\ \emph {et~al.}(2003)\citenamefont {Robinson}, \citenamefont {Kudryavtsev}, \citenamefont {Lüscher}, \citenamefont {McMillan}, \citenamefont {Lightfoot}, \citenamefont {Spooner}, \citenamefont {Smith},\ and\ \citenamefont {Liubarsky}}]{Robinson2003}%
  \BibitemOpen
  \bibfield  {author} {\bibinfo {author} {\bibfnamefont {M.}~\bibnamefont {Robinson}}, \bibinfo {author} {\bibfnamefont {V.}~\bibnamefont {Kudryavtsev}}, \bibinfo {author} {\bibfnamefont {R.}~\bibnamefont {Lüscher}}, \bibinfo {author} {\bibfnamefont {J.}~\bibnamefont {McMillan}}, \bibinfo {author} {\bibfnamefont {P.}~\bibnamefont {Lightfoot}}, \bibinfo {author} {\bibfnamefont {N.}~\bibnamefont {Spooner}}, \bibinfo {author} {\bibfnamefont {N.}~\bibnamefont {Smith}}, \ and\ \bibinfo {author} {\bibfnamefont {I.}~\bibnamefont {Liubarsky}},\ }\bibfield  {title} {\enquote {\bibinfo {title} {Measurements of muon flux at 1070m vertical depth in the boulby underground laboratory},}\ }\href {\doibase https://doi.org/10.1016/S0168-9002(03)01973-9} {\bibfield  {journal} {\bibinfo  {journal} {Nucl. Instrum. Methods Phys. Res. A}\ }\textbf {\bibinfo {volume} {511}},\ \bibinfo {pages} {347--353} (\bibinfo {year} {2003})}\BibitemShut {NoStop}%
\bibitem [{\citenamefont {Yeh}\ \emph {et~al.}(2011)\citenamefont {Yeh}, \citenamefont {Hans}, \citenamefont {Beriguete}, \citenamefont {Rosero}, \citenamefont {Hu}, \citenamefont {Hahn}, \citenamefont {Diwan}, \citenamefont {Jaffe}, \citenamefont {Kettell},\ and\ \citenamefont {Littenberg}}]{Yeh2011}%
  \BibitemOpen
  \bibfield  {author} {\bibinfo {author} {\bibfnamefont {M.}~\bibnamefont {Yeh}}, \bibinfo {author} {\bibfnamefont {S.}~\bibnamefont {Hans}}, \bibinfo {author} {\bibfnamefont {W.}~\bibnamefont {Beriguete}}, \bibinfo {author} {\bibfnamefont {R.}~\bibnamefont {Rosero}}, \bibinfo {author} {\bibfnamefont {L.}~\bibnamefont {Hu}}, \bibinfo {author} {\bibfnamefont {R.}~\bibnamefont {Hahn}}, \bibinfo {author} {\bibfnamefont {M.}~\bibnamefont {Diwan}}, \bibinfo {author} {\bibfnamefont {D.}~\bibnamefont {Jaffe}}, \bibinfo {author} {\bibfnamefont {S.}~\bibnamefont {Kettell}}, \ and\ \bibinfo {author} {\bibfnamefont {L.}~\bibnamefont {Littenberg}},\ }\bibfield  {title} {\enquote {\bibinfo {title} {A new water-based liquid scintillator and potential applications},}\ }\href {\doibase https://doi.org/10.1016/j.nima.2011.08.040} {\bibfield  {journal} {\bibinfo  {journal} {Nucl. Instrum. Methods Phys. Res. A}\ }\textbf {\bibinfo {volume} {660}},\ \bibinfo {pages} {51--56} (\bibinfo {year} {2011})}\BibitemShut {NoStop}%
\bibitem [{\citenamefont {Zsoldos}\ \emph {et~al.}(2022)\citenamefont {Zsoldos}, \citenamefont {Bagdasarian}, \citenamefont {Gann}, \citenamefont {Barna},\ and\ \citenamefont {Dye}}]{Zsoldos2022}%
  \BibitemOpen
  \bibfield  {author} {\bibinfo {author} {\bibfnamefont {S.}~\bibnamefont {Zsoldos}}, \bibinfo {author} {\bibfnamefont {Z.}~\bibnamefont {Bagdasarian}}, \bibinfo {author} {\bibfnamefont {G.~D.~O.}\ \bibnamefont {Gann}}, \bibinfo {author} {\bibfnamefont {A.}~\bibnamefont {Barna}}, \ and\ \bibinfo {author} {\bibfnamefont {S.}~\bibnamefont {Dye}},\ }\bibfield  {title} {\enquote {\bibinfo {title} {Geo- and reactor antineutrino sensitivity at {THEIA}},}\ }\href {\doibase 10.1140/epjc/s10052-022-11106-1} {\bibfield  {journal} {\bibinfo  {journal} {Eur. Phys. J. C}\ }\textbf {\bibinfo {volume} {82}} (\bibinfo {year} {2022}),\ 10.1140/epjc/s10052-022-11106-1}\BibitemShut {NoStop}%
\bibitem [{\citenamefont {Suzuki}(2019)}]{Suzuki2019}%
  \BibitemOpen
  \bibfield  {author} {\bibinfo {author} {\bibfnamefont {Y.}~\bibnamefont {Suzuki}},\ }\bibfield  {title} {\enquote {\bibinfo {title} {The {Super-Kamiokande} experiment},}\ }\href {\doibase 10.1140/epjc/s10052-019-6796-2} {\bibfield  {journal} {\bibinfo  {journal} {Eur. Phys. J. C}\ }\textbf {\bibinfo {volume} {79}} (\bibinfo {year} {2019}),\ 10.1140/epjc/s10052-019-6796-2}\BibitemShut {NoStop}%
\bibitem [{\citenamefont {{JUNO Collaboration}}(2022)}]{JUNO2022}%
  \BibitemOpen
  \bibfield  {author} {\bibinfo {author} {\bibnamefont {{JUNO Collaboration}}},\ }\bibfield  {title} {\enquote {\bibinfo {title} {{JUNO} physics and detector},}\ }\href {\doibase 10.1016/j.ppnp.2021.103927} {\bibfield  {journal} {\bibinfo  {journal} {Prog. Part. Nucl. Phys.}\ }\textbf {\bibinfo {volume} {123}},\ \bibinfo {pages} {103927} (\bibinfo {year} {2022})}\BibitemShut {NoStop}%
\bibitem [{\citenamefont {Askins}\ \emph {et~al.}(2020)\citenamefont {Askins}, \citenamefont {Bagdasarian}, \citenamefont {Barros}, \citenamefont {Beier}, \citenamefont {Blucher}, \citenamefont {Bonventre}, \citenamefont {Bourret}, \citenamefont {Callaghan}, \citenamefont {Caravaca}, \citenamefont {Diwan}, \citenamefont {Dye}, \citenamefont {Eisch}, \citenamefont {Elagin}, \citenamefont {Enqvist}, \citenamefont {Fischer}, \citenamefont {Frankiewicz}, \citenamefont {Grant}, \citenamefont {Guffanti}, \citenamefont {Hagner}, \citenamefont {Hallin}, \citenamefont {Jackson}, \citenamefont {Jiang}, \citenamefont {Kaptanoglu}, \citenamefont {Klein}, \citenamefont {Kolomensky}, \citenamefont {Kraus}, \citenamefont {Krennrich}, \citenamefont {Kutter}, \citenamefont {Lachenmaier}, \citenamefont {Land}, \citenamefont {Lande}, \citenamefont {Learned}, \citenamefont {Lozza}, \citenamefont {Ludhova}, \citenamefont {Malek}, \citenamefont {Manecki}, \citenamefont {Maneira}, \citenamefont {Maricic}, \citenamefont {Martyn},
  \citenamefont {Mastbaum}, \citenamefont {Mauger}, \citenamefont {Moretti}, \citenamefont {Napolitano}, \citenamefont {Naranjo}, \citenamefont {Nieslony}, \citenamefont {Oberauer}, \citenamefont {Orebi Gann}, \citenamefont {Ouellet}, \citenamefont {Pershing}, \citenamefont {Petcov}, \citenamefont {Pickard}, \citenamefont {Rosero}, \citenamefont {Sanchez}, \citenamefont {Sawatzki}, \citenamefont {Seo}, \citenamefont {Smiley}, \citenamefont {Smy}, \citenamefont {Stahl}, \citenamefont {Steiger}, \citenamefont {Stock}, \citenamefont {Sunej}, \citenamefont {Svoboda}, \citenamefont {Tiras}, \citenamefont {Trzaska}, \citenamefont {Tzanov}, \citenamefont {Vagins}, \citenamefont {Vilela}, \citenamefont {Wang}, \citenamefont {Wang}, \citenamefont {Wetstein}, \citenamefont {Wilking}, \citenamefont {Winslow}, \citenamefont {Wittich}, \citenamefont {Wonsak}, \citenamefont {Worcester}, \citenamefont {Wurm}, \citenamefont {Yang}, \citenamefont {Yeh}, \citenamefont {Zimmerman}, \citenamefont {Zsoldos},\ and\ \citenamefont
  {Zuber}}]{Askins2020}%
  \BibitemOpen
  \bibfield  {author} {\bibinfo {author} {\bibfnamefont {M.}~\bibnamefont {Askins}}, \bibinfo {author} {\bibfnamefont {Z.}~\bibnamefont {Bagdasarian}}, \bibinfo {author} {\bibfnamefont {N.}~\bibnamefont {Barros}}, \bibinfo {author} {\bibfnamefont {E.~W.}\ \bibnamefont {Beier}}, \bibinfo {author} {\bibfnamefont {E.}~\bibnamefont {Blucher}}, \bibinfo {author} {\bibfnamefont {R.}~\bibnamefont {Bonventre}}, \bibinfo {author} {\bibfnamefont {E.}~\bibnamefont {Bourret}}, \bibinfo {author} {\bibfnamefont {E.~J.}\ \bibnamefont {Callaghan}}, \bibinfo {author} {\bibfnamefont {J.}~\bibnamefont {Caravaca}}, \bibinfo {author} {\bibfnamefont {M.}~\bibnamefont {Diwan}}, \bibinfo {author} {\bibfnamefont {S.~T.}\ \bibnamefont {Dye}}, \bibinfo {author} {\bibfnamefont {J.}~\bibnamefont {Eisch}}, \bibinfo {author} {\bibfnamefont {A.}~\bibnamefont {Elagin}}, \bibinfo {author} {\bibfnamefont {T.}~\bibnamefont {Enqvist}}, \bibinfo {author} {\bibfnamefont {V.}~\bibnamefont {Fischer}}, \bibinfo {author} {\bibfnamefont
  {K.}~\bibnamefont {Frankiewicz}}, \bibinfo {author} {\bibfnamefont {C.}~\bibnamefont {Grant}}, \bibinfo {author} {\bibfnamefont {D.}~\bibnamefont {Guffanti}}, \bibinfo {author} {\bibfnamefont {C.}~\bibnamefont {Hagner}}, \bibinfo {author} {\bibfnamefont {A.}~\bibnamefont {Hallin}}, \bibinfo {author} {\bibfnamefont {C.~M.}\ \bibnamefont {Jackson}}, \bibinfo {author} {\bibfnamefont {R.}~\bibnamefont {Jiang}}, \bibinfo {author} {\bibfnamefont {T.}~\bibnamefont {Kaptanoglu}}, \bibinfo {author} {\bibfnamefont {J.~R.}\ \bibnamefont {Klein}}, \bibinfo {author} {\bibfnamefont {Y.~G.}\ \bibnamefont {Kolomensky}}, \bibinfo {author} {\bibfnamefont {C.}~\bibnamefont {Kraus}}, \bibinfo {author} {\bibfnamefont {F.}~\bibnamefont {Krennrich}}, \bibinfo {author} {\bibfnamefont {T.}~\bibnamefont {Kutter}}, \bibinfo {author} {\bibfnamefont {T.}~\bibnamefont {Lachenmaier}}, \bibinfo {author} {\bibfnamefont {B.}~\bibnamefont {Land}}, \bibinfo {author} {\bibfnamefont {K.}~\bibnamefont {Lande}}, \bibinfo {author} {\bibfnamefont
  {J.~G.}\ \bibnamefont {Learned}}, \bibinfo {author} {\bibfnamefont {V.}~\bibnamefont {Lozza}}, \bibinfo {author} {\bibfnamefont {L.}~\bibnamefont {Ludhova}}, \bibinfo {author} {\bibfnamefont {M.}~\bibnamefont {Malek}}, \bibinfo {author} {\bibfnamefont {S.}~\bibnamefont {Manecki}}, \bibinfo {author} {\bibfnamefont {J.}~\bibnamefont {Maneira}}, \bibinfo {author} {\bibfnamefont {J.}~\bibnamefont {Maricic}}, \bibinfo {author} {\bibfnamefont {J.}~\bibnamefont {Martyn}}, \bibinfo {author} {\bibfnamefont {A.}~\bibnamefont {Mastbaum}}, \bibinfo {author} {\bibfnamefont {C.}~\bibnamefont {Mauger}}, \bibinfo {author} {\bibfnamefont {F.}~\bibnamefont {Moretti}}, \bibinfo {author} {\bibfnamefont {J.}~\bibnamefont {Napolitano}}, \bibinfo {author} {\bibfnamefont {B.}~\bibnamefont {Naranjo}}, \bibinfo {author} {\bibfnamefont {M.}~\bibnamefont {Nieslony}}, \bibinfo {author} {\bibfnamefont {L.}~\bibnamefont {Oberauer}}, \bibinfo {author} {\bibfnamefont {G.~D.}\ \bibnamefont {Orebi Gann}}, \bibinfo {author} {\bibfnamefont
  {J.}~\bibnamefont {Ouellet}}, \bibinfo {author} {\bibfnamefont {T.}~\bibnamefont {Pershing}}, \bibinfo {author} {\bibfnamefont {S.~T.}\ \bibnamefont {Petcov}}, \bibinfo {author} {\bibfnamefont {L.}~\bibnamefont {Pickard}}, \bibinfo {author} {\bibfnamefont {R.}~\bibnamefont {Rosero}}, \bibinfo {author} {\bibfnamefont {M.~C.}\ \bibnamefont {Sanchez}}, \bibinfo {author} {\bibfnamefont {J.}~\bibnamefont {Sawatzki}}, \bibinfo {author} {\bibfnamefont {S.~H.}\ \bibnamefont {Seo}}, \bibinfo {author} {\bibfnamefont {M.}~\bibnamefont {Smiley}}, \bibinfo {author} {\bibfnamefont {M.}~\bibnamefont {Smy}}, \bibinfo {author} {\bibfnamefont {A.}~\bibnamefont {Stahl}}, \bibinfo {author} {\bibfnamefont {H.}~\bibnamefont {Steiger}}, \bibinfo {author} {\bibfnamefont {M.~R.}\ \bibnamefont {Stock}}, \bibinfo {author} {\bibfnamefont {H.}~\bibnamefont {Sunej}}, \bibinfo {author} {\bibfnamefont {R.}~\bibnamefont {Svoboda}}, \bibinfo {author} {\bibfnamefont {E.}~\bibnamefont {Tiras}}, \bibinfo {author} {\bibfnamefont {W.~H.}\
  \bibnamefont {Trzaska}}, \bibinfo {author} {\bibfnamefont {M.}~\bibnamefont {Tzanov}}, \bibinfo {author} {\bibfnamefont {M.}~\bibnamefont {Vagins}}, \bibinfo {author} {\bibfnamefont {C.}~\bibnamefont {Vilela}}, \bibinfo {author} {\bibfnamefont {Z.}~\bibnamefont {Wang}}, \bibinfo {author} {\bibfnamefont {J.}~\bibnamefont {Wang}}, \bibinfo {author} {\bibfnamefont {M.}~\bibnamefont {Wetstein}}, \bibinfo {author} {\bibfnamefont {M.~J.}\ \bibnamefont {Wilking}}, \bibinfo {author} {\bibfnamefont {L.}~\bibnamefont {Winslow}}, \bibinfo {author} {\bibfnamefont {P.}~\bibnamefont {Wittich}}, \bibinfo {author} {\bibfnamefont {B.}~\bibnamefont {Wonsak}}, \bibinfo {author} {\bibfnamefont {E.}~\bibnamefont {Worcester}}, \bibinfo {author} {\bibfnamefont {M.}~\bibnamefont {Wurm}}, \bibinfo {author} {\bibfnamefont {G.}~\bibnamefont {Yang}}, \bibinfo {author} {\bibfnamefont {M.}~\bibnamefont {Yeh}}, \bibinfo {author} {\bibfnamefont {E.~D.}\ \bibnamefont {Zimmerman}}, \bibinfo {author} {\bibfnamefont {S.}~\bibnamefont
  {Zsoldos}}, \ and\ \bibinfo {author} {\bibfnamefont {K.}~\bibnamefont {Zuber}},\ }\bibfield  {title} {\enquote {\bibinfo {title} {Theia: an advanced optical neutrino detector},}\ }\href {\doibase 10.1140/epjc/s10052-020-7977-8} {\bibfield  {journal} {\bibinfo  {journal} {Eur. Phys. J. C}\ }\textbf {\bibinfo {volume} {80}} (\bibinfo {year} {2020}),\ 10.1140/epjc/s10052-020-7977-8}\BibitemShut {NoStop}%
\bibitem [{\citenamefont {Beacom}\ and\ \citenamefont {Vagins}(2004)}]{Beacom2003}%
  \BibitemOpen
  \bibfield  {author} {\bibinfo {author} {\bibfnamefont {J.~F.}\ \bibnamefont {Beacom}}\ and\ \bibinfo {author} {\bibfnamefont {M.~R.}\ \bibnamefont {Vagins}},\ }\bibfield  {title} {\enquote {\bibinfo {title} {Antineutrino spectroscopy with large water \ifmmode \check{C}\else \v{C}\fi{}erenkov detectors},}\ }\href {\doibase 10.1103/PhysRevLett.93.171101} {\bibfield  {journal} {\bibinfo  {journal} {Phys. Rev. Lett.}\ }\textbf {\bibinfo {volume} {93}},\ \bibinfo {pages} {171101} (\bibinfo {year} {2004})}\BibitemShut {NoStop}%
\bibitem [{\citenamefont {Abe}\ \emph {et~al.}(2021)\citenamefont {Abe}, \citenamefont {Bronner}, \citenamefont {Hayato}, \citenamefont {Hiraide}, \citenamefont {Ikeda}, \citenamefont {Imaizumi}, \citenamefont {Kameda}, \citenamefont {Kanemura}, \citenamefont {Kataoka}, \citenamefont {Miki} \emph {et~al.}}]{ABE2021}%
  \BibitemOpen
  \bibfield  {author} {\bibinfo {author} {\bibfnamefont {K.}~\bibnamefont {Abe}}, \bibinfo {author} {\bibfnamefont {C.}~\bibnamefont {Bronner}}, \bibinfo {author} {\bibfnamefont {Y.}~\bibnamefont {Hayato}}, \bibinfo {author} {\bibfnamefont {K.}~\bibnamefont {Hiraide}}, \bibinfo {author} {\bibfnamefont {M.}~\bibnamefont {Ikeda}}, \bibinfo {author} {\bibfnamefont {S.}~\bibnamefont {Imaizumi}}, \bibinfo {author} {\bibfnamefont {J.}~\bibnamefont {Kameda}}, \bibinfo {author} {\bibfnamefont {Y.}~\bibnamefont {Kanemura}}, \bibinfo {author} {\bibfnamefont {Y.}~\bibnamefont {Kataoka}}, \bibinfo {author} {\bibfnamefont {S.}~\bibnamefont {Miki}},  \emph {et~al.} (\bibinfo {collaboration} {Super-Kamiokande Collaboration}),\ }\bibfield  {title} {\enquote {\bibinfo {title} {First gadolinium loading to {Super-Kamiokande}},}\ }\href {\doibase 10.1016/j.nima.2021.166248} {\bibfield  {journal} {\bibinfo  {journal} {Nucl. Instrum. Methods Phys. Res. A}\ }\textbf {\bibinfo {volume} {1027}},\ \bibinfo {pages} {166248} (\bibinfo
  {year} {2021})}\BibitemShut {NoStop}%
\bibitem [{\citenamefont {{EDF Energy}}(2022{\natexlab{a}})}]{EDFEnergy}%
  \BibitemOpen
  \bibfield  {author} {\bibinfo {author} {\bibnamefont {{EDF Energy}}},\ }\href {https://www.edfenergy.com/energy/nuclear-lifetime-management} {\enquote {\bibinfo {title} {{Nuclear Lifetime Management}},}\ } (\bibinfo {year} {2022}{\natexlab{a}})\BibitemShut {NoStop}%
\bibitem [{\citenamefont {{EDF Energy}}(2022{\natexlab{b}})}]{EDFhinkley}%
  \BibitemOpen
  \bibfield  {author} {\bibinfo {author} {\bibnamefont {{EDF Energy}}},\ }\href {https://www.edfenergy.com/energy/nuclear-new-build-projects/hinkley-point-c/about} {\enquote {\bibinfo {title} {{About Hinkley Point C}},}\ } (\bibinfo {year} {2022}{\natexlab{b}})\BibitemShut {NoStop}%
\bibitem [{\citenamefont {{EDF Energy}}(2022{\natexlab{c}})}]{EDFsizewell}%
  \BibitemOpen
  \bibfield  {author} {\bibinfo {author} {\bibnamefont {{EDF Energy}}},\ }\href {https://www.edfenergy.com/media-centre/news-releases/sizewell-b-starts-review-extend-operation-20-years} {\enquote {\bibinfo {title} {{Sizewell B starts review to extend operation by 20 years}},}\ } (\bibinfo {year} {2022}{\natexlab{c}})\BibitemShut {NoStop}%
\bibitem [{\citenamefont {{ASN}}(2021)}]{ASN}%
  \BibitemOpen
  \bibfield  {author} {\bibinfo {author} {\bibnamefont {{ASN}}},\ }\href {https://www.french-nuclear-safety.fr/asn-informs/news-releases/900-mwe-reactors-beyond-40-years} {\enquote {\bibinfo {title} {{ASN issues a position statement on the conditions for continued operation of the 900 MWe reactors beyond 40 years}},}\ } (\bibinfo {year} {2021})\BibitemShut {NoStop}%
\bibitem [{\citenamefont {{Google}}(2022)}]{Google2022}%
  \BibitemOpen
  \bibfield  {author} {\bibinfo {author} {\bibnamefont {{Google}}},\ }\href@noop {} {\enquote {\bibinfo {title} {{Google Maps satellite image showing location of nuclear power stations in the UK and northern France}},}\ } (\bibinfo {year} {2022})\BibitemShut {NoStop}%
\bibitem [{\citenamefont {Seibert}(2014)}]{Seibert2014}%
  \BibitemOpen
  \bibfield  {author} {\bibinfo {author} {\bibfnamefont {S.}~\bibnamefont {Seibert}},\ }\href@noop {} {\enquote {\bibinfo {title} {{RAT-PAC}},}\ } (\bibinfo {year} {2014})\BibitemShut {NoStop}%
\bibitem [{\citenamefont {Agostinelli}\ \emph {et~al.}(2003)\citenamefont {Agostinelli}, \citenamefont {Allison}, \citenamefont {Amako}, \citenamefont {Apostolakis}, \citenamefont {Araujo}, \citenamefont {Arce}, \citenamefont {Asai}, \citenamefont {Axen}, \citenamefont {Banerjee}, \citenamefont {Barrand} \emph {et~al.}}]{AGOSTINELLI2003}%
  \BibitemOpen
  \bibfield  {author} {\bibinfo {author} {\bibfnamefont {S.}~\bibnamefont {Agostinelli}}, \bibinfo {author} {\bibfnamefont {J.}~\bibnamefont {Allison}}, \bibinfo {author} {\bibfnamefont {K.}~\bibnamefont {Amako}}, \bibinfo {author} {\bibfnamefont {J.}~\bibnamefont {Apostolakis}}, \bibinfo {author} {\bibfnamefont {H.}~\bibnamefont {Araujo}}, \bibinfo {author} {\bibfnamefont {P.}~\bibnamefont {Arce}}, \bibinfo {author} {\bibfnamefont {M.}~\bibnamefont {Asai}}, \bibinfo {author} {\bibfnamefont {D.}~\bibnamefont {Axen}}, \bibinfo {author} {\bibfnamefont {S.}~\bibnamefont {Banerjee}}, \bibinfo {author} {\bibfnamefont {G.}~\bibnamefont {Barrand}},  \emph {et~al.},\ }\bibfield  {title} {\enquote {\bibinfo {title} {{GEANT4} — a simulation toolkit},}\ }\href {\doibase 10.1016/S0168-9002(03)01368-8} {\bibfield  {journal} {\bibinfo  {journal} {Nucl. Instrum. Methods Phys. Res. A}\ }\textbf {\bibinfo {volume} {506}},\ \bibinfo {pages} {250--303} (\bibinfo {year} {2003})}\BibitemShut {NoStop}%
\bibitem [{\citenamefont {Allison}\ \emph {et~al.}(2016)\citenamefont {Allison}, \citenamefont {Amako}, \citenamefont {Apostolakis}, \citenamefont {Arce}, \citenamefont {Asai}, \citenamefont {Aso}, \citenamefont {Bagli}, \citenamefont {Bagulya}, \citenamefont {Banerjee}, \citenamefont {Barrand} \emph {et~al.}}]{ALLISON2016}%
  \BibitemOpen
  \bibfield  {author} {\bibinfo {author} {\bibfnamefont {J.}~\bibnamefont {Allison}}, \bibinfo {author} {\bibfnamefont {K.}~\bibnamefont {Amako}}, \bibinfo {author} {\bibfnamefont {J.}~\bibnamefont {Apostolakis}}, \bibinfo {author} {\bibfnamefont {P.}~\bibnamefont {Arce}}, \bibinfo {author} {\bibfnamefont {M.}~\bibnamefont {Asai}}, \bibinfo {author} {\bibfnamefont {T.}~\bibnamefont {Aso}}, \bibinfo {author} {\bibfnamefont {E.}~\bibnamefont {Bagli}}, \bibinfo {author} {\bibfnamefont {A.}~\bibnamefont {Bagulya}}, \bibinfo {author} {\bibfnamefont {S.}~\bibnamefont {Banerjee}}, \bibinfo {author} {\bibfnamefont {G.}~\bibnamefont {Barrand}},  \emph {et~al.},\ }\bibfield  {title} {\enquote {\bibinfo {title} {Recent developments in {GEANT4}},}\ }\href {\doibase 10.1016/j.nima.2016.06.125} {\bibfield  {journal} {\bibinfo  {journal} {Nucl. Instrum. Methods Phys. Res. A}\ }\textbf {\bibinfo {volume} {835}},\ \bibinfo {pages} {186--225} (\bibinfo {year} {2016})}\BibitemShut {NoStop}%
\bibitem [{\citenamefont {Lönnblad}(1994)}]{LONNBLAD1994}%
  \BibitemOpen
  \bibfield  {author} {\bibinfo {author} {\bibfnamefont {L.}~\bibnamefont {Lönnblad}},\ }\bibfield  {title} {\enquote {\bibinfo {title} {Clhep—a project for designing a c++ class library for high energy physics},}\ }\href {\doibase 10.1016/0010-4655(94)90217-8} {\bibfield  {journal} {\bibinfo  {journal} {Computer Physics Communications}\ }\textbf {\bibinfo {volume} {84}},\ \bibinfo {pages} {307--316} (\bibinfo {year} {1994})}\BibitemShut {NoStop}%
\bibitem [{\citenamefont {Horton-Smith}(2005)}]{glg4sim}%
  \BibitemOpen
  \bibfield  {author} {\bibinfo {author} {\bibfnamefont {G.}~\bibnamefont {Horton-Smith}},\ }\href@noop {} {\enquote {\bibinfo {title} {{GLG4sim}},}\ } (\bibinfo {year} {2005})\BibitemShut {NoStop}%
\bibitem [{\citenamefont {Brun}\ and\ \citenamefont {Rademakers}(1997)}]{BRUN1997}%
  \BibitemOpen
  \bibfield  {author} {\bibinfo {author} {\bibfnamefont {R.}~\bibnamefont {Brun}}\ and\ \bibinfo {author} {\bibfnamefont {F.}~\bibnamefont {Rademakers}},\ }\bibfield  {title} {\enquote {\bibinfo {title} {{ROOT} — an object oriented data analysis framework v6.18/02},}\ }\href {\doibase 10.5281/zenodo.3895860} {\bibfield  {journal} {\bibinfo  {journal} {Nucl. Instrum. Methods Phys. Res. A}\ }\textbf {\bibinfo {volume} {389}},\ \bibinfo {pages} {81--86} (\bibinfo {year} {1997})}\BibitemShut {NoStop}%
\bibitem [{\citenamefont {Land}\ \emph {et~al.}(2021)\citenamefont {Land}, \citenamefont {Bagdasarian}, \citenamefont {Caravaca}, \citenamefont {Smiley}, \citenamefont {Yeh},\ and\ \citenamefont {Orebi~Gann}}]{MEV_WbLS_perf}%
  \BibitemOpen
  \bibfield  {author} {\bibinfo {author} {\bibfnamefont {B.~J.}\ \bibnamefont {Land}}, \bibinfo {author} {\bibfnamefont {Z.}~\bibnamefont {Bagdasarian}}, \bibinfo {author} {\bibfnamefont {J.}~\bibnamefont {Caravaca}}, \bibinfo {author} {\bibfnamefont {M.}~\bibnamefont {Smiley}}, \bibinfo {author} {\bibfnamefont {M.}~\bibnamefont {Yeh}}, \ and\ \bibinfo {author} {\bibfnamefont {G.~D.}\ \bibnamefont {Orebi~Gann}},\ }\bibfield  {title} {\enquote {\bibinfo {title} {Mev-scale performance of water-based and pure liquid scintillator detectors},}\ }\href {\doibase 10.1103/PhysRevD.103.052004} {\bibfield  {journal} {\bibinfo  {journal} {Phys. Rev. D}\ }\textbf {\bibinfo {volume} {103}},\ \bibinfo {pages} {052004} (\bibinfo {year} {2021})}\BibitemShut {NoStop}%
\bibitem [{\citenamefont {Caravaca}\ \emph {et~al.}(2020)\citenamefont {Caravaca}, \citenamefont {Land}, \citenamefont {Yeh},\ and\ \citenamefont {Gann}}]{CherenkovScintillation}%
  \BibitemOpen
  \bibfield  {author} {\bibinfo {author} {\bibfnamefont {J.}~\bibnamefont {Caravaca}}, \bibinfo {author} {\bibfnamefont {B.~J.}\ \bibnamefont {Land}}, \bibinfo {author} {\bibfnamefont {M.}~\bibnamefont {Yeh}}, \ and\ \bibinfo {author} {\bibfnamefont {G.~D.~O.}\ \bibnamefont {Gann}},\ }\bibfield  {title} {\enquote {\bibinfo {title} {Characterization of water-based liquid scintillator for cherenkov and scintillation separation},}\ }\href {\doibase 10.1140/epjc/s10052-020-8418-4} {\bibfield  {journal} {\bibinfo  {journal} {Eur. Phys. J. C}\ }\textbf {\bibinfo {volume} {80}},\ \bibinfo {pages} {867} (\bibinfo {year} {2020})}\BibitemShut {NoStop}%
\bibitem [{\citenamefont {Onken}\ \emph {et~al.}(2020)\citenamefont {Onken}, \citenamefont {Moretti}, \citenamefont {Caravaca}, \citenamefont {Yeh}, \citenamefont {Orebi~Gann},\ and\ \citenamefont {Bourret}}]{WbLSTimeResponse}%
  \BibitemOpen
  \bibfield  {author} {\bibinfo {author} {\bibfnamefont {D.~R.}\ \bibnamefont {Onken}}, \bibinfo {author} {\bibfnamefont {F.}~\bibnamefont {Moretti}}, \bibinfo {author} {\bibfnamefont {J.}~\bibnamefont {Caravaca}}, \bibinfo {author} {\bibfnamefont {M.}~\bibnamefont {Yeh}}, \bibinfo {author} {\bibfnamefont {G.~D.}\ \bibnamefont {Orebi~Gann}}, \ and\ \bibinfo {author} {\bibfnamefont {E.~D.}\ \bibnamefont {Bourret}},\ }\bibfield  {title} {\enquote {\bibinfo {title} {Time response of water-based liquid scintillator from x-ray excitation},}\ }\href {\doibase 10.1039/D0MA00055H} {\bibfield  {journal} {\bibinfo  {journal} {Mater. Adv.}\ }\textbf {\bibinfo {volume} {1}},\ \bibinfo {pages} {71--76} (\bibinfo {year} {2020})}\BibitemShut {NoStop}%
\bibitem [{Gab(2021)}]{Gabriel2022}%
  \BibitemOpen
  \href@noop {} {}\bibinfo {howpublished} {{Brookhaven National Laboratory and Lawrence Berkeley National Laboratory (private communication)}} (\bibinfo {year} {2021})\BibitemShut {NoStop}%
\bibitem [{\citenamefont {Marti}\ \emph {et~al.}(2020)\citenamefont {Marti}, \citenamefont {Ikeda}, \citenamefont {Kato}, \citenamefont {Kishimoto}, \citenamefont {Nakahata}, \citenamefont {Nakajima}, \citenamefont {Nakano}, \citenamefont {Nakayama}, \citenamefont {Okajima}, \citenamefont {Orii} \emph {et~al.}}]{Marti2020}%
  \BibitemOpen
  \bibfield  {author} {\bibinfo {author} {\bibfnamefont {L.}~\bibnamefont {Marti}}, \bibinfo {author} {\bibfnamefont {M.}~\bibnamefont {Ikeda}}, \bibinfo {author} {\bibfnamefont {Y.}~\bibnamefont {Kato}}, \bibinfo {author} {\bibfnamefont {Y.}~\bibnamefont {Kishimoto}}, \bibinfo {author} {\bibfnamefont {M.}~\bibnamefont {Nakahata}}, \bibinfo {author} {\bibfnamefont {Y.}~\bibnamefont {Nakajima}}, \bibinfo {author} {\bibfnamefont {Y.}~\bibnamefont {Nakano}}, \bibinfo {author} {\bibfnamefont {S.}~\bibnamefont {Nakayama}}, \bibinfo {author} {\bibfnamefont {Y.}~\bibnamefont {Okajima}}, \bibinfo {author} {\bibfnamefont {A.}~\bibnamefont {Orii}},  \emph {et~al.},\ }\bibfield  {title} {\enquote {\bibinfo {title} {Evaluation of gadolinium’s action on water {Cherenkov} detector systems with {EGADS}},}\ }\href {\doibase 10.1016/j.nima.2020.163549} {\bibfield  {journal} {\bibinfo  {journal} {Nucl. Instrum. Methods Phys. Res. A}\ }\textbf {\bibinfo {volume} {959}},\ \bibinfo {pages} {163549} (\bibinfo {year}
  {2020})}\BibitemShut {NoStop}%
\bibitem [{\citenamefont {Dye}\ and\ \citenamefont {Barna}(2015)}]{Dye2015}%
  \BibitemOpen
  \bibfield  {author} {\bibinfo {author} {\bibfnamefont {S.}~\bibnamefont {Dye}}\ and\ \bibinfo {author} {\bibfnamefont {A.}~\bibnamefont {Barna}},\ }\bibfield  {title} {\enquote {\bibinfo {title} {Global antineutrino modeling for a web application},}\ }\href {\doibase 10.48550/ARXIV.1510.05633} {\  (\bibinfo {year} {2015}),\ 10.48550/ARXIV.1510.05633},\ \Eprint {http://arxiv.org/abs/1510.05633} {arXiv:1510.05633 [physics.ins-det]} \BibitemShut {NoStop}%
\bibitem [{\citenamefont {{International Atomic Energy Agency}}(2022)}]{pris}%
  \BibitemOpen
  \bibfield  {author} {\bibinfo {author} {\bibnamefont {{International Atomic Energy Agency}}},\ }\href {https://pris.iaea.org/pris/home.aspx} {\enquote {\bibinfo {title} {{Power Reactor Information System (PRIS)}},}\ } (\bibinfo {year} {2022})\BibitemShut {NoStop}%
\bibitem [{\citenamefont {Li}\ and\ \citenamefont {Beacom}(2014)}]{Li2014}%
  \BibitemOpen
  \bibfield  {author} {\bibinfo {author} {\bibfnamefont {S.~W.}\ \bibnamefont {Li}}\ and\ \bibinfo {author} {\bibfnamefont {J.~F.}\ \bibnamefont {Beacom}},\ }\bibfield  {title} {\enquote {\bibinfo {title} {First calculation of cosmic-ray muon spallation backgrounds for {MeV} astrophysical neutrino signals in {Super-Kamiokande}},}\ }\href {\doibase 10.1103/PhysRevC.89.045801} {\bibfield  {journal} {\bibinfo  {journal} {Phys. Rev. C}\ }\textbf {\bibinfo {volume} {89}},\ \bibinfo {pages} {045801} (\bibinfo {year} {2014})}\BibitemShut {NoStop}%
\bibitem [{\citenamefont {Wang}\ \emph {et~al.}(2001)\citenamefont {Wang}, \citenamefont {Balic}, \citenamefont {Gratta}, \citenamefont {Fass\`o}, \citenamefont {Roesler},\ and\ \citenamefont {Ferrari}}]{Wang2001}%
  \BibitemOpen
  \bibfield  {author} {\bibinfo {author} {\bibfnamefont {Y.-F.}\ \bibnamefont {Wang}}, \bibinfo {author} {\bibfnamefont {V.}~\bibnamefont {Balic}}, \bibinfo {author} {\bibfnamefont {G.}~\bibnamefont {Gratta}}, \bibinfo {author} {\bibfnamefont {A.}~\bibnamefont {Fass\`o}}, \bibinfo {author} {\bibfnamefont {S.}~\bibnamefont {Roesler}}, \ and\ \bibinfo {author} {\bibfnamefont {A.}~\bibnamefont {Ferrari}},\ }\bibfield  {title} {\enquote {\bibinfo {title} {Predicting neutron production from cosmic-ray muons},}\ }\href {\doibase 10.1103/PhysRevD.64.013012} {\bibfield  {journal} {\bibinfo  {journal} {Phys. Rev. D}\ }\textbf {\bibinfo {volume} {64}},\ \bibinfo {pages} {013012} (\bibinfo {year} {2001})}\BibitemShut {NoStop}%
\bibitem [{\citenamefont {Mei}\ and\ \citenamefont {Hime}(2006)}]{Mei2006}%
  \BibitemOpen
  \bibfield  {author} {\bibinfo {author} {\bibfnamefont {D.-M.}\ \bibnamefont {Mei}}\ and\ \bibinfo {author} {\bibfnamefont {A.}~\bibnamefont {Hime}},\ }\bibfield  {title} {\enquote {\bibinfo {title} {Muon-induced background study for underground laboratories},}\ }\href {\doibase 10.1103/PhysRevD.73.053004} {\bibfield  {journal} {\bibinfo  {journal} {Phys. Rev. D}\ }\textbf {\bibinfo {volume} {73}},\ \bibinfo {pages} {053004} (\bibinfo {year} {2006})}\BibitemShut {NoStop}%
\bibitem [{\citenamefont {Zhang}\ \emph {et~al.}(2016{\natexlab{a}})\citenamefont {Zhang}, \citenamefont {Abe}, \citenamefont {Haga}, \citenamefont {Hayato}, \citenamefont {Ikeda}, \citenamefont {Iyogi}, \citenamefont {Kameda}, \citenamefont {Kishimoto}, \citenamefont {Miura}, \citenamefont {Moriyama} \emph {et~al.}}]{Zhang2016}%
  \BibitemOpen
  \bibfield  {author} {\bibinfo {author} {\bibfnamefont {Y.}~\bibnamefont {Zhang}}, \bibinfo {author} {\bibfnamefont {K.}~\bibnamefont {Abe}}, \bibinfo {author} {\bibfnamefont {Y.}~\bibnamefont {Haga}}, \bibinfo {author} {\bibfnamefont {Y.}~\bibnamefont {Hayato}}, \bibinfo {author} {\bibfnamefont {M.}~\bibnamefont {Ikeda}}, \bibinfo {author} {\bibfnamefont {K.}~\bibnamefont {Iyogi}}, \bibinfo {author} {\bibfnamefont {J.}~\bibnamefont {Kameda}}, \bibinfo {author} {\bibfnamefont {Y.}~\bibnamefont {Kishimoto}}, \bibinfo {author} {\bibfnamefont {M.}~\bibnamefont {Miura}}, \bibinfo {author} {\bibfnamefont {S.}~\bibnamefont {Moriyama}},  \emph {et~al.} (\bibinfo {collaboration} {Super-Kamiokande Collaboration}),\ }\bibfield  {title} {\enquote {\bibinfo {title} {First measurement of radioactive isotope production through cosmic-ray muon spallation in {Super-Kamiokande IV}},}\ }\href {\doibase 10.1103/PhysRevD.93.012004} {\bibfield  {journal} {\bibinfo  {journal} {Phys. Rev. D}\ }\textbf {\bibinfo {volume} {93}},\
  \bibinfo {pages} {012004} (\bibinfo {year} {2016}{\natexlab{a}})}\BibitemShut {NoStop}%
\bibitem [{\citenamefont {Haselschwardt}\ \emph {et~al.}(2019)\citenamefont {Haselschwardt}, \citenamefont {Shaw}, \citenamefont {Nelson}, \citenamefont {Witherell}, \citenamefont {Yeh}, \citenamefont {Lesko}, \citenamefont {Cole}, \citenamefont {Kyre},\ and\ \citenamefont {White}}]{HASELSCHWARDT2019}%
  \BibitemOpen
  \bibfield  {author} {\bibinfo {author} {\bibfnamefont {S.}~\bibnamefont {Haselschwardt}}, \bibinfo {author} {\bibfnamefont {S.}~\bibnamefont {Shaw}}, \bibinfo {author} {\bibfnamefont {H.}~\bibnamefont {Nelson}}, \bibinfo {author} {\bibfnamefont {M.}~\bibnamefont {Witherell}}, \bibinfo {author} {\bibfnamefont {M.}~\bibnamefont {Yeh}}, \bibinfo {author} {\bibfnamefont {K.}~\bibnamefont {Lesko}}, \bibinfo {author} {\bibfnamefont {A.}~\bibnamefont {Cole}}, \bibinfo {author} {\bibfnamefont {S.}~\bibnamefont {Kyre}}, \ and\ \bibinfo {author} {\bibfnamefont {D.}~\bibnamefont {White}},\ }\bibfield  {title} {\enquote {\bibinfo {title} {A liquid scintillation detector for radioassay of gadolinium-loaded liquid scintillator for the {LZ Outer Detector}},}\ }\href {\doibase https://doi.org/10.1016/j.nima.2019.05.055} {\bibfield  {journal} {\bibinfo  {journal} {Nucl. Instrum. Methods Phys. Res. A}\ }\textbf {\bibinfo {volume} {937}},\ \bibinfo {pages} {148--163} (\bibinfo {year} {2019})}\BibitemShut {NoStop}%
\bibitem [{Ham(2020)}]{Hamamatsu2020}%
  \BibitemOpen
  \href@noop {} {}\bibinfo {howpublished} {{Hamamatsu Photonics K. K. (private communication)}} (\bibinfo {year} {2020})\BibitemShut {NoStop}%
\bibitem [{\citenamefont {Zhang}\ \emph {et~al.}(2016{\natexlab{b}})\citenamefont {Zhang}, \citenamefont {Fu}, \citenamefont {Ji}, \citenamefont {Liu}, \citenamefont {Liu}, \citenamefont {Wang}, \citenamefont {Yao},\ and\ \citenamefont {Yuan}}]{Zhang2016b}%
  \BibitemOpen
  \bibfield  {author} {\bibinfo {author} {\bibfnamefont {T.}~\bibnamefont {Zhang}}, \bibinfo {author} {\bibfnamefont {C.}~\bibnamefont {Fu}}, \bibinfo {author} {\bibfnamefont {X.}~\bibnamefont {Ji}}, \bibinfo {author} {\bibfnamefont {J.}~\bibnamefont {Liu}}, \bibinfo {author} {\bibfnamefont {X.}~\bibnamefont {Liu}}, \bibinfo {author} {\bibfnamefont {X.}~\bibnamefont {Wang}}, \bibinfo {author} {\bibfnamefont {C.}~\bibnamefont {Yao}}, \ and\ \bibinfo {author} {\bibfnamefont {X.}~\bibnamefont {Yuan}},\ }\bibfield  {title} {\enquote {\bibinfo {title} {Low background stainless steel for the pressure vessel in the {PandaX-II} dark matter experiment},}\ }\href {\doibase 10.1088/1748-0221/11/09/t09004} {\bibfield  {journal} {\bibinfo  {journal} {JINST}\ }\textbf {\bibinfo {volume} {11}},\ \bibinfo {pages} {T09004} (\bibinfo {year} {2016}{\natexlab{b}})}\BibitemShut {NoStop}%
\bibitem [{\citenamefont {Araújo}\ \emph {et~al.}(2012)\citenamefont {Araújo}, \citenamefont {Akimov}, \citenamefont {Barnes}, \citenamefont {Belov}, \citenamefont {Bewick}, \citenamefont {Burenkov}, \citenamefont {Chepel}, \citenamefont {Currie}, \citenamefont {DeViveiros}, \citenamefont {Edwards} \emph {et~al.}}]{ARAUJO2012}%
  \BibitemOpen
  \bibfield  {author} {\bibinfo {author} {\bibfnamefont {H.}~\bibnamefont {Araújo}}, \bibinfo {author} {\bibfnamefont {D.~Y.}\ \bibnamefont {Akimov}}, \bibinfo {author} {\bibfnamefont {E.}~\bibnamefont {Barnes}}, \bibinfo {author} {\bibfnamefont {V.}~\bibnamefont {Belov}}, \bibinfo {author} {\bibfnamefont {A.}~\bibnamefont {Bewick}}, \bibinfo {author} {\bibfnamefont {A.}~\bibnamefont {Burenkov}}, \bibinfo {author} {\bibfnamefont {V.}~\bibnamefont {Chepel}}, \bibinfo {author} {\bibfnamefont {A.}~\bibnamefont {Currie}}, \bibinfo {author} {\bibfnamefont {L.}~\bibnamefont {DeViveiros}}, \bibinfo {author} {\bibfnamefont {B.}~\bibnamefont {Edwards}},  \emph {et~al.},\ }\bibfield  {title} {\enquote {\bibinfo {title} {Radioactivity backgrounds in {ZEPLIN–III}},}\ }\href {\doibase https://doi.org/10.1016/j.astropartphys.2011.11.001} {\bibfield  {journal} {\bibinfo  {journal} {Astropart. Phys.}\ }\textbf {\bibinfo {volume} {35}},\ \bibinfo {pages} {495--502} (\bibinfo {year} {2012})}\BibitemShut {NoStop}%
\bibitem [{\citenamefont {Chu}, \citenamefont {Ekström},\ and\ \citenamefont {Firestone}(1999)}]{TOI}%
  \BibitemOpen
  \bibfield  {author} {\bibinfo {author} {\bibfnamefont {S.}~\bibnamefont {Chu}}, \bibinfo {author} {\bibfnamefont {L.}~\bibnamefont {Ekström}}, \ and\ \bibinfo {author} {\bibfnamefont {R.}~\bibnamefont {Firestone}},\ }\href@noop {} {\enquote {\bibinfo {title} {The {Lund/LBNL Nuclear Data Search: Table of Radioactive Isotopes}},}\ } (\bibinfo {year} {1999})\BibitemShut {NoStop}%
\bibitem [{\citenamefont {{TUNL Nuclear Data Evaluation Project}}(2021)}]{TUNL}%
  \BibitemOpen
  \bibfield  {author} {\bibinfo {author} {\bibnamefont {{TUNL Nuclear Data Evaluation Project}}},\ }\href {https://nucldata.tunl.duke.edu/nucldata} {\enquote {\bibinfo {title} {{Energy Level Diagrams}},}\ } (\bibinfo {year} {2021})\BibitemShut {NoStop}%
\bibitem [{\citenamefont {Jollet}\ and\ \citenamefont {Meregaglia}(2020)}]{JOLLET2020}%
  \BibitemOpen
  \bibfield  {author} {\bibinfo {author} {\bibfnamefont {C.}~\bibnamefont {Jollet}}\ and\ \bibinfo {author} {\bibfnamefont {A.}~\bibnamefont {Meregaglia}},\ }\bibfield  {title} {\enquote {\bibinfo {title} {$^9${Li} and $^8${He} decays in {GEANT4}},}\ }\href {\doibase 10.1016/j.nima.2019.162904} {\bibfield  {journal} {\bibinfo  {journal} {Nucl. Instrum. Methods Phys. Res. A}\ }\textbf {\bibinfo {volume} {949}},\ \bibinfo {pages} {162904} (\bibinfo {year} {2020})}\BibitemShut {NoStop}%
\bibitem [{\citenamefont {Tang}\ \emph {et~al.}(2006)\citenamefont {Tang}, \citenamefont {Horton-Smith}, \citenamefont {Kudryavtsev},\ and\ \citenamefont {Tonazzo}}]{Tang2006}%
  \BibitemOpen
  \bibfield  {author} {\bibinfo {author} {\bibfnamefont {A.}~\bibnamefont {Tang}}, \bibinfo {author} {\bibfnamefont {G.}~\bibnamefont {Horton-Smith}}, \bibinfo {author} {\bibfnamefont {V.~A.}\ \bibnamefont {Kudryavtsev}}, \ and\ \bibinfo {author} {\bibfnamefont {A.}~\bibnamefont {Tonazzo}},\ }\bibfield  {title} {\enquote {\bibinfo {title} {Muon simulations for {Super-Kamiokande}, {KamLAND}, and {CHOOZ}},}\ }\href {\doibase 10.1103/PhysRevD.74.053007} {\bibfield  {journal} {\bibinfo  {journal} {Phys. Rev. D}\ }\textbf {\bibinfo {volume} {74}},\ \bibinfo {pages} {053007} (\bibinfo {year} {2006})}\BibitemShut {NoStop}%
\bibitem [{\citenamefont {Sutanto}\ \emph {et~al.}(2020)\citenamefont {Sutanto}, \citenamefont {Akindele}, \citenamefont {Askins}, \citenamefont {Bergevin}, \citenamefont {Bernstein}, \citenamefont {Bowden}, \citenamefont {Dazeley}, \citenamefont {Jaffke}, \citenamefont {Jovanovic}, \citenamefont {Quillin} \emph {et~al.}}]{Sutanto2020}%
  \BibitemOpen
  \bibfield  {author} {\bibinfo {author} {\bibfnamefont {F.}~\bibnamefont {Sutanto}}, \bibinfo {author} {\bibfnamefont {O.~A.}\ \bibnamefont {Akindele}}, \bibinfo {author} {\bibfnamefont {M.}~\bibnamefont {Askins}}, \bibinfo {author} {\bibfnamefont {M.}~\bibnamefont {Bergevin}}, \bibinfo {author} {\bibfnamefont {A.}~\bibnamefont {Bernstein}}, \bibinfo {author} {\bibfnamefont {N.~S.}\ \bibnamefont {Bowden}}, \bibinfo {author} {\bibfnamefont {S.}~\bibnamefont {Dazeley}}, \bibinfo {author} {\bibfnamefont {P.}~\bibnamefont {Jaffke}}, \bibinfo {author} {\bibfnamefont {I.}~\bibnamefont {Jovanovic}}, \bibinfo {author} {\bibfnamefont {S.}~\bibnamefont {Quillin}},  \emph {et~al.},\ }\bibfield  {title} {\enquote {\bibinfo {title} {Measurement of muon-induced high-energy neutrons from rock in an underground gd-doped water detector},}\ }\href {\doibase 10.1103/PhysRevC.102.034616} {\bibfield  {journal} {\bibinfo  {journal} {Phys. Rev. C}\ }\textbf {\bibinfo {volume} {102}},\ \bibinfo {pages} {034616} (\bibinfo {year}
  {2020})}\BibitemShut {NoStop}%
\bibitem [{\citenamefont {Battat}\ \emph {et~al.}(2014)\citenamefont {Battat}, \citenamefont {Brack}, \citenamefont {Daw}, \citenamefont {Dorofeev}, \citenamefont {Ezeribe}, \citenamefont {Fox}, \citenamefont {Gauvreau}, \citenamefont {Gold}, \citenamefont {Harmon}, \citenamefont {Harton}, \citenamefont {Landers}, \citenamefont {Lee}, \citenamefont {Loomba}, \citenamefont {Matthews}, \citenamefont {Miller}, \citenamefont {Monte}, \citenamefont {Murphy}, \citenamefont {Paling}, \citenamefont {Phan}, \citenamefont {Pipe}, \citenamefont {Robinson}, \citenamefont {Sadler}, \citenamefont {Scarff}, \citenamefont {Snowden-Ifft}, \citenamefont {Spooner}, \citenamefont {Telfer}, \citenamefont {Walker}, \citenamefont {Warner},\ and\ \citenamefont {Yuriev}}]{Battat2014}%
  \BibitemOpen
  \bibfield  {author} {\bibinfo {author} {\bibfnamefont {J.}~\bibnamefont {Battat}}, \bibinfo {author} {\bibfnamefont {J.}~\bibnamefont {Brack}}, \bibinfo {author} {\bibfnamefont {E.}~\bibnamefont {Daw}}, \bibinfo {author} {\bibfnamefont {A.}~\bibnamefont {Dorofeev}}, \bibinfo {author} {\bibfnamefont {A.}~\bibnamefont {Ezeribe}}, \bibinfo {author} {\bibfnamefont {J.}~\bibnamefont {Fox}}, \bibinfo {author} {\bibfnamefont {J.-L.}\ \bibnamefont {Gauvreau}}, \bibinfo {author} {\bibfnamefont {M.}~\bibnamefont {Gold}}, \bibinfo {author} {\bibfnamefont {L.}~\bibnamefont {Harmon}}, \bibinfo {author} {\bibfnamefont {J.}~\bibnamefont {Harton}}, \bibinfo {author} {\bibfnamefont {J.}~\bibnamefont {Landers}}, \bibinfo {author} {\bibfnamefont {E.}~\bibnamefont {Lee}}, \bibinfo {author} {\bibfnamefont {D.}~\bibnamefont {Loomba}}, \bibinfo {author} {\bibfnamefont {J.}~\bibnamefont {Matthews}}, \bibinfo {author} {\bibfnamefont {E.}~\bibnamefont {Miller}}, \bibinfo {author} {\bibfnamefont {A.}~\bibnamefont {Monte}}, \bibinfo
  {author} {\bibfnamefont {A.}~\bibnamefont {Murphy}}, \bibinfo {author} {\bibfnamefont {S.}~\bibnamefont {Paling}}, \bibinfo {author} {\bibfnamefont {N.}~\bibnamefont {Phan}}, \bibinfo {author} {\bibfnamefont {M.}~\bibnamefont {Pipe}}, \bibinfo {author} {\bibfnamefont {M.}~\bibnamefont {Robinson}}, \bibinfo {author} {\bibfnamefont {S.}~\bibnamefont {Sadler}}, \bibinfo {author} {\bibfnamefont {A.}~\bibnamefont {Scarff}}, \bibinfo {author} {\bibfnamefont {D.}~\bibnamefont {Snowden-Ifft}}, \bibinfo {author} {\bibfnamefont {N.}~\bibnamefont {Spooner}}, \bibinfo {author} {\bibfnamefont {S.}~\bibnamefont {Telfer}}, \bibinfo {author} {\bibfnamefont {D.}~\bibnamefont {Walker}}, \bibinfo {author} {\bibfnamefont {D.}~\bibnamefont {Warner}}, \ and\ \bibinfo {author} {\bibfnamefont {L.}~\bibnamefont {Yuriev}},\ }\bibfield  {title} {\enquote {\bibinfo {title} {Radon in the {DRIFT-II} directional dark matter {TPC}: emanation, detection and mitigation},}\ }\href {\doibase 10.1088/1748-0221/9/11/p11004} {\bibfield
  {journal} {\bibinfo  {journal} {J. Instrum.}\ }\textbf {\bibinfo {volume} {9}},\ \bibinfo {pages} {P11004--P11004} (\bibinfo {year} {2014})}\BibitemShut {NoStop}%
\bibitem [{\citenamefont {Smy}(2007)}]{Smy2007}%
  \BibitemOpen
  \bibfield  {author} {\bibinfo {author} {\bibfnamefont {M.}~\bibnamefont {Smy}} (\bibinfo {collaboration} {{Super-Kamiokande Collaboration}}),\ }\bibfield  {title} {\enquote {\bibinfo {title} {Low energy event reconstruction and selection in {Super-Kamiokande-III}},}\ }in\ \href@noop {} {\emph {\bibinfo {booktitle} {30th International Cosmic Ray Conference}}}\ (\bibinfo {year} {2007})\BibitemShut {NoStop}%
\bibitem [{\citenamefont {Pedregosa}\ \emph {et~al.}(2011)\citenamefont {Pedregosa}, \citenamefont {Varoquaux}, \citenamefont {Gramfort}, \citenamefont {Michel}, \citenamefont {Thirion}, \citenamefont {Grisel}, \citenamefont {Blondel}, \citenamefont {Prettenhofer}, \citenamefont {Weiss}, \citenamefont {Dubourg}, \citenamefont {Vanderplas}, \citenamefont {Passos}, \citenamefont {Cournapeau}, \citenamefont {Brucher}, \citenamefont {Perrot},\ and\ \citenamefont {Duchesnay}}]{scikit-learn}%
  \BibitemOpen
  \bibfield  {author} {\bibinfo {author} {\bibfnamefont {F.}~\bibnamefont {Pedregosa}}, \bibinfo {author} {\bibfnamefont {G.}~\bibnamefont {Varoquaux}}, \bibinfo {author} {\bibfnamefont {A.}~\bibnamefont {Gramfort}}, \bibinfo {author} {\bibfnamefont {V.}~\bibnamefont {Michel}}, \bibinfo {author} {\bibfnamefont {B.}~\bibnamefont {Thirion}}, \bibinfo {author} {\bibfnamefont {O.}~\bibnamefont {Grisel}}, \bibinfo {author} {\bibfnamefont {M.}~\bibnamefont {Blondel}}, \bibinfo {author} {\bibfnamefont {P.}~\bibnamefont {Prettenhofer}}, \bibinfo {author} {\bibfnamefont {R.}~\bibnamefont {Weiss}}, \bibinfo {author} {\bibfnamefont {V.}~\bibnamefont {Dubourg}}, \bibinfo {author} {\bibfnamefont {J.}~\bibnamefont {Vanderplas}}, \bibinfo {author} {\bibfnamefont {A.}~\bibnamefont {Passos}}, \bibinfo {author} {\bibfnamefont {D.}~\bibnamefont {Cournapeau}}, \bibinfo {author} {\bibfnamefont {M.}~\bibnamefont {Brucher}}, \bibinfo {author} {\bibfnamefont {M.}~\bibnamefont {Perrot}}, \ and\ \bibinfo {author} {\bibfnamefont
  {E.}~\bibnamefont {Duchesnay}},\ }\bibfield  {title} {\enquote {\bibinfo {title} {Scikit-learn: Machine learning in {Python}},}\ }\href@noop {} {\bibfield  {journal} {\bibinfo  {journal} {J. Mach. Learn. Res.}\ }\textbf {\bibinfo {volume} {12}},\ \bibinfo {pages} {2825--2830} (\bibinfo {year} {2011})}\BibitemShut {NoStop}%
\bibitem [{\citenamefont {Zhan}\ \emph {et~al.}(2009)\citenamefont {Zhan}, \citenamefont {Wang}, \citenamefont {Cao},\ and\ \citenamefont {Wen}}]{FCTFST}%
  \BibitemOpen
  \bibfield  {author} {\bibinfo {author} {\bibfnamefont {L.}~\bibnamefont {Zhan}}, \bibinfo {author} {\bibfnamefont {Y.}~\bibnamefont {Wang}}, \bibinfo {author} {\bibfnamefont {J.}~\bibnamefont {Cao}}, \ and\ \bibinfo {author} {\bibfnamefont {L.}~\bibnamefont {Wen}},\ }\bibfield  {title} {\enquote {\bibinfo {title} {Experimental requirements to determine the neutrino mass hierarchy using reactor neutrinos},}\ }\href {\doibase 10.1103/PhysRevD.79.073007} {\bibfield  {journal} {\bibinfo  {journal} {Phys. Rev. D}\ }\textbf {\bibinfo {volume} {79}},\ \bibinfo {pages} {073007} (\bibinfo {year} {2009})}\BibitemShut {NoStop}%
\bibitem [{\citenamefont {Ciuffoli}\ \emph {et~al.}(2014)\citenamefont {Ciuffoli}, \citenamefont {Evslin}, \citenamefont {Wang}, \citenamefont {Yang}, \citenamefont {Zhang},\ and\ \citenamefont {Zhong}}]{Ciuffoli2014}%
  \BibitemOpen
  \bibfield  {author} {\bibinfo {author} {\bibfnamefont {E.}~\bibnamefont {Ciuffoli}}, \bibinfo {author} {\bibfnamefont {J.}~\bibnamefont {Evslin}}, \bibinfo {author} {\bibfnamefont {Z.}~\bibnamefont {Wang}}, \bibinfo {author} {\bibfnamefont {C.}~\bibnamefont {Yang}}, \bibinfo {author} {\bibfnamefont {X.}~\bibnamefont {Zhang}}, \ and\ \bibinfo {author} {\bibfnamefont {W.}~\bibnamefont {Zhong}},\ }\bibfield  {title} {\enquote {\bibinfo {title} {Medium baseline reactor neutrino experiments with two identical detectors},}\ }\href {\doibase 10.1016/j.physletb.2014.07.007} {\bibfield  {journal} {\bibinfo  {journal} {Phys. Lett. B}\ }\textbf {\bibinfo {volume} {736}},\ \bibinfo {pages} {110--118} (\bibinfo {year} {2014})}\BibitemShut {NoStop}%
\bibitem [{\citenamefont {An}\ \emph {et~al.}(2016)\citenamefont {An}, \citenamefont {An}, \citenamefont {An}, \citenamefont {Antonelli}, \citenamefont {Baussan}, \citenamefont {Beacom}, \citenamefont {Bezrukov}, \citenamefont {Blyth}, \citenamefont {Brugnera}, \citenamefont {Avanzini}, \citenamefont {Busto}, \citenamefont {Cabrera}, \citenamefont {Cai}, \citenamefont {Cai}, \citenamefont {Cammi}, \citenamefont {Cao}, \citenamefont {Cao}, \citenamefont {Chang}, \citenamefont {Chen}, \citenamefont {Chen}, \citenamefont {Chen}, \citenamefont {Chiesa}, \citenamefont {Clemenza}, \citenamefont {Clerbaux}, \citenamefont {Conrad}, \citenamefont {D’Angelo}, \citenamefont {Kerret}, \citenamefont {Deng}, \citenamefont {Deng}, \citenamefont {Ding}, \citenamefont {Djurcic}, \citenamefont {Dornic}, \citenamefont {Dracos}, \citenamefont {Drapier}, \citenamefont {Dusini}, \citenamefont {Dye}, \citenamefont {Enqvist}, \citenamefont {Fan}, \citenamefont {Fang}, \citenamefont {Favart}, \citenamefont {Ford}, \citenamefont
  {Göger-Neff}, \citenamefont {Gan}, \citenamefont {Garfagnini}, \citenamefont {Giammarchi}, \citenamefont {Gonchar}, \citenamefont {Gong}, \citenamefont {Gong}, \citenamefont {Gonin}, \citenamefont {Grassi}, \citenamefont {Grewing}, \citenamefont {Guan}, \citenamefont {Guarino}, \citenamefont {Guo}, \citenamefont {Guo}, \citenamefont {Guo}, \citenamefont {Hagner}, \citenamefont {Han}, \citenamefont {He}, \citenamefont {Heng}, \citenamefont {Hsiung}, \citenamefont {Hu}, \citenamefont {Hu}, \citenamefont {Hu}, \citenamefont {Huang}, \citenamefont {Huang}, \citenamefont {Huo}, \citenamefont {Ioannisian}, \citenamefont {Jeitler}, \citenamefont {Ji}, \citenamefont {Jiang}, \citenamefont {Jollet}, \citenamefont {Kang}, \citenamefont {Karagounis}, \citenamefont {Kazarian}, \citenamefont {Krumshteyn}, \citenamefont {Kruth}, \citenamefont {Kuusiniemi}, \citenamefont {Lachenmaier}, \citenamefont {Leitner}, \citenamefont {Li}, \citenamefont {Li}, \citenamefont {Li}, \citenamefont {Li}, \citenamefont {Li},
  \citenamefont {Li}, \citenamefont {Li}, \citenamefont {Li}, \citenamefont {Li}, \citenamefont {Liang}, \citenamefont {Lin}, \citenamefont {Lin}, \citenamefont {Lin}, \citenamefont {Ling}, \citenamefont {Lippi}, \citenamefont {Liu}, \citenamefont {Liu}, \citenamefont {Liu}, \citenamefont {Liu}, \citenamefont {Liu}, \citenamefont {Liu}, \citenamefont {Liu}, \citenamefont {Liu}, \citenamefont {Liu}, \citenamefont {Lombardi}, \citenamefont {Long}, \citenamefont {Lu}, \citenamefont {Lu}, \citenamefont {Lu}, \citenamefont {Lu}, \citenamefont {Lubsandorzhiev}, \citenamefont {Ludhova}, \citenamefont {Luo}, \citenamefont {Lyashuk}, \citenamefont {Möllenberg}, \citenamefont {Ma}, \citenamefont {Mantovani}, \citenamefont {Mao}, \citenamefont {Mari}, \citenamefont {McDonough}, \citenamefont {Meng}, \citenamefont {Meregaglia}, \citenamefont {Meroni}, \citenamefont {Mezzetto}, \citenamefont {Miramonti}, \citenamefont {Mueller}, \citenamefont {Naumov}, \citenamefont {Oberauer}, \citenamefont {Ochoa-Ricoux}, \citenamefont
  {Olshevskiy}, \citenamefont {Ortica}, \citenamefont {Paoloni}, \citenamefont {Peng}, \citenamefont {Peng}, \citenamefont {Previtali}, \citenamefont {Qi}, \citenamefont {Qian}, \citenamefont {Qian}, \citenamefont {Qian}, \citenamefont {Qin}, \citenamefont {Raffelt}, \citenamefont {Ranucci}, \citenamefont {Ricci}, \citenamefont {Robens}, \citenamefont {Romani}, \citenamefont {Ruan}, \citenamefont {Ruan}, \citenamefont {Salamanna}, \citenamefont {Shaevitz}, \citenamefont {Sinev}, \citenamefont {Sirignano}, \citenamefont {Sisti}, \citenamefont {Smirnov}, \citenamefont {Soiron}, \citenamefont {Stahl}, \citenamefont {Stanco}, \citenamefont {Steinmann}, \citenamefont {Sun}, \citenamefont {Sun}, \citenamefont {Taichenachev}, \citenamefont {Tang}, \citenamefont {Tkachev}, \citenamefont {Trzaska}, \citenamefont {Waasen}, \citenamefont {Volpe}, \citenamefont {Vorobel}, \citenamefont {Votano}, \citenamefont {Wang}, \citenamefont {Wang}, \citenamefont {Wang}, \citenamefont {Wang}, \citenamefont {Wang}, \citenamefont
  {Wang}, \citenamefont {Wang}, \citenamefont {Wang}, \citenamefont {Wang}, \citenamefont {Wang}, \citenamefont {Wang}, \citenamefont {Wang}, \citenamefont {Wang}, \citenamefont {Wang}, \citenamefont {Wei}, \citenamefont {Wen}, \citenamefont {Wiebusch}, \citenamefont {Wonsak}, \citenamefont {Wu}, \citenamefont {Wulz}, \citenamefont {Wurm}, \citenamefont {Xi}, \citenamefont {Xia}, \citenamefont {Xie}, \citenamefont {zhong Xing}, \citenamefont {Xu}, \citenamefont {Yan}, \citenamefont {Yang}, \citenamefont {Yang}, \citenamefont {Yang}, \citenamefont {Yang}, \citenamefont {Yang}, \citenamefont {Yao}, \citenamefont {Yegin}, \citenamefont {Yermia}, \citenamefont {You}, \citenamefont {Yu}, \citenamefont {Yu}, \citenamefont {Yu}, \citenamefont {Zavatarelli}, \citenamefont {Zhan}, \citenamefont {Zhang}, \citenamefont {Zhang}, \citenamefont {Zhang}, \citenamefont {Zhang}, \citenamefont {Zhang}, \citenamefont {Zhang}, \citenamefont {Zhang}, \citenamefont {Zhao}, \citenamefont {Zheng}, \citenamefont {Zhong},
  \citenamefont {Zhou}, \citenamefont {Zhou}, \citenamefont {Zhou}, \citenamefont {Zhou}, \citenamefont {Zhou}, \citenamefont {Zhou}, \citenamefont {Zhou}, \citenamefont {Zhou}, \citenamefont {Zhou},\ and\ \citenamefont {Zou}}]{An2016}%
  \BibitemOpen
  \bibfield  {author} {\bibinfo {author} {\bibfnamefont {F.}~\bibnamefont {An}}, \bibinfo {author} {\bibfnamefont {G.}~\bibnamefont {An}}, \bibinfo {author} {\bibfnamefont {Q.}~\bibnamefont {An}}, \bibinfo {author} {\bibfnamefont {V.}~\bibnamefont {Antonelli}}, \bibinfo {author} {\bibfnamefont {E.}~\bibnamefont {Baussan}}, \bibinfo {author} {\bibfnamefont {J.}~\bibnamefont {Beacom}}, \bibinfo {author} {\bibfnamefont {L.}~\bibnamefont {Bezrukov}}, \bibinfo {author} {\bibfnamefont {S.}~\bibnamefont {Blyth}}, \bibinfo {author} {\bibfnamefont {R.}~\bibnamefont {Brugnera}}, \bibinfo {author} {\bibfnamefont {M.~B.}\ \bibnamefont {Avanzini}}, \bibinfo {author} {\bibfnamefont {J.}~\bibnamefont {Busto}}, \bibinfo {author} {\bibfnamefont {A.}~\bibnamefont {Cabrera}}, \bibinfo {author} {\bibfnamefont {H.}~\bibnamefont {Cai}}, \bibinfo {author} {\bibfnamefont {X.}~\bibnamefont {Cai}}, \bibinfo {author} {\bibfnamefont {A.}~\bibnamefont {Cammi}}, \bibinfo {author} {\bibfnamefont {G.}~\bibnamefont {Cao}}, \bibinfo {author}
  {\bibfnamefont {J.}~\bibnamefont {Cao}}, \bibinfo {author} {\bibfnamefont {Y.}~\bibnamefont {Chang}}, \bibinfo {author} {\bibfnamefont {S.}~\bibnamefont {Chen}}, \bibinfo {author} {\bibfnamefont {S.}~\bibnamefont {Chen}}, \bibinfo {author} {\bibfnamefont {Y.}~\bibnamefont {Chen}}, \bibinfo {author} {\bibfnamefont {D.}~\bibnamefont {Chiesa}}, \bibinfo {author} {\bibfnamefont {M.}~\bibnamefont {Clemenza}}, \bibinfo {author} {\bibfnamefont {B.}~\bibnamefont {Clerbaux}}, \bibinfo {author} {\bibfnamefont {J.}~\bibnamefont {Conrad}}, \bibinfo {author} {\bibfnamefont {D.}~\bibnamefont {D’Angelo}}, \bibinfo {author} {\bibfnamefont {H.~D.}\ \bibnamefont {Kerret}}, \bibinfo {author} {\bibfnamefont {Z.}~\bibnamefont {Deng}}, \bibinfo {author} {\bibfnamefont {Z.}~\bibnamefont {Deng}}, \bibinfo {author} {\bibfnamefont {Y.}~\bibnamefont {Ding}}, \bibinfo {author} {\bibfnamefont {Z.}~\bibnamefont {Djurcic}}, \bibinfo {author} {\bibfnamefont {D.}~\bibnamefont {Dornic}}, \bibinfo {author} {\bibfnamefont {M.}~\bibnamefont
  {Dracos}}, \bibinfo {author} {\bibfnamefont {O.}~\bibnamefont {Drapier}}, \bibinfo {author} {\bibfnamefont {S.}~\bibnamefont {Dusini}}, \bibinfo {author} {\bibfnamefont {S.}~\bibnamefont {Dye}}, \bibinfo {author} {\bibfnamefont {T.}~\bibnamefont {Enqvist}}, \bibinfo {author} {\bibfnamefont {D.}~\bibnamefont {Fan}}, \bibinfo {author} {\bibfnamefont {J.}~\bibnamefont {Fang}}, \bibinfo {author} {\bibfnamefont {L.}~\bibnamefont {Favart}}, \bibinfo {author} {\bibfnamefont {R.}~\bibnamefont {Ford}}, \bibinfo {author} {\bibfnamefont {M.}~\bibnamefont {Göger-Neff}}, \bibinfo {author} {\bibfnamefont {H.}~\bibnamefont {Gan}}, \bibinfo {author} {\bibfnamefont {A.}~\bibnamefont {Garfagnini}}, \bibinfo {author} {\bibfnamefont {M.}~\bibnamefont {Giammarchi}}, \bibinfo {author} {\bibfnamefont {M.}~\bibnamefont {Gonchar}}, \bibinfo {author} {\bibfnamefont {G.}~\bibnamefont {Gong}}, \bibinfo {author} {\bibfnamefont {H.}~\bibnamefont {Gong}}, \bibinfo {author} {\bibfnamefont {M.}~\bibnamefont {Gonin}}, \bibinfo {author}
  {\bibfnamefont {M.}~\bibnamefont {Grassi}}, \bibinfo {author} {\bibfnamefont {C.}~\bibnamefont {Grewing}}, \bibinfo {author} {\bibfnamefont {M.}~\bibnamefont {Guan}}, \bibinfo {author} {\bibfnamefont {V.}~\bibnamefont {Guarino}}, \bibinfo {author} {\bibfnamefont {G.}~\bibnamefont {Guo}}, \bibinfo {author} {\bibfnamefont {W.}~\bibnamefont {Guo}}, \bibinfo {author} {\bibfnamefont {X.-H.}\ \bibnamefont {Guo}}, \bibinfo {author} {\bibfnamefont {C.}~\bibnamefont {Hagner}}, \bibinfo {author} {\bibfnamefont {R.}~\bibnamefont {Han}}, \bibinfo {author} {\bibfnamefont {M.}~\bibnamefont {He}}, \bibinfo {author} {\bibfnamefont {Y.}~\bibnamefont {Heng}}, \bibinfo {author} {\bibfnamefont {Y.}~\bibnamefont {Hsiung}}, \bibinfo {author} {\bibfnamefont {J.}~\bibnamefont {Hu}}, \bibinfo {author} {\bibfnamefont {S.}~\bibnamefont {Hu}}, \bibinfo {author} {\bibfnamefont {T.}~\bibnamefont {Hu}}, \bibinfo {author} {\bibfnamefont {H.}~\bibnamefont {Huang}}, \bibinfo {author} {\bibfnamefont {X.}~\bibnamefont {Huang}}, \bibinfo
  {author} {\bibfnamefont {L.}~\bibnamefont {Huo}}, \bibinfo {author} {\bibfnamefont {A.}~\bibnamefont {Ioannisian}}, \bibinfo {author} {\bibfnamefont {M.}~\bibnamefont {Jeitler}}, \bibinfo {author} {\bibfnamefont {X.}~\bibnamefont {Ji}}, \bibinfo {author} {\bibfnamefont {X.}~\bibnamefont {Jiang}}, \bibinfo {author} {\bibfnamefont {C.}~\bibnamefont {Jollet}}, \bibinfo {author} {\bibfnamefont {L.}~\bibnamefont {Kang}}, \bibinfo {author} {\bibfnamefont {M.}~\bibnamefont {Karagounis}}, \bibinfo {author} {\bibfnamefont {N.}~\bibnamefont {Kazarian}}, \bibinfo {author} {\bibfnamefont {Z.}~\bibnamefont {Krumshteyn}}, \bibinfo {author} {\bibfnamefont {A.}~\bibnamefont {Kruth}}, \bibinfo {author} {\bibfnamefont {P.}~\bibnamefont {Kuusiniemi}}, \bibinfo {author} {\bibfnamefont {T.}~\bibnamefont {Lachenmaier}}, \bibinfo {author} {\bibfnamefont {R.}~\bibnamefont {Leitner}}, \bibinfo {author} {\bibfnamefont {C.}~\bibnamefont {Li}}, \bibinfo {author} {\bibfnamefont {J.}~\bibnamefont {Li}}, \bibinfo {author} {\bibfnamefont
  {W.}~\bibnamefont {Li}}, \bibinfo {author} {\bibfnamefont {W.}~\bibnamefont {Li}}, \bibinfo {author} {\bibfnamefont {X.}~\bibnamefont {Li}}, \bibinfo {author} {\bibfnamefont {X.}~\bibnamefont {Li}}, \bibinfo {author} {\bibfnamefont {Y.}~\bibnamefont {Li}}, \bibinfo {author} {\bibfnamefont {Y.}~\bibnamefont {Li}}, \bibinfo {author} {\bibfnamefont {Z.-B.}\ \bibnamefont {Li}}, \bibinfo {author} {\bibfnamefont {H.}~\bibnamefont {Liang}}, \bibinfo {author} {\bibfnamefont {G.-L.}\ \bibnamefont {Lin}}, \bibinfo {author} {\bibfnamefont {T.}~\bibnamefont {Lin}}, \bibinfo {author} {\bibfnamefont {Y.-H.}\ \bibnamefont {Lin}}, \bibinfo {author} {\bibfnamefont {J.}~\bibnamefont {Ling}}, \bibinfo {author} {\bibfnamefont {I.}~\bibnamefont {Lippi}}, \bibinfo {author} {\bibfnamefont {D.}~\bibnamefont {Liu}}, \bibinfo {author} {\bibfnamefont {H.}~\bibnamefont {Liu}}, \bibinfo {author} {\bibfnamefont {H.}~\bibnamefont {Liu}}, \bibinfo {author} {\bibfnamefont {J.}~\bibnamefont {Liu}}, \bibinfo {author} {\bibfnamefont
  {J.}~\bibnamefont {Liu}}, \bibinfo {author} {\bibfnamefont {J.}~\bibnamefont {Liu}}, \bibinfo {author} {\bibfnamefont {Q.}~\bibnamefont {Liu}}, \bibinfo {author} {\bibfnamefont {S.}~\bibnamefont {Liu}}, \bibinfo {author} {\bibfnamefont {S.}~\bibnamefont {Liu}}, \bibinfo {author} {\bibfnamefont {P.}~\bibnamefont {Lombardi}}, \bibinfo {author} {\bibfnamefont {Y.}~\bibnamefont {Long}}, \bibinfo {author} {\bibfnamefont {H.}~\bibnamefont {Lu}}, \bibinfo {author} {\bibfnamefont {J.}~\bibnamefont {Lu}}, \bibinfo {author} {\bibfnamefont {J.}~\bibnamefont {Lu}}, \bibinfo {author} {\bibfnamefont {J.}~\bibnamefont {Lu}}, \bibinfo {author} {\bibfnamefont {B.}~\bibnamefont {Lubsandorzhiev}}, \bibinfo {author} {\bibfnamefont {L.}~\bibnamefont {Ludhova}}, \bibinfo {author} {\bibfnamefont {S.}~\bibnamefont {Luo}}, \bibinfo {author} {\bibfnamefont {V.}~\bibnamefont {Lyashuk}}, \bibinfo {author} {\bibfnamefont {R.}~\bibnamefont {Möllenberg}}, \bibinfo {author} {\bibfnamefont {X.}~\bibnamefont {Ma}}, \bibinfo {author}
  {\bibfnamefont {F.}~\bibnamefont {Mantovani}}, \bibinfo {author} {\bibfnamefont {Y.}~\bibnamefont {Mao}}, \bibinfo {author} {\bibfnamefont {S.~M.}\ \bibnamefont {Mari}}, \bibinfo {author} {\bibfnamefont {W.~F.}\ \bibnamefont {McDonough}}, \bibinfo {author} {\bibfnamefont {G.}~\bibnamefont {Meng}}, \bibinfo {author} {\bibfnamefont {A.}~\bibnamefont {Meregaglia}}, \bibinfo {author} {\bibfnamefont {E.}~\bibnamefont {Meroni}}, \bibinfo {author} {\bibfnamefont {M.}~\bibnamefont {Mezzetto}}, \bibinfo {author} {\bibfnamefont {L.}~\bibnamefont {Miramonti}}, \bibinfo {author} {\bibfnamefont {T.}~\bibnamefont {Mueller}}, \bibinfo {author} {\bibfnamefont {D.}~\bibnamefont {Naumov}}, \bibinfo {author} {\bibfnamefont {L.}~\bibnamefont {Oberauer}}, \bibinfo {author} {\bibfnamefont {J.~P.}\ \bibnamefont {Ochoa-Ricoux}}, \bibinfo {author} {\bibfnamefont {A.}~\bibnamefont {Olshevskiy}}, \bibinfo {author} {\bibfnamefont {F.}~\bibnamefont {Ortica}}, \bibinfo {author} {\bibfnamefont {A.}~\bibnamefont {Paoloni}}, \bibinfo
  {author} {\bibfnamefont {H.}~\bibnamefont {Peng}}, \bibinfo {author} {\bibfnamefont {J.-C.}\ \bibnamefont {Peng}}, \bibinfo {author} {\bibfnamefont {E.}~\bibnamefont {Previtali}}, \bibinfo {author} {\bibfnamefont {M.}~\bibnamefont {Qi}}, \bibinfo {author} {\bibfnamefont {S.}~\bibnamefont {Qian}}, \bibinfo {author} {\bibfnamefont {X.}~\bibnamefont {Qian}}, \bibinfo {author} {\bibfnamefont {Y.}~\bibnamefont {Qian}}, \bibinfo {author} {\bibfnamefont {Z.}~\bibnamefont {Qin}}, \bibinfo {author} {\bibfnamefont {G.}~\bibnamefont {Raffelt}}, \bibinfo {author} {\bibfnamefont {G.}~\bibnamefont {Ranucci}}, \bibinfo {author} {\bibfnamefont {B.}~\bibnamefont {Ricci}}, \bibinfo {author} {\bibfnamefont {M.}~\bibnamefont {Robens}}, \bibinfo {author} {\bibfnamefont {A.}~\bibnamefont {Romani}}, \bibinfo {author} {\bibfnamefont {X.}~\bibnamefont {Ruan}}, \bibinfo {author} {\bibfnamefont {X.}~\bibnamefont {Ruan}}, \bibinfo {author} {\bibfnamefont {G.}~\bibnamefont {Salamanna}}, \bibinfo {author} {\bibfnamefont
  {M.}~\bibnamefont {Shaevitz}}, \bibinfo {author} {\bibfnamefont {V.}~\bibnamefont {Sinev}}, \bibinfo {author} {\bibfnamefont {C.}~\bibnamefont {Sirignano}}, \bibinfo {author} {\bibfnamefont {M.}~\bibnamefont {Sisti}}, \bibinfo {author} {\bibfnamefont {O.}~\bibnamefont {Smirnov}}, \bibinfo {author} {\bibfnamefont {M.}~\bibnamefont {Soiron}}, \bibinfo {author} {\bibfnamefont {A.}~\bibnamefont {Stahl}}, \bibinfo {author} {\bibfnamefont {L.}~\bibnamefont {Stanco}}, \bibinfo {author} {\bibfnamefont {J.}~\bibnamefont {Steinmann}}, \bibinfo {author} {\bibfnamefont {X.}~\bibnamefont {Sun}}, \bibinfo {author} {\bibfnamefont {Y.}~\bibnamefont {Sun}}, \bibinfo {author} {\bibfnamefont {D.}~\bibnamefont {Taichenachev}}, \bibinfo {author} {\bibfnamefont {J.}~\bibnamefont {Tang}}, \bibinfo {author} {\bibfnamefont {I.}~\bibnamefont {Tkachev}}, \bibinfo {author} {\bibfnamefont {W.}~\bibnamefont {Trzaska}}, \bibinfo {author} {\bibfnamefont {S.~v.}\ \bibnamefont {Waasen}}, \bibinfo {author} {\bibfnamefont {C.}~\bibnamefont
  {Volpe}}, \bibinfo {author} {\bibfnamefont {V.}~\bibnamefont {Vorobel}}, \bibinfo {author} {\bibfnamefont {L.}~\bibnamefont {Votano}}, \bibinfo {author} {\bibfnamefont {C.-H.}\ \bibnamefont {Wang}}, \bibinfo {author} {\bibfnamefont {G.}~\bibnamefont {Wang}}, \bibinfo {author} {\bibfnamefont {H.}~\bibnamefont {Wang}}, \bibinfo {author} {\bibfnamefont {M.}~\bibnamefont {Wang}}, \bibinfo {author} {\bibfnamefont {R.}~\bibnamefont {Wang}}, \bibinfo {author} {\bibfnamefont {S.}~\bibnamefont {Wang}}, \bibinfo {author} {\bibfnamefont {W.}~\bibnamefont {Wang}}, \bibinfo {author} {\bibfnamefont {Y.}~\bibnamefont {Wang}}, \bibinfo {author} {\bibfnamefont {Y.}~\bibnamefont {Wang}}, \bibinfo {author} {\bibfnamefont {Y.}~\bibnamefont {Wang}}, \bibinfo {author} {\bibfnamefont {Z.}~\bibnamefont {Wang}}, \bibinfo {author} {\bibfnamefont {Z.}~\bibnamefont {Wang}}, \bibinfo {author} {\bibfnamefont {Z.}~\bibnamefont {Wang}}, \bibinfo {author} {\bibfnamefont {Z.}~\bibnamefont {Wang}}, \bibinfo {author} {\bibfnamefont
  {W.}~\bibnamefont {Wei}}, \bibinfo {author} {\bibfnamefont {L.}~\bibnamefont {Wen}}, \bibinfo {author} {\bibfnamefont {C.}~\bibnamefont {Wiebusch}}, \bibinfo {author} {\bibfnamefont {B.}~\bibnamefont {Wonsak}}, \bibinfo {author} {\bibfnamefont {Q.}~\bibnamefont {Wu}}, \bibinfo {author} {\bibfnamefont {C.-E.}\ \bibnamefont {Wulz}}, \bibinfo {author} {\bibfnamefont {M.}~\bibnamefont {Wurm}}, \bibinfo {author} {\bibfnamefont {Y.}~\bibnamefont {Xi}}, \bibinfo {author} {\bibfnamefont {D.}~\bibnamefont {Xia}}, \bibinfo {author} {\bibfnamefont {Y.}~\bibnamefont {Xie}}, \bibinfo {author} {\bibfnamefont {Z.}~\bibnamefont {zhong Xing}}, \bibinfo {author} {\bibfnamefont {J.}~\bibnamefont {Xu}}, \bibinfo {author} {\bibfnamefont {B.}~\bibnamefont {Yan}}, \bibinfo {author} {\bibfnamefont {C.}~\bibnamefont {Yang}}, \bibinfo {author} {\bibfnamefont {C.}~\bibnamefont {Yang}}, \bibinfo {author} {\bibfnamefont {G.}~\bibnamefont {Yang}}, \bibinfo {author} {\bibfnamefont {L.}~\bibnamefont {Yang}}, \bibinfo {author}
  {\bibfnamefont {Y.}~\bibnamefont {Yang}}, \bibinfo {author} {\bibfnamefont {Y.}~\bibnamefont {Yao}}, \bibinfo {author} {\bibfnamefont {U.}~\bibnamefont {Yegin}}, \bibinfo {author} {\bibfnamefont {F.}~\bibnamefont {Yermia}}, \bibinfo {author} {\bibfnamefont {Z.}~\bibnamefont {You}}, \bibinfo {author} {\bibfnamefont {B.}~\bibnamefont {Yu}}, \bibinfo {author} {\bibfnamefont {C.}~\bibnamefont {Yu}}, \bibinfo {author} {\bibfnamefont {Z.}~\bibnamefont {Yu}}, \bibinfo {author} {\bibfnamefont {S.}~\bibnamefont {Zavatarelli}}, \bibinfo {author} {\bibfnamefont {L.}~\bibnamefont {Zhan}}, \bibinfo {author} {\bibfnamefont {C.}~\bibnamefont {Zhang}}, \bibinfo {author} {\bibfnamefont {H.-H.}\ \bibnamefont {Zhang}}, \bibinfo {author} {\bibfnamefont {J.}~\bibnamefont {Zhang}}, \bibinfo {author} {\bibfnamefont {J.}~\bibnamefont {Zhang}}, \bibinfo {author} {\bibfnamefont {Q.}~\bibnamefont {Zhang}}, \bibinfo {author} {\bibfnamefont {Y.-M.}\ \bibnamefont {Zhang}}, \bibinfo {author} {\bibfnamefont {Z.}~\bibnamefont {Zhang}},
  \bibinfo {author} {\bibfnamefont {Z.}~\bibnamefont {Zhao}}, \bibinfo {author} {\bibfnamefont {Y.}~\bibnamefont {Zheng}}, \bibinfo {author} {\bibfnamefont {W.}~\bibnamefont {Zhong}}, \bibinfo {author} {\bibfnamefont {G.}~\bibnamefont {Zhou}}, \bibinfo {author} {\bibfnamefont {J.}~\bibnamefont {Zhou}}, \bibinfo {author} {\bibfnamefont {L.}~\bibnamefont {Zhou}}, \bibinfo {author} {\bibfnamefont {R.}~\bibnamefont {Zhou}}, \bibinfo {author} {\bibfnamefont {S.}~\bibnamefont {Zhou}}, \bibinfo {author} {\bibfnamefont {W.}~\bibnamefont {Zhou}}, \bibinfo {author} {\bibfnamefont {X.}~\bibnamefont {Zhou}}, \bibinfo {author} {\bibfnamefont {Y.}~\bibnamefont {Zhou}}, \bibinfo {author} {\bibfnamefont {Y.}~\bibnamefont {Zhou}}, \ and\ \bibinfo {author} {\bibfnamefont {J.}~\bibnamefont {Zou}},\ }\bibfield  {title} {\enquote {\bibinfo {title} {Neutrino physics with {JUNO}},}\ }\href {\doibase 10.1088/0954-3899/43/3/030401} {\bibfield  {journal} {\bibinfo  {journal} {J. Phys. G: Nucl. Part. Phys.}\ }\textbf {\bibinfo {volume}
  {43}},\ \bibinfo {pages} {030401} (\bibinfo {year} {2016})}\BibitemShut {NoStop}%
\end{thebibliography}%

\end{document}